\documentclass[%
mathleft,%
final,%
]{an}
\usepackage{graphicx}
\usepackage{times}
\newcommand{\Halpha}{H$\alpha$}
\newcommand{\kms}{km\,s$^{-1}$}
\newcommand{\ms}{m\,s$^{-1}$}

\begin{document}

\Pagespan{1}{}
\Yearpublication{2012}%
\Yearsubmission{2012}%
\Month{8}%
\Volume{333}%
\Issue{8}%
\DOI{This.is/not.aDOI}%

\title{Rotation, activity, and lithium abundance in cool binary stars\thanks{Based on data obtained with the STELLA robotic
telescopes in Tenerife, an AIP facility jointly operated with IAC,
and the Automatic Photoelectric Telescopes in Arizona, jointly
operated with Fairborn Observatory.}}

\author{K.~G.~Strassmeier\thanks{Corresponding author.
\email{kstrassmeier@aip.de}}, M.~Weber, T.~Granzer, \and S.~J\"arvinen}
\authorrunning{K. G. Strassmeier, M.~Weber, T.~Granzer, S.~J\"arvinen}
\institute{Leibniz-Institut f\"ur Astrophysik Potsdam (AIP), An der
Sternwarte 16, D-14482 Potsdam, Germany}

\received{2012}
\accepted{2012}
\publonline{2012}


\abstract{%
We have used two robotic telescopes to obtain time-series
high-resolution optical echelle spectroscopy and $VI$ and/or $by$
photometry for a sample of 60 active stars, mostly binaries.
Orbital solutions are presented for 26 double-lined systems and
for 19 single-lined systems, seven of them for the first time but
all of them with unprecedented phase coverage and accuracy.
Eighteen systems turned out to be single stars. The total of 6,609 $R$=55,000 echelle spectra are also used to systematically
determine effective temperatures, gravities, metallicities,
rotational velocities, lithium abundances and absolute \Halpha
-core fluxes as a function of time. The photometry is used to
infer unspotted brightness, $V-I$ and/or $b-y$ colors,
spot-induced brightness amplitudes and precise rotation periods.
An extra 22 radial-velocity standard stars were monitored
throughout the science observations and yield a new barycentric
zero point for our STELLA/SES robotic system. Our data are
complemented by literature data and are used to determine
rotation-temperature-activity relations for active binary
components. We also relate lithium abundance to rotation and
surface temperature. We find that 74\% of all known
rapidly-rotating active binary stars are synchronized and in
circular orbits but 26\% (61 systems) are rotating asynchronously
of which half have $P_{\rm rot}>P_{\rm orb}$ and $e>0$. Because
rotational synchronization is predicted to occur before orbital
circularization active binaries should undergo an extra spin-down
besides tidal dissipation. We suspect this to be due to a
magnetically channeled wind with its subsequent braking torque.
We find a steep increase of rotation period with decreasing
effective temperature for active stars, $P_{\rm rot} \propto
T_{\rm eff}^{-7}$, for both single and binaries, main sequence and
evolved. For inactive, single giants with $P_{\rm rot}>100$~d, the
relation is much weaker, $P_{\rm rot} \propto T_{\rm
eff}^{-1.12}$. Our data also indicate a period-activity
relation for \Halpha\ of the form $R_{\rm H\alpha} \propto P_{\rm
rot}^{-0.24}$ for binaries and $R_{\rm H\alpha} \propto P_{\rm
rot}^{-0.14}$ for singles. Its power-law difference is possibly
significant. Lithium abundances in our (field-star) sample
generally increase with effective temperature and are
paralleled with an increase of the dispersion. The
dispersion for binaries can be 1--2 orders of magnitude larger than
for singles, peaking at an absolute spread of 3 orders of
magnitude near $T_{\rm eff}\approx$5000~K. On average, binaries of comparable effective temperature appear to exhibit 0.25~dex less surface lithium than singles, as expected if the depletion mechanism is rotation dependent. We also find a trend of increased Li abundance with rotational period of form $\log n (\mathrm{Li}) \propto -0.6 \ \log P_{\rm rot}$ but again with a dispersion of as large as 3--4 orders of magnitude.}

\maketitle


\section{Introduction}

In a previous paper (Strassmeier et al.~\cite{ca2}; paper~I) we
reported on radial and rotational velocities, chromospheric emission-line fluxes, lithium abundances, and
rotation periods of a total sample of 1,058 G5--K2 dwarfs,
subgiants, and giants based on 1,429 moderate-resolution KPNO coud\'e
spectra and 8,038 Str\"omgren $y$ photometric data points. The aim
of this survey was to detect new candidates for Doppler imaging
but, besides the discovery of 170 new variable stars and 36 new
spectroscopic binaries, the more intriguing result was that 74\%\
of the G-K stars with Ca\,{\sc ii} H\&K emission also showed
significant lithium on their surface. However, G-K giants should
have very few lithium on their surface because of convective
mixing. Theoretical models predict that surface lithium has to be
diluted by many factors once a star arrives at the bottom of the
red giant branch (Iben \cite{iben67}, Charbonnel \& Balachandran
\cite{cha:bal}). Out of the 21 Doppler imaging candidates found,
just four stars were single stars, three of them evolved, the rest
were spectroscopic binaries, but all four single stars had very
strong lithium.

Despite that it is generally acknowledged that higher than normal
lithium abundance is common among magnetically active stars, no
unique correlation with rotation rate was found after Skumanich's
(\cite{sku}) original discovery. Recently, White et al.
(\cite{white2007}) revisited this issue in their sample of
solar-type dwarfs but no such correlation was found. Almost all
surveys just revealed trends, if at all, and even these appear to be
of different quality (e.g. L\'ebre et al.~\cite{lebre2006};
B\"ohm-Vitense~\cite{boehm}; do Nascimento et al.~\cite{dona2000},
\cite{dona2003}; De Medeiros et al.~\cite{dem2000}; De Laverny et
al.~\cite{dela2003}; Randich et al. \cite{ran:gra}). The
comprehensive survey of nearby giants by Luck \& Heiter
(\cite{luc:hei}) did not even show a trend. However, the line
broadening in their stellar sample was just 3--7~\kms\ and likely
too narrow a range to see a trend. B\"ohm-Vitense~(\cite{boehm})
suggested that the steep decrease of $v\sin i$ in early G giants as
well as in Hyades F dwarfs at effective temperatures cooler than
$\approx$6,450~K, i.e. at their lithium dip at about the same temperature,
are the result of deep mixing and related to the merging of the
hydrogen and the helium convection zones. More recently, Takeda et
al. (\cite{tak:hon}) announced evidence for a (positive) correlation
of Li abundance with rotational velocity in a sample of solar-analog stars.

The spectroscopic survey of 390 solar-like dwarf stars by White et
al.~(\cite{white2007}) included 28 of the stars in our survey.
Relating Ca\,{\sc ii} H\&K radiative losses to stellar rotation,
White et al.~(\cite{white2007}) found a saturation of chromospheric
emission for rotational velocities above approximately 30~\kms . In
an earlier paper, Strassmeier et al.~(\cite{str:han}) verified that
evolved stars obey qualitatively the same scaling of Ca\,{\sc
ii}-K-line flux with stellar rotational velocity or period as do
main-sequence stars (see, e.g., Mamajek \& Hillenbrand
\cite{mam:hil}, Pace \& Pasquini~\cite{pac:pas}, or Pizzolato et
al.~\cite{piz:mag} for a summary). No qualitative difference was
found between single evolved stars and their equally rapidly
rotating counterparts in a spectroscopic binary. However, large
scatter indicated that rotation might not be the only relevant
parameter. Based on a sample of 22 intermediate-mass G and K giants
in close binaries, Gondoin (\cite{gon}) not only verified the
rotational dependency of (coronal) X-ray surface flux but also found
a dependency on surface gravity. Such a dependence could stem from
the effect of gravity on coronal electron density and on the overall
sizes of coronal loops.

Massarotti et al. (\cite{mas:lat}) reported rotational and radial
velocities for 761 giants within 100~pc of the Sun. They found that
all binaries in their sample with periods less than 20 days have
circular orbits while about half the orbits with periods between
20--100 days still showed significant eccentricity. They also found
evidence that the rotational velocity of horizontal branch stars is
larger than that of first-ascend giants by a few \kms . Earlier,
De~Medeiros et al. (\cite{dem2002}) presented a study of 134
late-type giants in spectroscopic binaries and found a considerable
number of G-K giant stars with moderate to moderately-high rotation
rates. These rotators have orbital periods shorter than 250 days and
circular or nearly circular orbits and appear to be synchronized
with the orbit.

The present paper follows up on the newly identified spectroscopic
binaries with active components from our paper~I. Its direct aim
is to determine their orbits on the basis of high-precision radial
velocities and to separate their component's rotation and activity
tracers along with other absolute astrophysical parameters. Only
with precise stellar parameters can we directly compare binary
components with single stars and then be aware of the spectrum
contamination from unknown secondaries or even tertiary stars. We
recall that an unknown continuum contribution from a secondary
star impacts on the determination of the effective temperature,
gravity etc. and could together drastically alter the derived
lithium abundances and thereby mask any relation if present. In
Sect.~\ref{S2} we restate our sample selection criteria and give a
summary of the target stars. In Sect.~\ref{S3} we describe the new
observations and in Sect.~\ref{S4} we derive basic quantities from
the spectra and the light curves. These include radial velocities,
orbital parameters, rotational velocities and photometric periods,
stellar atmospheric parameters like temperature, gravity and
metallicity, lithium abundances, and absolute \Halpha -core
fluxes. Sect.~\ref{S5} lists notes to individual stars.
Sect.~\ref{S6} presents the analysis in terms of rotation,
temperature, activity, and lithium-abundance relations. Finally,
Sect.~\ref{S7} summarizes our findings and conclusions.

\section{Sample selection}\label{S2}

Our sample selection is based on the 1,058 stars from the KPNO
Doppler-imaging candidate survey in paper~I. It itself was drawn
from a total of 6,440 stars from the Hipparcos catalog (ESA
\cite{hip}, van Leeuwen \cite{nhip}) for the brightness range
7\fm0--9\fm5 and declination $-30^\circ$ through +70$^\circ$, $B-V$
colors between 0\fm67 and 1\fm0 for stars with parallaxes
$\pi>20$~mas (i.e. G5--K3 dwarfs) and between 0\fm87 and 1\fm2 for
$3<\pi<20$~mas (i.e. G5--K2 giants and subgiants). These criteria
were imposed to select stars with a significant outer convective
envelope where the likelihood of detecting magnetic activity is
highest. Out of the 1,058 stars, 371 (35\%) were found with Ca\,{\sc
ii} H\&K emission but only 78 (7.3\%) with $v\sin i\ge$ 10 \kms . On
the contrary, a lithium line was detected in 283 (74\%) of all stars
that had Ca\,{\sc ii} emission (with 58\%\ of the stars with lithium
above 10 m\AA ). Out of a subsample of 172 stars with moderate to
strong Ca\,{\sc ii} emission, 168 (97.7\%) turned out to be
photometric variable and for 134 a photometric (rotational) period
could be obtained. Finally, 36 targets were single-lined
spectroscopic binaries (SB1), of which 17 were new detections. A
further 16 targets were found ``possible SB1s''. An additional 30
targets were double-lined spectroscopic binaries (SB2), of which 19
were new detections. Two targets were even triple lined (SB3) of
which one was a new detection and another four were new candidates.
All along, there were a few misidentifications and
misinterpretations as well as unrecognized literature entries.
Whenever recognized, we try to clarify these in the present paper.

\begin{figure*}[tb]
\center
\includegraphics[angle=0,width=165mm,clip]{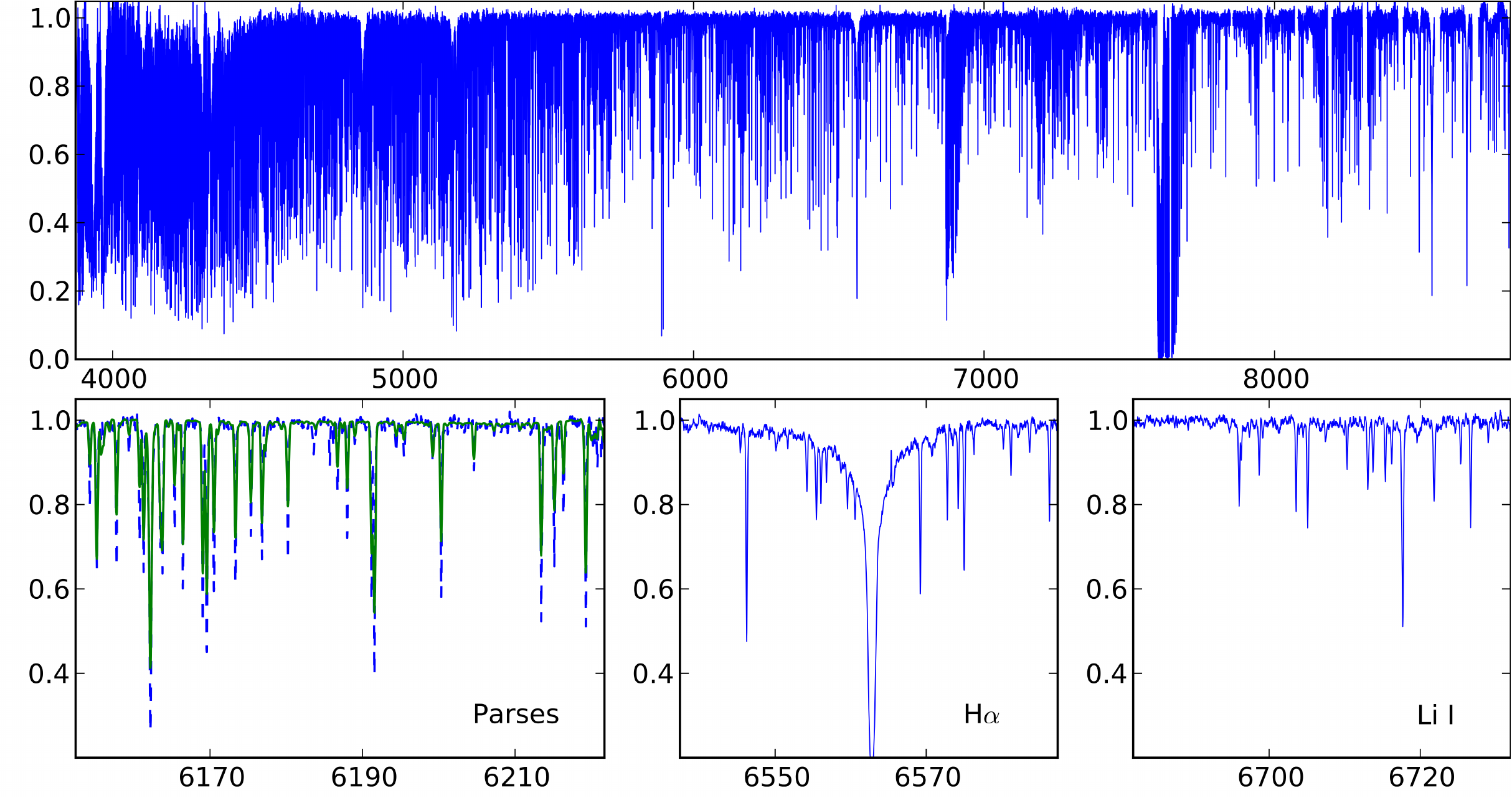}
\caption{HD~171067 as an example for a STELLA/SES spectrum. The top
panel shows the entire wavelength coverage from 3880--8820~\AA .
Note the increasing inter-order gaps starting around 7340~\AA .
The lower panels zoom into three wavelength regions employed in
this paper for analysis; from left to right, one of the five
\'echelle orders that are used for the PARSES spectrum-synthesis
fit (thick line), the Balmer \Halpha\ line, and the region around
the neutral lithium line at 6708~\AA . HD~171067 ($V$=7\fm2) is an
inactive single, slightly evolved, G8 star with $v\sin i<3$~\kms .
Exposure time was 1200~s. }\label{F1}
\end{figure*}

We now present time-series observations of all those targets in
the survey that were previously unknown or suspected spectroscopic
binaries in 2000. This sample comprised 59 stars presented in this
paper. One additional star was added because of its comparable
uncertain orbit despite being a fourth magnitude star
($\epsilon$~UMi). During the final observational stages for the
present paper we learned that Griffin (\cite{griff09}), Griffin
(\cite{griff10}) and Griffin \& Filiz~Ak (\cite{gri:fil}) had
picked up many of our original SB candidate stars from paper~I
that found their way into the new edition of the CABS catalog
(Eker et al.~\cite{cabs3}), and independently determined orbital
elements. We compare our results whenever possible. Stellar
identifications and some basic observable properties for all our
target stars are summarized in Table~\ref{T1}. Note that three of
the stars actually do not show significant Ca\,{\sc ii} H\&K
emission and would not be dubbed magnetically active but are left
in the sample because they were monitored initially in order to
search for signs of binarity. All three turned out to be single
stars though. Altogether, 18 of the SB1 candidates were found to
be single with some of them still members of a visual-binary
system. These stars are also in Table~\ref{T1} but are summarized
in a separate table later in the paper. In one case the two
components of a wide visual binary (BD+11\,2052AB = ADS\,7406AB)
were treated as separate stars throughout the paper. One
double-lined binary (HD~16884) turned out to be actually a
quadruple system with both pairs being SB1. First orbit
determinations are presented for seven systems (HD~50255,
HD~82841, HD~106855, HD~147866, HD~190642, HD~199967, and
HD~226099).

\section{New observations and data reductions}\label{S3}

\subsection{High-resolution optical spectroscopy}

Time-series high-resolution echelle spectroscopy was taken with
the 1.2\,m STELLA-I telescope between June 2006 and May 2012. Most
spectra were exposed just long enough to measure a precise radial
velocity and had S/N of between 40--80:1 but several spectra per
target were exposed to reach S/N well above 100:1. A total of
6,609 spectra for a total of 60 stars were obtained over the
course of approximately six years. STELLA-I is a fully robotic
telescope that, together with STELLA-II, makes up the STELLA
observatory at the Iz\~{a}na ridge on Tenerife in the Canary
islands (Strassmeier et al.~\cite{stella}, \cite{malaga}). The
fiber-fed STELLA Echelle Spectrograph (SES) is the telescope's
only instrument. It is a white-pupil design with an R2 grating
with two off-axis collimators, a prism cross disperser and a
folded Schmidt camera with an E2V 2k$\times$2k CCD as the
detector. All spectra have a fixed format on the CCD and cover the
wavelength range from 388--882~nm with increasing inter-order gaps
near the red end starting at 734~nm towards 882~nm. The resolving
power is $R$=55,000 corresponding to a spectral resolution of
0.12~\AA \ at 650~nm (3-pixel sampling). An example spectrum is
shown in Fig.~\ref{F1}. We note that the SES received a major
upgrade in summer 2012 with a new cross disperser, a new optical
camera, and a new CCD. A bit earlier, the SES fiber was moved to
the prime focus of the second STELLA telescope in 2011. Further
details of the performance of the system were reported by Weber et
al. (\cite{spie}) and Granzer et al. (\cite{malaga2}).

\begin{table*}
\caption{Summary of program stars and name aliases. Listed are the
\emph{Tycho} $V$ brightness and the $B-V$ color in magnitudes, the spectral
classification if available, the binarity according to this paper
(S=single star, SB=spectrosopic binary, SB1=single-lined SB,
SB2=double-lined SB, SB3=triple-lined, VB=visual binary), the number
of STELLA spectra $N_{\rm STELLA}$ obtained and used for the orbit
determination, its time span covered ($\Delta t$), and a primary
reference if existent. }\label{T1}
\begin{tabular}{lllllllllll}
\hline \noalign{\smallskip}
HD & Var. name & HIP & SAO & $V$ &
$B-V$ & Sp.type & SB & $N_{\rm STELLA}$ & $\Delta t$ & Ref. \\
\noalign{\smallskip}\hline \noalign{\smallskip}
\object{HD 553}   & V741 Cas &  834 &  11013 & 8.17 & 1.03 & K0 & SB2 & 121 & 1320& s,34 \\
         & \object{LN Peg}   &  999 &  91772 & 8.59 & 0.81 & K0 & SB2+1 & 135 & 1302& r \\
\object{HD 8997}  & EO Psc &  6917 &  74742 & 7.74 & 0.96 & K2/K1-K6V & SB2 & 165 & 1453&  b \\
\object{HD 9902}  & BG Psc &      &  74827 & 8.71 & 0.63 & F5-6V/G9-K0IV & SB2 & 95 & 748& a,c\\
\object{HD 16884} &        &      &  110699 & 8.94 & 1.37 & K4III & 2$\times$SB1 & 83 & 1046& a,e\\
\object{HD 18645} & FU Cet &  13968 &  130230 & 7.86 & 0.75 & G2III-IV & S & 70 & 784& f\\
\object{HD 18955} & IR Eri &  14157 &  148731 & 8.45 & 0.82 & K0V/K2-3V & SB2 & 67 & 1446& g \\
\object{HD 23551} & MM Cam &  18012 &  12924 & 7.11 & 0.91 & K0III & S & 101 & 836& \\
\object{HD 24053} & &  17936 &  111446 & 7.7 & 0.8 & G0 & S & 77 & 803& \\
         & \object{AI Lep} & &  150676 & 8.97 & 0.57 & G6IV/G0V & S & 93 & 1460& l \\
         & \object{HY CMa} & &  151224 & 9.33 & 1.00 & K0-3V-IV/K1V-IV & SB2 & 87 & 1782& l \\
\object{HD 40891} & &  28935 &  13714 & 8.40 & 0.85 & G5 & SB1 & 85 & 1380&  h\\
\object{HD 43516} & &  29750 &  151290 & 7.37 & 0.85 & G8III & S & 57 & 830& \\
\object{HD 45762} & V723 Mon &  30891 &  133321 & 8.30 & 0.87 & G0 & SB2+1 & 89 & 1231& z\\
\object{HD 50255} & &  32971 &  152024 & 7.43 & 0.68 & G4.5V & SB2 & 66 & 1471& \\
\object{HD 61994} & &  38018 &  6310 & 7.08 & 0.67 & G0 & SB2 & 105 & 899& t \\
\object{HD 62668} & BM Lyn &  38003 &  41995 & 7.73 & 1.10 & K0III & SB1 & 137 & 2000& a,f,i \\
\object{HD 66553} & &  39515 &  97536 & 8.48 & 0.85 & G5 & SB1 & 92 & 2001& h \\
\object{HD 73512} & &  42418 &  116990 & 7.91 & 0.90 & K0V/K4V & SB2 & 74 & 892& a\\
\object{HD 76799} & &  44007 &  176747 & 7.11 & 0.99 & K0III & S & 75 &1442 & \\
         & & 46634 &  98614 & 8.76 & 0.86 & G5 & VB,S & 96 & 1129& y \\
\object{HD 82159} & GS Leo &  46637 &  98615 & 8.85 & 0.92 & G9V & VB,SB1 & 107 & 1458& u,32 \\
\object{HD 82286} & FF UMa &  46919 &  14919 & 7.89 & 0.96 & G5 & SB2 & 161 & 1446& i,31,35\\
\object{HD 82841} & OS Hya &  46987 &  136965 & 8.45 & 1.08 & K2 & SB1 & 61 & 1447& \\
         & \object{EQ Leo} &  50072 &  99011 & 9.39 & 1.09 & K1III & SB1 & 62 & 1169& a\\
\object{HD 93915} & &  53051 & 43492 & 8.07 & 0.68 & G5V/G6V & SB2 & 85 & 1169& a\\
\object{HD 95188} & XZ LMi &  53747 &  81610 & 8.45 & 0.74 & G8V & S & 85 & 891& v,j\\
\object{HD 95559} & GZ Leo &  53923 &  81634 & 8.83 & 0.96 & K0V/K2V & SB2 & 113 & 1162& k,30 \\
\object{HD 95724} & YY LMi &  54028 &  62375 & 8.96 & 0.94 & G5V & S & 76 & 856& v \\
\object{HD 104067} & &  58451 & 180353 & 7.93 & 0.99 & K2V & S & 76 & 887& \\
\object{HD 105575} & QY Hya &  59259 &  180519 & 9.04 & 0.93 & (K5/M1)/G4 & SB3 & 51 & 1164& 29\\
\object{HD 106855} & UV Crv &  59914 &  180648 & 9.59 & 0.81 & K1V & SB2,VB & 88 & 1227& \\
\object{HD 108564} & &  60853 & & 9.45 & 0.98 & K5V & S & 70 & 867& \\
\object{HD 109011} & NO UMa &  61100 &  28414 & 8.10 & 0.94 & K2V & SB2 & 184 & 1629& o \\
\object{HD 111487} & IM Vir & &  138983 & 9.69 & 0.64 & G5 & SB2 & 106 & 1292& p \\
\object{HD 112099} & &  62942 &  119652 & 8.23 & 0.86 & K1V & SB1 & 80 & 1142& a\\
          & &  63322 &  63275 & 9.27 & 0.84 & G6V & VB,SB1 & 104 & 937& y,32\\
\object{HD 112859} & BQ CVN &  63368 &  44410 & 8.09 & 0.92 & F5V/K0III-IV & SB2 & 93 & 1249& a,w\\
          & \object{CD CVn} &  63442 &  44421 & 9.39 & 1.19 & K0III & SB1 & 85 & 1165& a\\
\object{HD 120205} & &  67344 &  158174 & 8.3 & 0.9 & G5 & S & 93 & 849& \\
\object{HD 127068} & HK Boo &  70826 &  101044 & 8.43 & 0.89 & G8V/G5-8IV & SB2 & 99 & 1257& a\\
\object{HD 136655} & &  75132 &  83780 & 9.01 & 0.9 & K0 & S & 122 & 842& z\\
\object{HD 138157} & OX Ser &  75861 &  101580 & 7.14 & 1.02 & K0III & SB1 & 148 & 1086& u,32 \\
          & \object{V381 Ser} &  77210 &  121177 & 9.16 & 0.83 & K2V & SB2 & 96 & 1114& 33\\
\object{HD 142680} & V383 Ser &  77963 &  101816 & 8.71 & 0.95 & K0-2V/K7V & SB2 & 90 & 781& a,v \\
\object{HD 143937} & V1044 Sco &  78708 &  184077 & 8.65 & 0.91 & G9V/M0V & SB2 & 86 & 1131& l\\
\object{HD 147866} & V894 Her &  80302 &  84343 & 8.1 & 1.1 & K0 & SB1 & 115 & 1269& \\
\object{HD 150202} & GI Dra &  81284 &  30015 & 7.97 & 0.93 & K0III & SB1 & 179 & 1112& a\\
\object{HD 153525} & V1089 Her &  83006 &  46403 & 7.88 & 1.04 & K0V & S & 196 & 1349& x \\
\object{HD 153751} & $\epsilon$~UMi & 82080 & 2770 & 4.22 & 0.88 &
G5III & SB1 & 220 & 2144& d\\
\object{HD 155802} & &  84303 &  141567 & 8.51 & 0.89 & K3V & S & 114 & 857& \\
\object{HD 171067} & &  90864 &  103819 & 7.19 & 0.696 & G8V & S & 143 & 857& \\
\object{HD 184591} & &  96280 &  104991 & 7.37 & 0.85 & G5 & S & 155 & 842& x,z \\
\object{HD 190642} & V4429 Sgr &  99042 &  163264 & 8.08 & 0.99 & F6-8V/K1III-IV & SB1 & 196 & 1274& \\
\object{HD 197913} & OR Del& 102490& 106466 & 7.55 & 0.76 & G6V/G8V & SB2 & 194 & 1848& m \\
\object{HD 199967} & &  103454 & & 8.01 & 0.58 & G5 & SB2 & 121 & 945& \\
\noalign{\smallskip}\hline
\end{tabular}
\end{table*}
\setcounter{table}{0}
\begin{table*}
\caption{(continued)}
\begin{tabular}{lllllllllll}
\hline \noalign{\smallskip} HD & Var. name & HIP & SAO & $V$ &
$B-V$ & Sp.type & SB & $N_{\rm STELLA}$ & $\Delta t$ & Ref. \\
\noalign{\smallskip}\hline \noalign{\smallskip}
\object{HD 202109} & $\zeta$~Cyg &  104732 &  71070 & 3.20 & 0.99 & G8III & SB1 & 150 & 1288& n \\
\object{HD 218739} & &  114385 &  52754 & 7.14 & 0.61 & G5 & S & 107 & 791& \\
\object{HD 226099} & &  97640 &  68946 & 8.01 & 0.783 & G5 & SB2 & 209 & 1334& \\
\object{HD 237944} & &  53209 &  27812 & 9.36 & 0.71 & (G8V+/G8V+)/? & SB3 & 119 & 1431& a,q \\
\noalign{\smallskip}\hline
\end{tabular}

\vspace{1mm} References.  a: Griffin (2009), b: Griffin (1987), c:
Cutispoto et al. (2003), d: Climenhaga et al. (1951), e: Favata et
al. (1993), f: Fekel (1997), g: Fekel et al. (2004), h: Latham et
al. (2002), i: Henry et al. (1995), j: Wright et al. (2004), k:
Karatas et al. (2004), l: Cutispoto et al. (1999), m: Griffin
(2005), n: Griffin \& Keenan (1992), o: Halbwachs et al. (2003),
p: Morales et al. (2009), q: Otero \& Dubovsky (2004), r: Fekel et
al. (1999), s: Duemmler et al. (2002), t: Duquennoy \& Mayor
(1988), u: G\'alvez et al. (2006), v: Barnes (2007), w: Montes et
al. (2000), x: Strassmeier (1994), y: G\'alvez et al. (2005), z:
Griffin (2010), 29: Szalai et al. (2007), 30: G\'alvez et al.
(2009), 31: G\'alvez et al. (2007), 32: Griffin \& Filiz Ak
(2010), 33: Goldberg et al. (2002), 34: Griffin (2003), 35: Griffin (2012).

\end{table*}

\begin{table*}
\caption{STELLA radial velocities of standard stars. $N$ is the
total number of spectra. $\langle v_{\rm STELLA}\rangle$ is the
average radial velocity from STELLA spectra, $\sigma_{\rm STELLA}$
its standard deviation, and $\langle v_{\rm CORAVEL}\rangle$ the
CORAVEL velocity from Famaey et al. (\cite{famaey}) or, if not in
Famaey et al., from Udry et al. (\cite{udry}). The velocity
difference STELLA-minus-CORAVEL is given in the last column. The
averaged offsets from all stars weighted by the number of data
points is given in the last line. All velocities are in \kms .
}\label{T2}
\begin{tabular}{llllllll}
\hline \noalign{\smallskip} HD & Name & Class. & $N$ & $\langle
v_{\rm STELLA}\rangle$ & $\sigma_{\rm STELLA}$ & $\langle v_{\rm
CORAVEL}\rangle$ &
$v_{\rm STELLA-CORAVEL}$ \\
\noalign{\smallskip}\hline \noalign{\smallskip}
\object{HD 3712}   & $\alpha$ Cas & K0 II-III & 230 &  --3.863 & 0.094 & --4.31 & +0.447 \\
\object{HD 4128}   & $\beta$ Cet  & K0 III    & 143 & +13.686 & 0.107 & +13.1 & +0.586 \\
\object{HD 12929}  & $\alpha$ Ari & K2 III    & 440 &  --14.138 & 0.118 & --14.64 & +0.502 \\
\object{HD 18884}  & $\alpha$ Cet & M2 III    & 60  & --25.688 & 0.293 & --26.08 & +0.392 \\
\object{HD 20902}  & $\alpha$ Per & F5 Ib     & 152 & --1.800 & 0.138 & --2 & +0.200 \\
\object{HD 25025}  & $\gamma$ Eri & M0 III    & 26  & +61.159 & 0.217 & +61.1 & +0.059 \\
\object{HD 29139}  & $\alpha$ Tau & K5 III    & 503 & +54.646 & 0.188 & +54.26 & +0.386 \\
\object{HD 36673}  & $\alpha$ Lep & F0 Ib     & 53  & +25.871 & 0.413 & +25.2 & +0.671 \\
\object{HD 62509}  & $\beta$ Gem  & K0 III    & 442 &  +3.783 & 0.076 & +3.23 & +0.553 \\
\object{HD 81797}  & $\alpha$ Hya & K3 III    & 203 & --4.033 & 0.076 & --4.7 & +0.667 \\
\object{HD 84441}  & $\epsilon$ Leo&G0 II     & 165 & +5.001 & 0.046 & +4.5 & +0.501 \\
\object{HD 107328} & 16~Vir       & K0.5 III  & 332 & +37.117 & 0.160 & +36.4 & +0.717 \\
\object{HD 109379} & $\beta$ Crv  & G5 II     & 49  &  -6.947 & 0.069 & --7.6 & +0.653 \\
\object{HD 124897} & $\alpha$ Boo & K1.5 III  & 787 &  --4.871 & 0.149 & --5.3 & +0.429 \\
\object{HD 146051} & $\delta$ Oph & M0.5 IIIab& 146 &  --19.133 & 0.146 & --19.6 & +0.467 \\
\object{HD 156014}$^a$ & $\alpha$ Her & M5Ib-II   & 303 & --31.168 & 0.694 & --32.09 & +0.922 \\
\object{HD 161096} & $\beta$ Oph  & K2 III    & 763 &  --11.844 & 0.095 & --12.53 & +0.686 \\
\object{HD 168454} & $\delta$ Sgr & K2 III    & 18  &  --19.798 & 0.035 & --20.4 & +0.602 \\
\object{HD 186791} & $\gamma$ Aql & K3 II     & 538 & --2.430 & 0.261 & --2.79 & +0.360 \\
\object{HD 204867} & $\beta$ Aqr  & G0 Ib     & 80  &  +6.808 & 0.151 & +6.3 & +0.508 \\
\object{HD 206778} & $\epsilon$ Peg&K2 Ib     & 572 & +3.588 & 0.262 & +3.39 & +0.198 \\
\object{HD 212943} & 35 Peg       & K0 III    & 402 &  +54.740 & 0.111 & +54.16 & +0.580 \\
\noalign{\smallskip} \hline \noalign{\smallskip}
Weighted average  & & & & & & & +0.503 \\
\noalign{\smallskip}\hline
\end{tabular}

\vspace{1mm} $^a$Double star with the A component being a
``semiregular pulsating variable''.

\end{table*}

SES spectra are automatically reduced using the IRAF\footnote{The
Image Reduction and Analysis Facility is hosted by the National
Optical Astronomy Observatories in Tucson, Arizona at URL
iraf.noao.edu.}-based STELLA data-reduction pipeline (Weber et al.
\cite{spie11}). The images were corrected for bad pixels and
cosmic-ray impacts. Bias levels were removed by subtracting the
average overscan from each image followed by the subtraction of
the mean of the (already overscan subtracted) master bias frame.
The target spectra were flattened by dividing by a nightly master
flat which has been normalized to unity. The nightly master flat
itself is constructed from around 50 individual flats observed
during dusk, dawn, and around midnight. After removal of the
scattered light, the one-dimensional spectra were extracted with
the standard IRAF optimal extraction routine (Horne \cite{horne}).
The blaze function was then removed from the target spectra,
followed by a wavelength calibration using consecutively recorded
Th-Ar spectra. Finally, the extracted spectral orders were
continuum normalized by dividing them with a flux-normalized
synthetic spectrum of the same spectral classification as the
target in question.

\subsection{VI and by photometry}

Johnson-Cousins $V(I)_C$ and/or Str\"omgren $by$ photometry of 56 of
the targets were obtained with the {\sl Amadeus} (T7) and/or the
{\sl Wolfgang} (T6) automatic photoelectric telescope (APT) at
Fairborn Observatory in southern Arizona, respectively. These 56
stars are listed in Table~\ref{T6} along with the analysis results.
A total of 16,591 observations in either $V(I)_C$ or $by$ pairs were
obtained. Typically, one APT observation consists of three
ten-second (30~s for Str\"omgren) integrations on the variable, four
integrations on the comparison star, two integrations on the check
star, and two integrations on the sky. A 30\arcsec \ diaphragm was
used for all targets. The standard error of a nightly mean for {\sl
Amadeus} from the overall seasonal mean changed over the past decade
but was typically 4--6~mmag in $V$ and 6--8~mmag in $I_C$ for the
brightness range of stars in this paper. The observing seasons
between 2007--2009 showed increasing scatter due to a slowly but
systematic malfunction of the acquisition CCD camera. By late 2009
it got so bad that we had to exchange the entire camera plus its CCD
and after that, starting with HJD\,2,455,143, the scatter was back
to the original values quoted above. The standard error of a nightly
mean for {\sl Wolfgang} was between 1.2--3~mmag, depending on system
brightness. For further details we refer to Strassmeier et al.
(\cite{apt}) and Granzer et al. (\cite{granzer01}).

From concurrent observations of Johnson standards in $V(I)_C$, we
also deduce an all-sky solution and apply it to the differential
values whenever feasible. Its accuracy never significantly exceeds
0\fm01 in $V$ though. In case $B$ data were taken, $B-V$ can be
determined only relative to the comparison star because our
all-sky standards were not observed in the $B$ band due to time
constraints. Absolute errors are typically around 0\fm01 for
$\Delta(V-I_C)$ except for the time period 2007--2009, as
mentioned above, and was then likely 0\fm02.

\section{Data products}\label{S4}

\subsection{Radial velocity precision and zeropoint}

Radial velocities were determined from an order-by-order cross
correlation with a synthetic template spectrum from an ATLAS9
atmosphere (Kurucz~\cite{kur}) that roughly matches the target
spectral classification. A two-dimensional cross correlation is
performed in case the target is double lined. Each of 19 selected
echelle orders gives one relative radial velocity. These velocities
are then weighted according to the spectral region depending on the
amount of available spectral lines and are then averaged. The true
internal error is likely close to the rms of these relative
velocities. Numerical simulations of the cross correlation errors
from synthetic spectra and various two-Gaussian fits to its peaks
are discussed in the paper by Weber \& Strassmeier (\cite{capella}).
Note that the external rms values were significantly larger during
the initial year of STELLA operation in 2006/07
($\approx$120~m\,s$^{-1}$), compared to thereafter
($\approx$30~m\,s$^{-1}$). The final radial velocities of our
program stars are barycentric and corrected for Earth rotation. No
gravitational redshift corrections are applied.

\begin{table}
\caption{Average radial velocities for the single stars. $\Delta t$ is the time span of
STELLA observations in days from Table~\ref{T1}. $\langle v_{\rm
r}\rangle $ is the average radial velocity and its standard
deviation. Radial-velocity variations are detected for some of the
stars due to stellar rotation and an asymmetrically spotted
surface. Rotation periods, $P$, if existent, are given in days.
}\label{T3}
\begin{tabular}{llll}
\hline \noalign{\smallskip}
Name  & $\Delta t$ & $\langle v_{\rm r}\rangle$ & Note \\
      & (days)     &  (\kms )        &   \\
\noalign{\smallskip}\hline \noalign{\smallskip}
HD 18645  & 784 & --2.151$\pm$0.052 & $P$=21.5d \\
HD 23551  & 836 & --6.925$\pm$0.043 & $P$=81.7d \\
HD 24053  & 803 & +4.852$\pm$0.040 &  \\
SAO150676& 1460& +25.69$\pm$0.53 & $P$=1.79d \\
HD 43516  & 830 & +21.846$\pm$0.059 &  \\
HD 76799  & 1442& --31.015$\pm$0.187 & $P$=50.13d \\
\object{HIP 46634}  & 1129 & +27.927$\pm$0.047 &  Var. but no $P$\\
HD 95188  & 891 & +6.439$\pm$0.051 & $P$=7.00d \\
HD 95724  & 856 & +3.365$\pm$0.055 & $P$=11.49d \\
HD 104067 & 887 & +15.355$\pm$0.066 &  \\
HD 108564 & 867 & +111.363$\pm$0.009 &  \\
HD 120205 & 849 & --28.870$\pm$0.048 &  \\
HD 136655$^a$ & 842 & --31.921$\pm$0.035 &  \\
HD 153525 & 1349 & --6.844$\pm$0.039 & $P$=15.4d \\
HD 155802 & 857 & --32.711$\pm$0.033 &  \\
HD 171067$^a$ & 857 & --45.771$\pm$0.042 &  \\
HD 184591$^a$ & 842 & --37.635$\pm$0.044 & $P$=48.9d:: \\
HD 218739 & 791 & --5.201$\pm$0.040 &  \\
\noalign{\smallskip}\hline
\end{tabular}
\vspace{1mm}

$^a$Without Ca\,{\sc ii} H\&K emission according to
our paper~I and thus not an active star.
\end{table}


Table~\ref{T2} lists our standard-star measurements. A total of 22
radial-velocity standards were monitored since mid 2006. Their
velocities were derived similarly from cross correlations with a
synthetic template spectrum that fitted the star's classification.
Among these standards are often observed bright giant stars like
$\beta$~Oph, $\alpha$~Ari, $\beta$~Gem, $\alpha$~Tau, 16~Vir, a.o.,
which are all IAU radial-velocity standards (see Scarfe et al.
\cite{sbf90}). These spectra are also used to identify two epochs in
our data when all stars on the observing menu showed a constant
offset in velocity. These epochs are identified with two of our
hardware maintenance episodes where the fiber injection unit had
been readjusted. The average offsets determined from above standard
stars were $-0.244$~\kms\ and $-0.366$~\kms\ for the epochs
2,453,902--4,085 and 2,454,192--435, respectively. These corrections
were applied to all data in this paper.

The preliminary zero point determination in Strassmeier et
al.~(\cite{hd1}) from four standard stars ($\beta$~Oph,
$\alpha$~Cas, $\beta$~Gem, $\beta$~Cet) showed an offset of the
STELLA spectra by +0.46~\kms\ with respect to the CORAVEL system
defined by Udry et al.~(\cite{udry}). In the present paper, we
revise this value to +0.503~\kms\ from 22 standard stars. The
weights applied for this average shift are the total number of
spectra per radial-velocity standard.  Note that an absolute
heliocentric zero point for STELLA has not been determined yet,
e.g. by using asteroids. All velocities in the present paper are
on the STELLA system.

\begin{figure*}
\includegraphics[angle=0,width=17.0cm,clip]{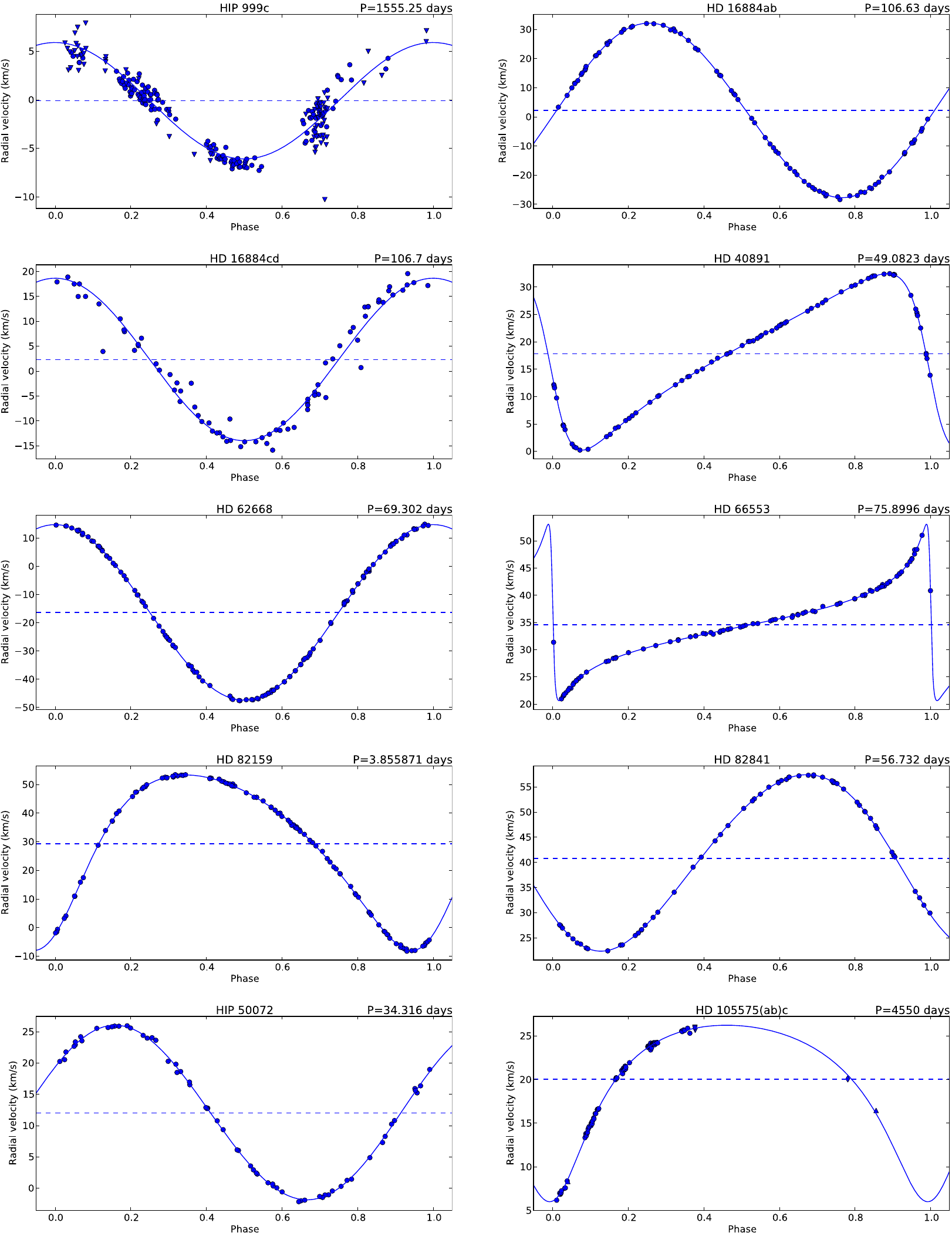}
\caption{Velocity curves for the STELLA single-lined orbital
solutions. The dots are the observations and the lines are the
orbital fits. Velocities are in \kms \ and the horizontal axes are
in orbital phase. The horizontal dashed line in each plot
indicates the systemic velocity. STELLA data are shown as filled
dots. For HIP~999, the triangles are from Fekel et al. (\cite{velo}). For HD\,105575, values from paper~I are shown as down-ward
pointing triangles, the one FEROS point as a diamond, and the two
HARPS points as upward-pointing triangles. For the other
long-period system, HD\,202109, data from Griffin \& Keenan
(\cite{gri:kee}) are added and shown as triangles.}\label{F2}
\end{figure*}
\setcounter{figure}{1}
\begin{figure*}
\includegraphics[angle=0,width=17.0cm,clip]{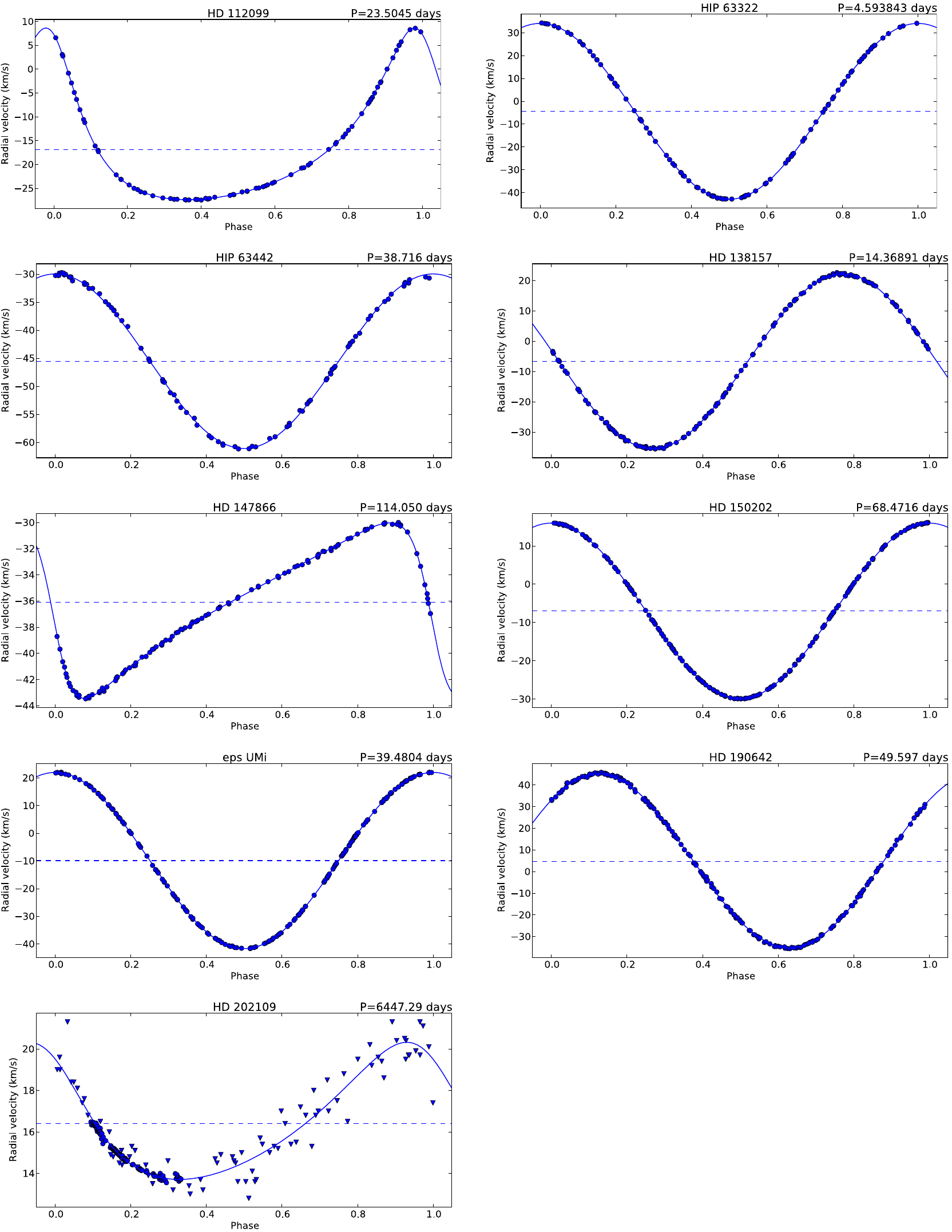}
\caption{(continued).}
\end{figure*}

%
\begin{figure*}
\includegraphics[angle=0,width=17.0cm,clip]{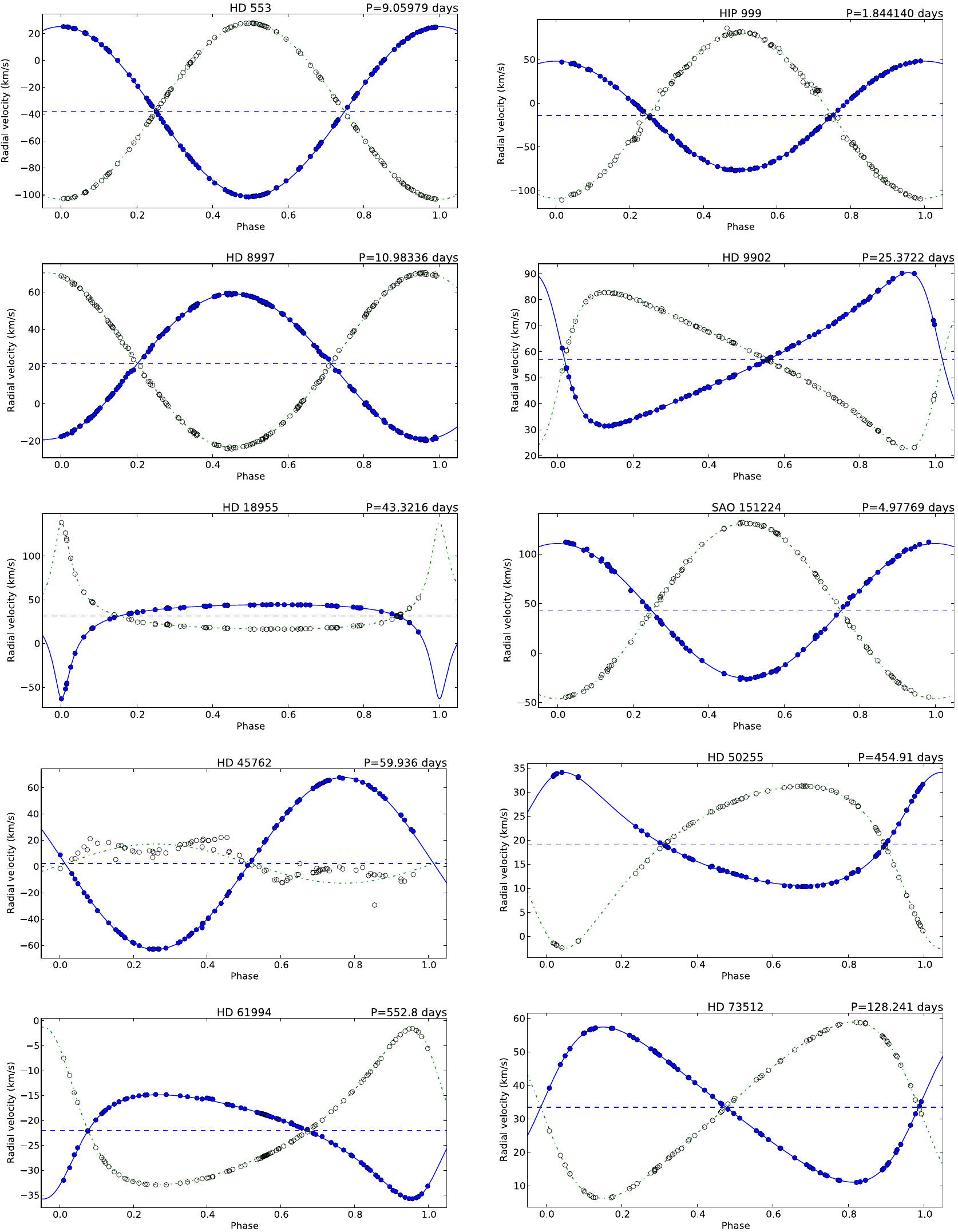}
\caption{Velocity curves for the STELLA double-lined orbital
solutions. The primary velocities are plotted as filled dots, the
secondaries as open circles. Otherwise as in Fig.~\ref{F2}. For
HD\,237944 the triangles mark the tertiary component. }\label{F3}
\end{figure*}
\setcounter{figure}{2}
\begin{figure*}
\includegraphics[angle=0,width=17.0cm,clip]{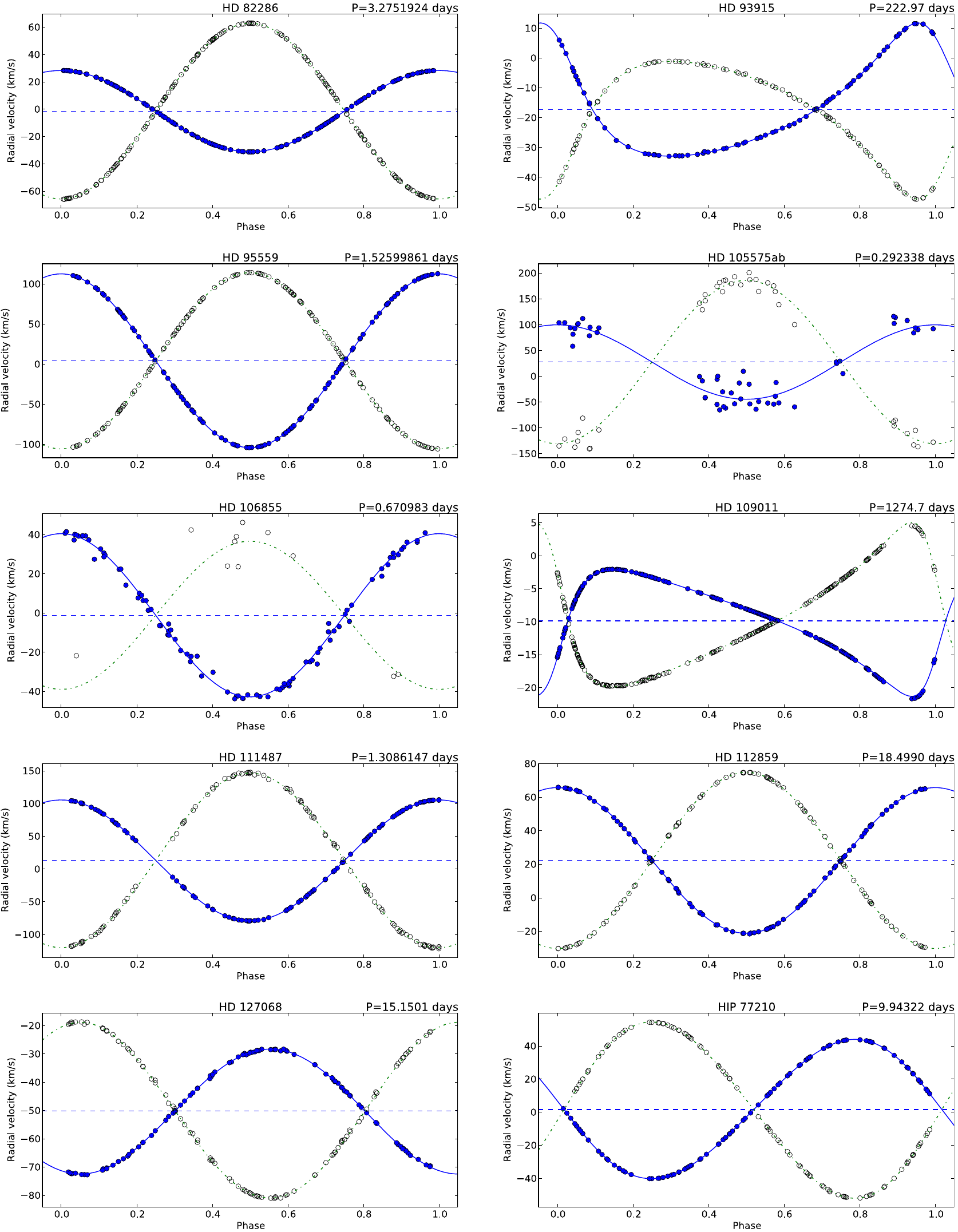}
\caption{(continued).}
\end{figure*}
\setcounter{figure}{2}
\begin{figure*}
\includegraphics[angle=0,width=17.0cm,clip]{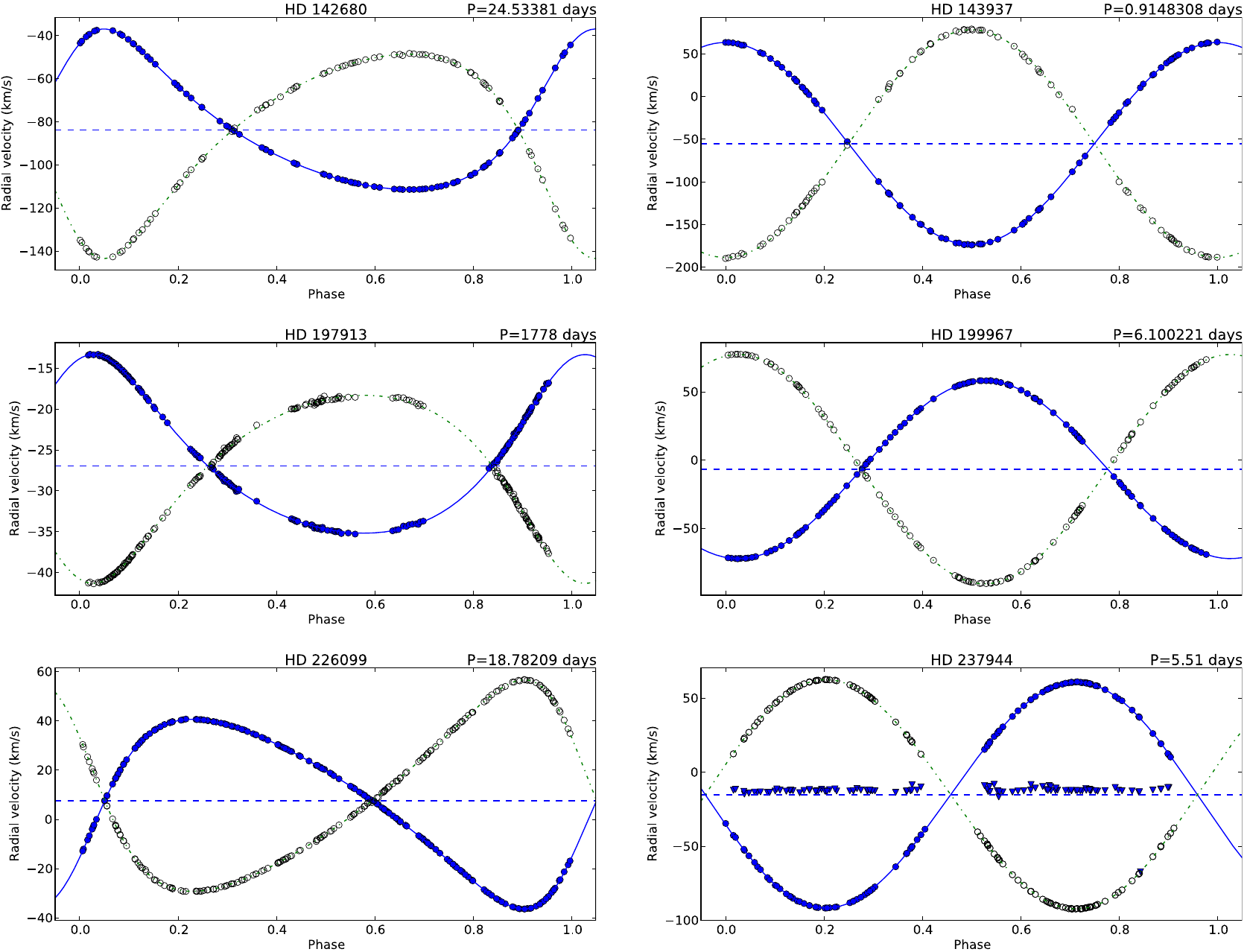}
\caption{(continued).}
\end{figure*}

Table~\ref{T3} lists the stars in our sample that were found to
exhibit no orbital-induced radial velocity variations, i.e. are
presumably single stars. These stars are mostly still very active
stars and further investigation with the present time-series data
are forthcoming. Several of them, e.g. SAO~150676 or HD~76799,
show periodic radial-velocity variations with amplitudes of the
order of 1~\kms , which we interpret to be due to stellar
rotation. Few of our single-star survey targets in paper~I were
also revisited by Griffin (\cite{griff10}), most notably HD~136655
and HD~184591, for which we recommend to read his ``Notes on the
six constant-velocity stars'' for their observational history.

\subsection{Binary orbits}

For single-lined (SB1) as well as double-lined (SB2) binaries we
solve for the components using the general least-square fitting
algorithm {\em MPFIT} (Markwardt \cite{mpfit}). For solutions with
non-zero eccentricity we use the prescription from Danby \&
Burkardt (\cite{dan:bur}) to calculate the eccentric anomaly.
Usually, we first determine the orbit for both components
separately and then combine the two and give rms values for both.
Eclipsing binaries were treated slightly separately due to the
Rossiter-McLaughlin effect in the radial-velocity curve. Some of
the velocities around the two conjunctions were discarded by
applying 3-$\sigma$ clipping. Most of the orbits are sampled
nearly perfectly due to robotic scheduling (Granzer et
al.~\cite{granzer01}) but the unexpected high eccentricity in
three systems left some phase gaps during periastron.  The STELLA time
coverage does not exceed 5.9 years, typical values are around
3.3 years. Although partly compensated for by the high velocity
precision and the dense sampling, systems with periods longer
than, say, 200 days, have comparably uncertain orbital periods.
Whenever there were literature data for the eight systems with
orbital periods in excess of $\approx$100~d that expanded the
total time range and improved the orbit, these are included in the
period determination. This was the case for HIP~999, HD\,105575(ab)c and
HD\,202109. Otherwise, the orbits are computed from STELLA radial
velocities alone. The resulting elements and their errors are summarized in
Table~\ref{T4} and Table~\ref{T5} for the single-lined and the
double-lined systems, respectively. Assuming that the fit is of good quality, we derive the element uncertainties by scaling the formal one-sigma errors from the covariance matrix using the measured $\chi^2$ values. $T_{\rm per}$ is a time of
periastron or, if the orbit is circular, a time of maximum
positive radial velocity. Note that we give orbital periods as
observed and not corrected for the rest frame of the system.

The computed radial-velocity curves are compared with our observed
velocities in the series of panels in Fig.~\ref{F2} and
Fig.~\ref{F3} for the single-lined orbits and the double-lined
orbits, respectively. Note that two of the four triple systems,
HIP~999 and HD~105575, appear in both figures. The latter system
is even triple lined. The third triple system, HD\,237944, is also
triple lined and consists of an SB2 in a short orbit but with the
third component in a wide thousands-of-years orbit around the
close pair. The fourth triple system is HD\,45762 where the secondary is again an SB1 but with a period exactly 1/3 of the wide SB2 pair. HD\,106855 is a close visual double with its primary
A-component to appear double lined, its bona-fide more massive Ab
component being again an SB1 for itself. HD\,16884 appears to be a
quadruple system with two SB1 pairs in a bound 106.65-d orbit.
More comments to individual systems are collected in
Sect.~\ref{S5}. Note that these comments do not intend to give a
comprehensive overview of CDS/Simbad-collected literature but are
meant to draw attention to some specific system features.

\begin{table*}
\caption{STELLA single-lined orbital solutions. Column ``rms''
denotes the quality of the orbital fit for a point of unit weight in
\kms .}\label{T4}
\begin{tabular}{llllllllll}
\hline \noalign{\smallskip}
Name & $P_{\rm orb}$ & $T_{\rm Per}^a$ & K  & $\gamma$ & $e$ & $\omega$ & $a_1\sin i$ & $f(M)$ & rms \\
     & (days) & (HJD 245+)      & (\kms) & (\kms)   & & (deg) & (10$^6$ km) & &  \\
\noalign{\smallskip}\hline \noalign{\smallskip}
HIP999(ab)c$^b$ & 1555. & 0885.7 & 5.99 & --0.10 & 0 & \dots & 128. & 0.0348 & 1.154 \\
     & $\pm 2.6$ & $\pm 5.5$ & $\pm 0.14$ & $\pm 0.08$ & \dots & \dots & $\pm 3.1$ & $\pm 0.0025$ & \\
HD 16884ab$^c$  & 106.63& 4088.3& 29.97 & 2.23 & 0.0161 & 266.  & 43.93 & 0.2978 & 0.261 \\
            & $\pm$0.011  & $\pm$1.6  & $\pm$0.045 &$\pm$0.031 & $\pm$0.0014 & $\pm$5.2 & $\pm$0.066  & $\pm$0.0013 &  \\
HD 16884cd$^c$  & 106.7& 4115.6& 16.3 & 2.3 & 0 & \dots  & 24.0 & 0.0482 & 2.215 \\
            & $\pm$0.17& $\pm$0.77& $\pm$0.38 & $\pm$0.26 & \dots & \dots & $\pm$0.56  & $\pm$0.0034 &  \\
HD 40891 & $49.0823$ & $4092.901$ & $16.122$ & $17.791$ & $0.5105$ & $100.3$ & $9.357$ & $0.01358$ & 0.080 \\
     & $\pm 0.00071$ & $\pm 0.012$ & $\pm 0.015$ & $\pm 0.0098$ & $\pm 0.00069$ & $\pm 0.11$ & $\pm 0.0098$ & $\pm 0.000043$ & \\
HD 62668 & $69.3023$ & $4115.271$ & $31.10$ & $-16.34$ & $0$ & \dots & $29.64$ & $0.2165$ & 0.294 \\
     & $\pm 0.0016$ & $\pm 0.022$ & $\pm 0.037$ & $\pm 0.028$ & \dots & \dots & $\pm 0.036$ & $\pm 0.00080$ & \\
HD 66553 & $75.89957$ & $4110.790$ & $16.2$ & $34.564$ & $0.8719$ & $80.71$ & $8.30$ & $0.00396$ & 0.130 \\
     & $\pm 0.00048$ & $\pm 0.0075$ & $\pm 0.19$ & $\pm 0.015$ & $\pm 0.0028$ & $\pm 0.32$ & $\pm 0.13$ & $\pm 0.00018$ & \\
HD 82159 & $3.855871$ & $4064.9028$ & $30.68$ & $29.261$ & $0.2640$ & $215.2$ & $1.569$ & $0.01038$ & 0.242 \\
     & $\pm$0.000006 & $\pm 0.0025$ & $\pm 0.036$ & $\pm 0.025$ & $\pm 0.0012$ & $\pm 0.25$ & $\pm 0.0019$ & $\pm 0.00004$ & \\
HD 82841 & $56.7316$ & $4064.48$ & $17.552$ & $40.782$ & $0.0874$ & $126.2$ & $13.640$ & $0.03149$ & 0.092 \\
     & $\pm 0.0017$ & $\pm 0.11$ & $\pm 0.017$ & $\pm 0.013$ & $\pm 0.0010$ & $\pm 0.72$ & $\pm 0.013$ & $\pm 0.00009$ & \\
HIP50072 & $34.3157$ & $4220.7$ & $13.95$ & $11.97$ & $0.014$ & 301. & $6.582$ & $0.00967$ & 0.310 \\
     & $\pm 0.0027$ & $\pm 1.7$ & $\pm 0.060$ & $\pm 0.042$ & $\pm$0.0043 & $\pm$18 & $\pm 0.029$ & $\pm 0.00013$ & \\
HD105575$^d$ & 4550. & 4061. & 10.11 & 20.03 & $0.391$ & 186.8 & 582. & 0.381 & 0.160 \\
     & $\pm$52 & $\pm 13$ & $\pm 0.09$ & $\pm 0.075$ & $\pm 0.0059$ & $\pm 1.2$ & $\pm 8.7$ & $\pm 0.012$ & \\
HD112099 & $23.50447$ & $4082.691$ & $18.017$ & $-16.858$ & $0.4466$ & $21.90$ & $5.2105$ & $0.01023$ & 0.067 \\
     & $\pm 0.00013$ & $\pm 0.0051$ & $\pm 0.015$ & $\pm 0.0079$ & $\pm 0.00057$ & $\pm 0.094$ & $\pm 0.0046$ & $\pm 0.000027$ & \\
HIP63322 & $4.593843$ & $4078.3032$ & $38.530$ & $-4.429$ & $0$ & \dots & $2.4340$ & $0.02729$ & 0.119 \\
     & $\pm$0.000005 & $\pm 0.00056$ & $\pm 0.017$ & $\pm 0.012$ & \dots & \dots & $\pm 0.0011$ & $\pm 0.000036$ & \\
HIP63442 & $38.7156$ & $4071.693$ & $15.56$ & $-45.538$ & $0$ & \dots & $8.282$ & $0.01514$ & 0.266 \\
     & $\pm 0.0026$ & $\pm 0.031$ & $\pm 0.042$ & $\pm 0.030$ & \dots & \dots & $\pm 0.023$ & $\pm 0.00012$ & \\
HD138157 & $14.36891$ & $4150.70$ & $28.83$ & $-6.642$ & $0.0051$ & 83. & $5.696$ & $0.03574$ & 0.216 \\
     & $\pm 0.00010$ & $\pm 0.40$ & $\pm 0.047$ & $\pm 0.018$ & $\pm$0.0009 & $\pm$9.9 & $\pm 0.0093$ & $\pm 0.00018$ & \\
HD147866 & $114.050$ & $4234.12$ & $6.713$ & $-36.108$ & $0.5141$ & $100.5$ & $9.030$ & $0.002261$ & 0.076 \\
     & $\pm 0.0073$ & $\pm 0.043$ & $\pm 0.012$ & $\pm 0.0075$ & $\pm 0.0013$ & $\pm 0.22$ & $\pm 0.018$ & $\pm 0.000014$ & \\
HD150202 & $68.47160$ & $4162.339$ & $22.913$ & $-6.9620$ & $0$ & \dots & $21.573$ & $0.08553$ & 0.083 \\
     & $\pm 0.00097$ & $\pm 0.0081$ & $\pm 0.0090$ & $\pm 0.0064$ & \dots & \dots & $\pm 0.0085$ & $\pm 0.00010$ & \\
$\epsilon$~UMi & $39.48042$ & $3933.392$ & $31.849$ & $-9.8623$ & $0$ & \dots & $17.290$ & $0.1325$ & 0.121 \\
     & $\pm 0.00012$ & $\pm 0.0047$ & $\pm 0.013$ & $\pm 0.0085$ & \dots & \dots & $\pm 0.0068$ & $\pm 0.00016$ & \\
HD190642 & $49.5969$ & $4166.0$ & $40.20$ & $4.631$ & $0.0126$ & 314. & $27.42$ & $0.3346$ & 0.445 \\
     & $\pm 0.0012$ & $\pm 0.73$ & $\pm 0.045$ & $\pm 0.033$ & $\pm 0.0011$ & $\pm 5.3$ & $\pm 0.031$ & $\pm 0.0011$ & \\
HD202109$^e$ & 6446. & 2421363. & 3.30 & $16.41$ & $0.244$ & 40.7 & 284. & $0.0220$ & 0.413 \\
     & $\pm 14$ & $\pm 82$ & $\pm 0.057$ & $\pm 0.043$ & $\pm$0.016 & $\pm$3.3 & $\pm 5.1$ & $\pm 0.0012$ & \\
\noalign{\smallskip}\hline
\end{tabular}

\vspace{1mm} $^a$Time of periastron, or ascending node for
circular orbits.\\
$^b$Triple system. Orbit (ab) around c is given.\\
$^c$Quadruple system with two SB1 pairs in a synchronized orbit. \\
$^d$SB1 orbit of the tertiary component, $c$, around the close
(eclipsing) $ab$ pair. \\
$^e$SB1 with long period. The orbit given is from a combination of
STELLA data and selected published values. $T_{\rm Per}$ is 2421363. See individual notes.

\end{table*}

At the time of the start of our monitoring program in 2006, a total
of 17 stars were already either known binaries or had even an orbit
computed, or at least had an orbital period known. However, we
accumulated enough observations for all systems to allow for orbital
solutions just using our own data for a certain epoch. This was
motivated not only by consistency arguments but also by detections
of orbital-period variations in active binaries and particularly in
one of our targets (FF~UMa; G\'alvez et
al.~\cite{galvez07})\footnote{See our individual notes for this star
though.}. Orbital-period variations of magnetically active binaries
are known for a long time (e.g. Hall \& Kreiner \cite{hal:kre}; see
also Lanza et al.~\cite{lan:pil}) and were partly explained by mild
mass exchange and mass loss due to a stellar wind. However,
Applegate (\cite{apple}) proposed a connection between the
gravitational quadrupole moment and the shape of a magnetically
active component in the system as it goes through its magnetic
activity cycle. Therefore, it appears rewarding to redetermine the
orbits of such binaries from time to time.

\begin{table*}
\caption{STELLA double-lined orbital solutions. Suffices 1 and 2
denote the primary and the secondary star, respectively. }\label{T5}
\begin{tabular}{llllllllll}
\hline \noalign{\smallskip}
Name & Period & $T_{\rm Per}^a$ & $K_1$ & $\gamma$ & $e$ & $\omega$ & $a_1\sin i$ & $M_1\sin^3i$ & rms$_1$ \\
     &        &                 & $K_2$ & & &                       & $a_2\sin i$ & $M_2\sin^3i$ & rms$_2$ \\
     & (days) & (HJD245+)      & (\kms) & (\kms)  & & (deg)    & (10$^6$ km) & (M$_\odot$)& (\kms) \\
\noalign{\smallskip}\hline \noalign{\smallskip}
HD 553 & $9.059792$ & $4087.1233$ & $63.312$ & $-38.014$ & $0$ & \dots & $7.8874$ & $1.019$ & 0.411 \\
   & $\pm 0.000032$ & $\pm 0.0019$ & $\pm 0.037$ & $\pm 0.027$ & \dots & \dots & $\pm 0.0047$ & $\pm 0.0073$ & \\
   &  &  & $65.46$ &  &  &  & $8.155$ & $0.9854$ & 0.390 \\
   &  &  & $\pm 0.23$ &  &  &  & $\pm 0.029$ & $\pm 0.0037$ & \\
HIP 999$^b$ & 1.8441405  & 4063.6374 & $62.40$ & $-14.24$ & $0$ & \dots & $1.582$ & $0.447$ & 0.519 \\
    & $\pm 1.6\,10^{-6}$ & $\pm$0.00055 & $\pm 0.063$ & $\pm 0.041$ & \dots & \dots & $\pm 0.0016$ & $\pm 0.0071$ & \\
   &  &  & $94.77$ &  &  &  & $2.403$ & $0.2945$ & 3.292 \\
   &  &  & $\pm 0.68$ &  &  &  & $\pm 0.017$ & $\pm 0.0026$ & \\
HD 8997 & $10.98336$ & $4072.091$ & $39.124$ & $21.546$ & $0.0412$ & $195.2$ & $5.9040$ & $0.3956$ & 0.365 \\
   & $\pm 0.00004$ & $\pm 0.029$ & $\pm 0.027$ & $\pm 0.021$ & $\pm 0.0007$ & $\pm 0.93$ & $\pm 0.0041$ & $\pm 0.0011$ & \\
   &  &  & $47.00$ &  &  &  & $7.092$ & $0.3294$ & 0.400 \\
   &  &  & $\pm 0.060$ &  &  &  & $\pm 0.0090$ & $\pm 0.00063$ & \\
HD 9902 & $25.3722$ & $4079.516$ & $29.46$ & $56.956$ & $0.5068$ & $74.33$ & $8.861$ & $0.1813$ & 0.201 \\
   & $\pm 0.00026$ & $\pm 0.0060$ & $\pm 0.035$ & $\pm 0.014$ & $\pm 0.00074$ & $\pm 0.11$ & $\pm 0.012$& $\pm 0.00066$ & \\
   &  &  & $30.22$ &  &  &  & $9.089$ & $0.1767$ & 0.154 \\
   &  &  & $\pm 0.046$ &  &  &  & $\pm 0.015$ & $\pm 0.00057$ & \\
HD 18955 & $43.32161$ & $4086.710$ & $53.83$ & $31.273$ & $0.75874$ & $174.84$ & $20.89$ & $0.9982$ & 0.237 \\
   & $\pm 0.00022$ & $\pm 0.0045$ & $\pm 0.090$ & $\pm 0.024$ & $\pm 0.00050$ & $\pm 0.080$ & $\pm 0.039$ & $\pm 0.0052$ & \\
   &  &  & $61.02$ &  &  &  & $23.68$ & $0.8807$ & 0.480 \\
   &  &  & $\pm 0.12$ &  &  &  & $\pm 0.053$ & $\pm 0.0042$ & \\
SAO151224 & $4.977691$ & $4069.9382$ & $68.07$ & $42.67$ & $0$ & \dots & $4.659$ & $1.130$ & 1.130 \\
   & $\pm 0.000012$ & $\pm 0.0021$ & $\pm 0.085$ & $\pm 0.057$ & \dots & \dots & $\pm 0.0058$ & $\pm 0.0041$ & \\
   &  &  & $88.92$ &  &  &  & $6.087$ & $0.8652$ & 0.869 \\
   &  &  & $\pm 0.14$ &  &  &  & $\pm 0.0099$ & $\pm 0.0026$ & \\
HD 45762$^c$ & $59.9363$ & $4110.40$ & $65.447$ & $2.23$ & $0.0150$ & 84.6 & $53.935$ & $0.599$ & 0.533 \\
     & $\pm 0.0016$ & $\pm 0.61$ & $\pm 0.064$ & $\pm 0.048$ & $\pm$0.0009 & $\pm$3.7 & $\pm 0.053$ & $\pm 0.058$ & \\
   &  &  & $14.9$ &  &  &  & $12.3$ & $2.63$ & 6.4 \\
   &  &  & $\pm 1.1$ &  &  &  & $\pm 0.87$ & $\pm 0.069$ & \\
HD 50255 & $454.910$ & $4152.91$ & $11.77$ & $19.059$ & $0.3440$ & $324.42$ & $69.11$ & $0.5414$ & 0.150 \\
   & $\pm 0.038$ & $\pm 0.28$ & $\pm 0.023$ & $\pm 0.011$ & $\pm 0.0011$ & $\pm 0.23$ & $\pm 0.14$ & $\pm 0.0045$ & \\
   &  &  & $16.89$ &  &  &  & $99.23$ & $0.3771$ & 0.176 \\
   &  &  & $\pm 0.063$ &  &  &  & $\pm 0.37$ & $\pm 0.0022$ & \\
HD 61994 & $552.82$ & $4397.45$ & $10.46$ & $-22.038$ & $0.4230$ & $222.3$ & $72.07$ & $0.465$ & 0.118 \\
     & $\pm 0.25$ & $\pm 0.26$ & $\pm 0.018$ & $\pm 0.0088$ & $\pm 0.0012$ & $\pm 0.18$ & $\pm 0.013$ & $\pm 0.0044$ & \\
   &  &  & $15.81$ &  &  &  & $108.9$ & $0.3120$ & 0.117 \\
   &  &  & $\pm 0.065$ &  &  &  & $\pm 0.46$ & $\pm 0.0019$ & \\
HD 73512 & $128.241$ & $4145.44$ & $23.200$ & $33.478$ & $0.26240$ & $277.27$ & $39.478$ & $0.7722$ & 0.087 \\
   & $\pm 0.0046$ & $\pm 0.045$ & $\pm 0.014$ & $\pm 0.0086$ & $\pm 0.00045$ & $\pm 0.13$ & $\pm 0.025$ & $\pm 0.0016$ & \\
   &  &  & $26.349$ &  &  &  & $44.836$ & $0.6799$ & 0.170 \\
   &  &  & $\pm 0.025$ &  &  &  & $\pm 0.043$ & $\pm 0.0011$ & \\
HD 82286 & $3.2751924$ & $4066.0088$ & $29.775$ & $-1.473$ & $0$ & \dots & $1.3410$ & $0.19296$ & 0.025 \\
   & $\pm 4\,10^{-7}$ & $\pm 0.00012$ & $\pm 0.0043$ & $\pm 0.0026$ & \dots & \dots & $\pm 0.00019$ & $\pm 0.00010$ & \\
   &  &  & $64.282$ &  &  &  & $2.8951$ & $0.089378$ & 0.174 \\
   &  &  & $\pm 0.014$ &  &  &  & $\pm 0.00062$ & $\pm 0.00003$ & \\
HD 93915 & $222.965$ & $4182.78$ & $22.35$ & $-17.288$ & $0.3795$ & $37.80$ & $63.40$ & $0.8732$ & 0.180 \\
   & $\pm 0.022$ & $\pm 0.086$ & $\pm 0.033$ & $\pm 0.015$ & $\pm 0.00088$ & $\pm 0.16$ & $\pm 0.097$ & $\pm 0.0032$ & \\
   &  &  & $23.10$ &  &  &  & $65.54$ & $0.8447$ & 0.196 \\
   &  &  & $\pm 0.037$ &  &  &  & $\pm 0.11$ & $\pm 0.0030$ & \\
HD 95559 & $1.5259986$ & $4076.8100$ & $108.369$ & $4.212$ & $0$ & \dots & $2.27402$ & $0.8282$ & 0.106 \\
   & $\pm 7.5\,10^{-8}$ & $\pm 0.00003$ & $\pm 0.011$ & $\pm 0.0064$ & \dots & \dots & $\pm 0.00022$ & $\pm 0.0004$ & \\
   &  &  & $109.93$ &  &  &  & $2.3068$ & $0.81644$ & 0.165 \\
   &  &  & $\pm 0.025$ &  &  &  & $\pm 0.0005$ & $\pm 0.00025$ & \\
HD105575$^d$ & $0.2923378$ & $4160.768$ & $87.4$ & $27.4$ & $0$ & \dots & $0.351$ & $0.288$ & 8.6 \\
   & $\pm 1.7\,10^{-6}$ & $\pm 0.0044$ & $\pm 4.5$ & $\pm 2.3$ & \dots & \dots & $\pm 0.018$ & $\pm 0.027$ & \\
   &  &  & $158.$ &  &  &  & $0.635$ & $0.16$ & 13 \\
   &  &  & $\pm 6$ &  &  &  & $\pm 0.024$ & $\pm 0.016$ & \\
\noalign{\smallskip}\hline
\end{tabular}
\end{table*}
\setcounter{table}{4}
\begin{table*}
\caption{(continued)}
\begin{tabular}{llllllllll}
\hline \noalign{\smallskip}
Name & Period & $T_{\rm Per}^a$ & $K_1$ & $\gamma$ & $e$ & $\omega$ & $a_1\sin i$ & $M_1\sin^3i$ & rms$_1$ \\
     &        &                 & $K_2$ & & &                       & $a_2\sin i$ & $M_2\sin^3i$ & rms$_2$ \\
     & (days) & (HJD245+)      & (\kms) & (\kms)  & & (deg)    & (10$^6$ km) & (M$_\odot$)& (\kms) \\
\noalign{\smallskip}\hline \noalign{\smallskip}
HD106855 & $0.6709843$ & $4097.2944$ & $41.8$ & $-1.1$ & $0$ & \dots & $0.385$ & $0.0166$ & 1.811 \\
   & $\pm$0.0000026 & $\pm 0.0031$ & $\pm 0.65$ & $\pm 0.42$ & \dots & \dots & $\pm 0.006$ & $\pm 0.0012$ & \\
   &  &  & $37.7$ &  &  &  & $0.348$ & $0.0183$ & 9.92 \\
   &  &  & $\pm 1.4$ &  &  &  & $\pm 0.013$ & $\pm 0.0009$ & \\
HD109011 & $1274.7$ & $5213.63$ & $9.582$ & $-9.8616$ & $0.5071$ & $247.36$ & $144.8$ & $0.5130$ & 0.094 \\
   & $\pm 0.86$ & $\pm 0.23$ & $\pm 0.014$ & $\pm 0.0044$ & $\pm 0.00060$ & $\pm 0.13$ & $\pm 0.24$ & $\pm 0.0021$ & \\
   &  &  & $12.47$ &  &  &  & $188.4$ & $0.3941$ & 0.147 \\
   &  &  & $\pm 0.021$ &  &  &  & $\pm 0.35$ & $\pm 0.0014$ & \\
HD111487 & $1.3086147$ & $4072.3403$ & $92.350$ & $13.17$ & $0$ & \dots & $1.6618$ & $0.9217$ & 0.268 \\
   & $\pm 0.0000002$ & $\pm 0.00014$ & $\pm 0.027$ & $\pm 0.021$ & \dots & \dots & $\pm 0.0005$ & $\pm 0.0028$ & \\
   &  &  & $133.4$ &  &  &  & $2.401$ & $0.6381$ & 1.776 \\
   &  &  & $\pm 0.19$ &  &  &  & $\pm 0.0033$ & $\pm 0.0011$ & \\
HD112859 & $18.49896$ & $4072.681$ & $43.466$ & $22.267$ & $0$ & \dots & $11.057$ & $0.9293$ & 0.300 \\
   & $\pm 0.00014$ & $\pm 0.0038$ & $\pm 0.031$ & $\pm 0.020$ & \dots & \dots & $\pm 0.0078$ & $\pm 0.0027$ & \\
   &  &  & $52.57$ &  &  &  & $13.37$ & $0.7683$ & 0.265 \\
   &  &  & $\pm 0.07$ &  &  &  & $\pm 0.018$ & $\pm 0.0015$ & \\
HD127068 & $15.15011$ & $4108.69$ & $22.09$ & $-50.188$ & $0.0098$ & 161. & $4.601$ & $0.1377$ & 0.190 \\
   & $\pm 0.00011$ & $\pm 0.23$ & $\pm 0.026$ & $\pm 0.014$ & $\pm$0.00087 & $\pm$5.5 & $\pm 0.0055$ & $\pm 0.00069$ & \\
   &  &  & $31.06$ &  &  &  & $6.471$ & $0.09790$ & 0.364 \\
   &  &  & $\pm 0.071$ &  &  &  & $\pm 0.015$ & $\pm 0.00034$ & \\
HIP77210 & $9.943224$ & $4106.9724$ & $42.096$ & $1.609$ & $0.06069$ & $83.32$ & $5.7451$ & $0.4962$ & 0.052 \\
   & $\pm 0.000011$ & $\pm 0.0041$ & $\pm 0.007$ & $\pm 0.0047$ & $\pm 0.00015$ & $\pm 0.15$ & $\pm 0.001$ & $\pm 0.00060$ & \\
   &  &  & $53.262$ &  &  &  & $7.2691$ & $0.39213$ & 0.203 \\
   &  &  & $\pm 0.030$ &  &  &  & $\pm 0.0041$ & $\pm 0.00028$ & \\
HD142680 & $24.53381$ & $4169.6573$ & $37.257$ & $-83.830$ & $0.3158$ & $324.81$ & $11.926$ & $0.7375$ & 0.042 \\
   & $\pm 0.00006$ & $\pm 0.0022$ & $\pm 0.006$ & $\pm 0.0041$ & $\pm 0.00015$ & $\pm 0.031$ & $\pm 0.0021$ & $\pm 0.0010$ & \\
   &  &  & $47.398$ &  &  &  & $15.172$ & $0.5797$ & 0.205 \\
   &  &  & $\pm 0.031$ &  &  &  & $\pm 0.010$ & $\pm 0.00047$ & \\
HD143937 & $0.91483076$ & $4150.5197$ & $118.56$ & $-55.307$ & $0$ & \dots & $1.4914$ & $0.8036$ & 0.285 \\
   & $\pm 0.0000002$ & $\pm 0.0001$ & $\pm 0.04$ & $\pm 0.030$ & \dots & \dots & $\pm 0.0005$ & $\pm 0.0019$ & \\
   &  &  & $133.5$ &  &  &  & $1.679$ & $0.7137$ & 1.258 \\
   &  &  & $\pm 0.16$ &  &  &  & $\pm 0.0020$ & $\pm 0.0010$ & \\
HD197913 & $1778.5$ & $6285.5$ & $10.941$ & $-26.969$ & $0.2600$ & $342.57$ & $258.4$ & $0.9649$ & 0.100 \\
   & $\pm 1.3$ & $\pm 2.0$ & $\pm 0.010$ & $\pm 0.0060$ & $\pm 0.00072$ & $\pm 0.24$ & $\pm 0.30$ & $\pm 0.0026$ & \\
   &  &  & $11.52$ &  &  &  & $272.1$ & $0.9161$ & 0.161 \\
   &  &  & $\pm 0.013$ &  &  &  & $\pm 0.37$ & $\pm 0.0021$ & \\
HD199967 & $6.1002208$ & $4169.06$ & $65.12$ & $-6.830$ & $0.00244$ & 170. & $5.463$ & $1.177$ & 0.102 \\
   & $\pm 0.000005$ & $\pm 0.06$ & $\pm 0.076$ & $\pm 0.007$ & $\pm$0.00015 & $\pm$3.5 & $\pm 0.0064$ & $\pm 0.0032$ & \\
   &  &  & $83.89$ &  &  &  & $7.037$ & $0.9139$ & 0.149 \\
   &  &  & $\pm 0.10$ &  &  &  & $\pm 0.0085$ & $\pm 0.0024$ & \\
HD226099 & $18.78209$ & $4162.0653$ & $38.528$ & $7.5772$ & $0.30961$ & $242.63$ & $9.4618$ & $0.47739$ & 0.086 \\
   & $\pm 0.000019$ & $\pm 0.0016$ & $\pm 0.0068$ & $\pm 0.0042$ & $\pm 0.00014$ & $\pm 0.031$ & $\pm 0.0017$ & $\pm 0.00032$ & \\
   &  &  & $42.968$ &  &  &  & $10.552$ & $0.42807$ & 0.182 \\
   &  &  & $\pm 0.013$ &  &  &  & $\pm 0.0033$ & $\pm 0.00021$ & \\
HD237944$^e$ & $5.5076262$ & $4070.103$ & $76.237$ & $-15.25$ & $0.0141$ & 105. & $5.7733$ & $1.042$ & 0.264 \\
   & $\pm 0.0000047$ & $\pm 0.019$ & $\pm 0.048$ & $\pm 0.017$ & $\pm$0.0003 & $\pm$1.3 & $\pm 0.0036$ & $\pm 0.0015$ & \\
   &  &  & $77.410$ &  &  &  & $5.8621$ & $1.027$ & 0.223 \\
   &  &  & $\pm 0.048$ &  &  &  & $\pm 0.0036$ & $\pm 0.0014$ & \\
\noalign{\smallskip}\hline
\end{tabular}

\vspace{1mm}
$^a$Time of periastron, or ascending node for circular orbits.\\
$^b$SB2, but with a third component determined from the $\gamma$-residuals.
The third-component orbit is given in the SB1 list in Table~\ref{T4}. See also the individual notes.\\
$^c$Hierarchial triple system with lines from components $a$ and
$c$. The orbit $ab$ is SB1 with $P$=19.933~d. See individual notes. \\
$^d$SB3 with a third component in the spectrum denoted
$c$. Its orbit around the (eclipsing) $ab$ pair is given in Table~\ref{T4}. \\
$^e$SB3. The third component, $c$, has a median velocity of
--12.25$\pm$1.3(stdev)~\kms\ for the time of our observations.

\end{table*}

%
\begin{table*}
\caption{Summary of photometric data and results. Listed are the
comparison (``cmp'') star used for the differential observations,
the start date of observations (``HJD start''), the time span of
observations ($\Delta t$) in days, the telescope and photometric
bandpasses used (Johnson-Cousins $VI$ for the T7 APT and
Str\"omgren $by$ for the T6 APT), and the number of actual data
points $N$, each the mean of three individual readings. The
results are summarizes in the subsequent columns and list $V_{\rm
max}$, the maximum, i.e. least spotted, brightness during our
observations; $\Delta V$, the largest observed amplitude due to
spots and, if eclipsing, due to primary eclipse;
$\langle$C.I.$\rangle$, the average color index ($V$-$I$)$_C$ for
T7 observations and/or ($b$-$y$) for T6; $P_{\rm phtm}$, the
photometric period and its likely error obtained from a refitting
to synthetic data; and a note if applicable.}\label{T6}
\begin{tabular}{lllllllllll}
\noalign{\smallskip}\hline \noalign{\smallskip}
HD/name & HD/cmp & HJD start & $\Delta t$ &Tel. & $N$ & $V_{\rm max}$ & $\Delta V_{\rm max}$ & $<$C.I.$>$ & $P_{\rm phtm}$ & Note \\
& star & (2,4+)& & & & (mag) & (mag) & (mag) & (days)& \\
\noalign{\smallskip}\hline \noalign{\smallskip}
HD 553 & 443 & 51107 &  253 &T6+7&  217 & 8.15 & 0.10 & \dots &$\approx P_{\rm orb}$ &  rotation \\
HD 553 &  &  &  &  &  &  & 0.28 & & 9.061$\pm$0.002 &  eclipsing \\
HIP 999 & 977 & 50395 &  700 &T6&  134 & 8.53 & 0.08 & 0.56 & 1.84475$\pm$0.00008 &   \\
HD 8997 & 9472 & 51098 &   95 &T6&   69 & 7.75 & 0.02 & \dots & 10.54$\pm$0.11 &   \\
HD 9902 & HIP7910 & 51098 &  112 &T6&  122 & 8.68 & 0.05 & \dots & 7.445$\pm$0.02 &  \\
HD 16884 & 17030 & 51434 & 3327 &T6+7&  286 & 8.87 & 0.16 & 1.58 & 65.44$\pm$0.05 &  \\
HD 18645 & 18668 & 51098 &  727 &T6&  241 & 7.85 & 0.04& 0.55 & 21.5$\pm$0.1 &  \\
HD 18955 & 19223 & 51099 &   50 &T6&   29 & 8.46 & \dots & \dots & \dots &  var at $2.3\sigma$ \\
HD 23551 & 24513 & 51447 & 3729 &T6+7&  242 & 7.08& 0.035 & 0.94 & 81.69$\pm$0.03 & $P$ from T6 \\
HD 24053 & 23570 & 55570 & 74 & T7 & 130 & 7.72 & 0.06 & 0.55 & \dots & \\
SAO 150676& 37792 & 51434 &  745 &T6+7&  389 & 9.09 & 0.08 & 0.79 & 1.7889$\pm$0.0003 &  \\
HD 40891 & 42756 & 55570 & 60 & T7 & 71 & 8.42 & 0.06 & 0.76 & \dots & \\
SAO 151224& 43306 & 51079 & 4160 &T7& 1352 & 9.27 & 0.39 & 1.27 & $\approx P_{\rm orb}$ & rotation \\
SAO 151224& & & & & & & 0.68 & & 4.98242$\pm$0.00003 &  eclipsing \\
HD 43516 & 44396 & 55576 & 67 & T7 & 82 & 7.40 & $<$0.03 & 0.85 & \dots & \\
HD 45762 & 45168 & 55576 & 49 & T7 & 61 & 8.35 & 0.15 & 0.98 & 62.0$\pm$1.6 & +ellipticity\\
HD 50255 & 50674 & 55576 & 49 & T7 & 38 & 7.48 & 0.05 & 0.62 & \dots & \\
HD 61994 & 63748 & 55576 & 49 & T7 & 43 & 7.05 & 0.03 & 0.64 & \dots & \\
HD 62668 & 64106 & 51457 & 3782 &T7&  568 & 7.46 & 0.40 & 1.18 & 67.470$\pm$0.007 &   \\
HD 66553 & 65257 & 55576 & 29 & T7 & 6 & 8.52 & $<$0.02 & 0.72 & \dots & \\
HD 73512 & 73764 & 55572 & 92 & T7 & 131 & 7.92 & $<$0.02 & 0.80 & \dots & \\
HD 76799 & 77361 & 55572 & 92 & T7 & 133 & 7.25 & 0.06 & 0.92 & \dots & \\
HD 82159 & 82410 & 51242 &   89 &T6&   52 & 7.97 & 0.05 & \dots & 3.055$\pm$0.003 & polluted \\
HD 82286 & 82719 & 51459 & 3780 &T7&  609 & 7.84 & 0.16 & 1.12 & 3.27629$\pm$0.00003 &  \\
HD 82841 & 83046 & 51476 & 3763 &T7&  914 & 8.22 & 0.07 & 1.15 & 56.58$\pm$0.04 &  \\
HIP 50072 & 88853 & 51242 &   96 &T6&   58 & 9.44 & 0.12 & \dots & 33.782$\pm$0.009 &  \\
HD 93915 & 95379 & 55572 & 138 & T7 & 156 & 8.10 & 0.20 & 0.65 & \dots & \\
HD 95188 & 94996 & 51249 &  102 &T6&   77 & 8.51 & 0.025 & \dots & 7.00$\pm$0.06 &  \\
HD 95559 & 95242 & 51551 & 3686 &T6+7& 1061 & 8.83 & 0.08 & 0.93 & 1.517000$\pm$4\,10$^{-6}$ & \\
HD 95724 & 96778 & 51239 &  112 &T6&   78 & 8.97 & 0.05 & \dots & 11.49$\pm$0.02&  \\
HD 104067 & 104414 & 55572 & 168 & T7 & 158 & 7.98 & 0.05 & 0.92 & \dots & \\
HD 105575 & 105814 & 51239 &   20 &T6&  101 & 8.98 & \dots & 0.62 & $\approx P_{\rm orb}$ &  rotation\\
HD 105575 & & & & & & & 0.36 & & 0.29238$\pm$0.00009 &  eclipsing\\
HD 106855 & 106991 & 51551 & 3685 &T7&  992 & 9.35 & 0.11 & 1.27 & 0.6705258$\pm$6\,10$^{-7}$ &  \\
HD 108564 & 108599 & 55572 & 168 & T7 & 203 & 9.50 & 0.06 & 1.02 & \dots & \\
HD 109011 & 109894 & 51225 &  135 &T6&   60 & 8.10 & 0.02 & \dots & 8.4$\pm$0.2 &   \\
HD 111487 & 111276 & 51142 & 4116 &T7& 1571 & 9.62 & 0.15 & 0.86 &$\approx P_{\rm orb}$  &  rotation \\
HD 111487 &  &  &  &  &  & & 0.76 & & 1.30859$\pm$0.000007 &  eclipsing \\
HIP 63322 & 113168 & 51240 &   65 &T6&   35 & 8.89 & 0.08 & \dots & 2.227$\pm$0.003 & FAP 20\% \\
HD 112099 & 111733 & 55569 & 171 & T7 & 171 & 8.22 & 0.04 & 0.70& \dots &  \\
HD 112859 & 112220 & 51225 &  135 &T6&   76 & 8.23 & 0.06 & \dots & 18.34$\pm$0.05 &  \\
HIP 63442 & 113828 & 51240 &  120 &T6&   80 & 9.39 & 0.12 & \dots & 40.29$\pm$0.08 &  \\
HD 120205 & 120969 & 55572 & 168 & T7 & 199 & 8.35 & 0.06 & 0.75 & 14.3$\pm$0.7 & \\
HD 127068 & 126583 & 51537 & 3700 &T7&  972 & 8.47 & 0.18 & 1.05 & 18.187$\pm$0.0001 &  \\
HD 136655 & 136643 & 50492 &  355 &T7&  465 & 8.81 & $<$0.02 & 1.00 & \dots &  constant\\
HD 138157 & 138085 & 51540 & 3697 &T6+7& 1159 & 7.11 & 0.14 & 1.10 & 7.18357$\pm$0.00005 & $P/2$+ell.  \\
HIP 77210 & 140750 & 51260 &  100 &T6&   49 & 9.14 & 0.02 & \dots & 13.7$\pm$0.1 &  \\
HD 142680 & 142245 & 51260 &  100 &T6&   64 & 8.70 & 0.03 & \dots & 33.4$\pm$1.5 &   \\
HD 143937 & 143438 & 51559 &   48 &T7&   57 & 8.67 & 0.10 & 1.16 &$\approx P_{\rm orb}$  &  rotation\\
HD 143937 &  &  &  & &  &  & 0.41 &  & 0.91479$\pm$0.00003 &  eclipsing\\
HD 147866 & 147266 & 51447 &  728 &T6+7&   92 & 8.08 & 0.03 & 1.14 & 80.3$\pm$2.2 & \\
\noalign{\smallskip}\hline
\end{tabular}
\end{table*}
\setcounter{table}{5}
\begin{table*}[t]
\caption{(continued)}
\begin{tabular}{lllllllllll}
\hline \noalign{\smallskip}
HD/name & HD/cmp & HJD start & $\Delta t$ &Tel. & $N$ & $V_{\rm max}$ & $\Delta V_{\rm max}$ & $<$C.I.$>$ & $P_{\rm phtm}$ & Note \\
& star & (2,4+)& & & & (mag) & (mag) & (mag) & (days)& \\
\noalign{\smallskip}\hline \noalign{\smallskip}
HD 150202 & 149843 & 55572 & 168 & T7 & 210 & 7.94 & 0.05 & 0.73 & 36.4$\pm$0.5 & 1/2 $P_{\rm rot}$\\
HD 153525 & 153286 & 51260 &  100 &T6&   53 & 7.94 & 0.02 & \dots & 15.4$\pm$0.1 &   \\
HD 155802 & 156227 & 55580 & 160 & T7 & 149 & 8.50 & 0.03 & 0.85 & \dots & \\
HD 171067 & 170651 & 55602 & 138 & T7 & 194 & 7.17 & 0.02 & 0.55 & \dots & \\
HD 184591 & 183849 & 51099 & & T6 & 263 & 7.34 & $<$0.01& 0.62 & (48.9$\pm$2.2) & uncertain \\
HD 226099 & 188149 & 55634 & 106 & T7 & 210 & 7.98 & 0.05 & 0.65 & \dots & \\
HD 190642 & 189893 & 51434 & 3668 &T7&  791 & 8.15 & 0.32 & 1.26 & $\approx P_{\rm orb}$ &  rotation\\
HD 190642 &  &  &  &  &  &  & 0.43 &  & 24.8339$\pm$0.0009 &  eclipsing\\
HD 197913 & 198109 & 51098 &   51 &T6&   33 & 7.13 & 0.02 & \dots & 6.556$\pm$0.07 &   \\
HD 199967 & 199354 & 55652 & 88 & T7 & 160 & 7.65 & 0.03 & 0.40 & \dots & \\
HD 237944 & 93470 & 55569 & 169 & T7 & 675 & 9.27 & 0.07 & 0.50 & 5.41$\pm$0.013 & rotation\\
HD 237944 &  &  &  &  &  &  & 0.25 &  & 5.5075$\pm$0.0003 &   eclipsing \\
\noalign{\smallskip}\hline
\end{tabular}
\end{table*}

\subsection{Rotational periods and spot amplitudes from photometry}

Table~\ref{T6} is the summary of the results from our photometric
analysis. A total of 56 stars out of the total sample of 60 targets
were observed, of which 39 have a detectable period. The time
coverage and the sampling are very different from target to target,
ranging from a minimum of 20 nights for QY~Hya (HD\,105575) to 4,160
nights ($>$11 yrs) for HY~CMa (SAO\,151224) with a sampling of
all-night-long monitoring (QY~Hya) to one observation per night as
for most targets. This diversity of data wealth makes a coherent
spot-modelling analysis not feasible for the present paper and we
just focus on the determination of average light-curve parameters. A
more detailed analysis per target is planned for the future whenever
feasible (see, e.g. Strassmeier et al. \cite{hd123351} for the most
recent example of HD~123351, excluded from this paper).

Most relevant for this paper is the determination of a precise
photometric period, $P_{\rm phtm}$, that can be interpreted as the
stellar rotation period. This allows a superior measurement of
rotation with respect to spectroscopic $v\sin i$ measurements
because they are independent of the (unknown) inclination of the
rotational axis and can be determined much more precisely. To
proceed, we first pre-filtered all photometric data by excluding
data points with an rms of $\ge$0.02~mag in order to remove data
grossly affected by clouds. Periods were then searched for either
in Johnson $V$ or in Str\"omgren $y$ or, for the stars where we
have both passbands, in the combined $y+V$ series. For
light-curves with a highly harmonic content, the Lomb periodogram
(Lomb \cite{lomb}) in the formulation of Scargle (\cite{sca}) was
applied. In cases where a period could have been affected by the
window function, the CLEAN algorithm (Roberts et
al.~\cite{rob:etal}) was employed to verify, but not to alter, the
period identification from the Lomb periodogram. For light curves
with a highly non-harmonic content, i.e. for eclipsing binaries, a
string-length minimization, a variant of a Lafler-Kinman (Lafler
\& Kinman~\cite{laf:kin}) statistic described in Dworetsky
(\cite{msl}), was used to determine the final photometric period.
Some of the stars had already had a period determined in our
paper~I but the period in the present paper supersede these
values. In Fig.~\ref{F4}, a phased light curve from all data for
each of the variable targets is shown. These light curves are
phased with the photometric period indicated in the graph and the
respective time of periastron listed in Tables~\ref{T4} or
\ref{T5}, or ascending node for circular orbits. Single-star zero
epochs are arbitrary.

Errors for the periods were obtained by using a method sometimes
referred to as ``refitting to synthetic data sets'' (e.g.
Ford~\cite{ford}). This method estimates confidence intervals by
synthesizing a large number of data sets (typically 10$^5$) out from
the original data by adding Gaussian random values to the
measurements proportional to the actual rms of the data. The
respective period-search algorithm is then applied to these
synthetic data and the resulting standard deviation assumed to be
the standard deviation of the original period. This value is given
in Table~\ref{T6} as the error for $P_{\rm phtm}$.

Note that Table~\ref{T6} strictly lists photometric periods. We
interpret these usually as stellar rotational periods but emphasize
that care must be taken because, e.g., in case the star is a close
binary, an ellipticity effect could mimic a spot light curve with a
$P_{\rm orb}/2$ photometric period that is easily misinterpreted as
the stellar rotation period. For one system, HD\,138157, we
interpret the 55-mmag amplitude partly due to the ellipticity
effect. Its true rotational period is close to the orbital period.
For five of our six eclipsing binaries the rotational modulation
appears to have the same period as the orbital motion. For one case,
HD\,190642, a newly discovered eclipsing pair, the spot wave appears
doubled humped and gives half of the true rotation period.

The typical error of the magnitude zero point is around
0.01--0.02~mag, but never higher than 0.05~mag. The wave
amplitude, $\Delta V$, in Table~\ref{T6} is the maximum observed
amplitude within our data coverage. For the eclipsing systems, we
cut out the data points during eclipse phases and then
redetermined the maximum amplitude. The color index, C.I., is the
average color index over all observations. In case of T7
observations it is ($V$-$I$)$_C$, for T6 it is $b$-$y$, if
available. Variations in the C.I. are present but are in general
too noisy for period analysis. Most of the observations of
HD~136655 had to be discarded due to obvious instrumental
problems. Its remaining data shows no evidence for photometric
variability. Orbital period and $T_0$ of the eclipsing binary HD~105575
are equally constrainable from photometry than from radial
velocities. If $e$=0 is adopted, we obtain $T_0$=2,454,160.772 and
$P_{\rm orb}$=0.29238$\pm$0.00009~d just from our 20-day long time
series. However, because the photometry was obtained around
JD2,451,250, i.e. almost 10,000 eclipses prior to the spectroscopic
data, the combination of the two data sets still results in a loss
of the cycle count and could not be combined. Table~\ref{T5} gives
the orbit computed only with the radial velocities.

\subsection{Rotational velocities}

Selected spectra of our program stars were subjected to a
de-noising procedure based on a principal component analysis
developed for Doppler imaging (Carroll et al.~\cite{carr:an},
\cite{carr:iau}). This procedure employs typically a total of around 1100
spectral lines in the wavelength range 480--850~nm to determine
the noise spectrum. This noise spectrum can then be used to de-noise an arbitrary wavelength section. The only
free parameter is the number of principal components. No strict
rule can be given for the choice of this value but we followed the
prescription laid out in Mart\'inez Gonz\'alez et al.
(\cite{pca}). It suggests 15 components for our typical late
spectral types and wavelength coverage. It is then used to
de-noise a well-known and unblended spectral line from which we
measure $v\sin i$ under the assumption of a temperature and
gravity dependent macroturbulence. For this paper, we have chosen
the well-known and unblended Fe\,{\sc i} 549.7516~nm line with an excitation energy of
1.011~eV, a transition probability of $\log gf = -2.849$, and a
typical microturbulence of 2.0~\kms . The radial-tangential
macroturbulences, $\zeta_{\rm RT}$, are taken from Gray
(\cite{gray}), as listed in Fekel (\cite{fcf97}). This
macroturbulence is then subtracted from the measured total line
broadening. At this point, we note that Gray (\cite{gray}) gave most probable
macroturbulent velocities, which are 1.414 times greater than the
rms values given here. The total line broadening is measured independently from fits with synthetic spectra over a large range of wavelengths (see next section) and the resulting $v\sin i$ measures were then compared and found to agree within their estimated errors. The final values of $v\sin i$ and $\zeta_{\rm RT}$ for all our program stars are tabulated in Table~\ref{T7} for further reference.

Up to three nightly (and two daily) Th-Ar calibration frames are
used to monitor the spectrograph focus throughout the year. The
line widths of Th-Ar emission lines are determined automatically
and an unweighted average stored in the STELLA data base. No focus
drifts were evident during the time in question. However, manual
refocusing was done after every maintenance run. The telescope
focus is being controlled and adjusted daily by inserting a focus
pyramid into the telescope beam. It splits a stellar image into
four equally distant images if the telescope is in focus (see
Granzer et al. \cite{malaga2}). The STELLA control system issues a
message to the operator in Potsdam if this had to be corrected.

\begin{figure*}
\center
\includegraphics[angle=0,width=170mm,clip]{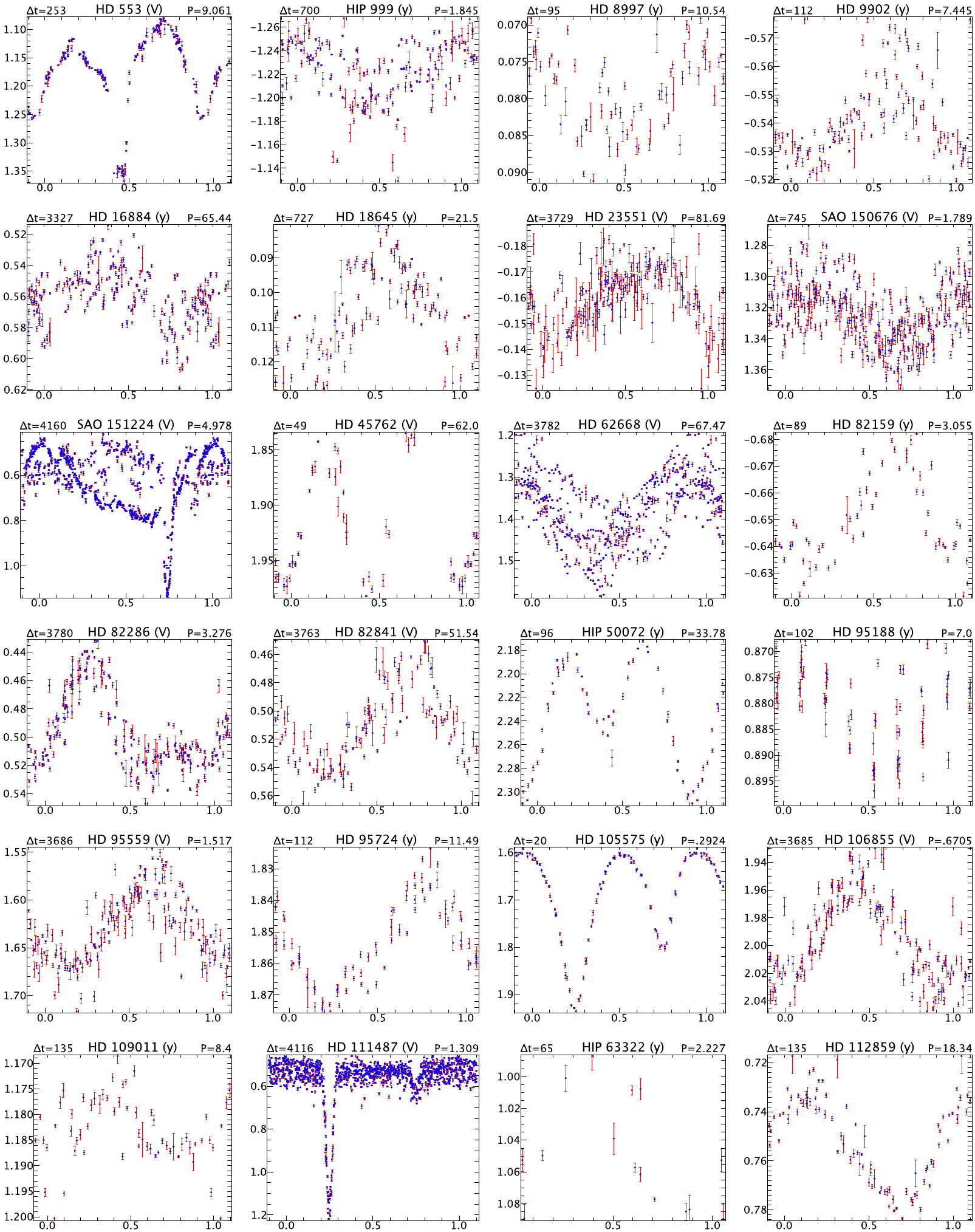}
\caption{Light curves for the stars in Table~\ref{T6}. The time
coverage of the observations, $\Delta t$, and the photometric
period, $P$, are indicated on the top of each panel, both in days. The dots are
the $V$ or $y$ observations from the full epochs given in
Table~\ref{T6}. Error bars indicate the standard deviation from
three individual measurements. Photometric phase has been computed
with the photometric period and the respective time of periastron or
ascending node for circular orbits. Single star light curves are
plotted with an arbitrary starting time. Note that most of the
scatter is due to intrinsic spot changes and not due to instrumental
origin. The standard error of a mean was 4--6~mmag for the $V$ data
and 1.2--3~mmag for the $y$ data. }\label{F4}
\end{figure*}
\setcounter{figure}{3}
\begin{figure*}[t]
\center
\includegraphics[angle=0,width=170mm,clip]{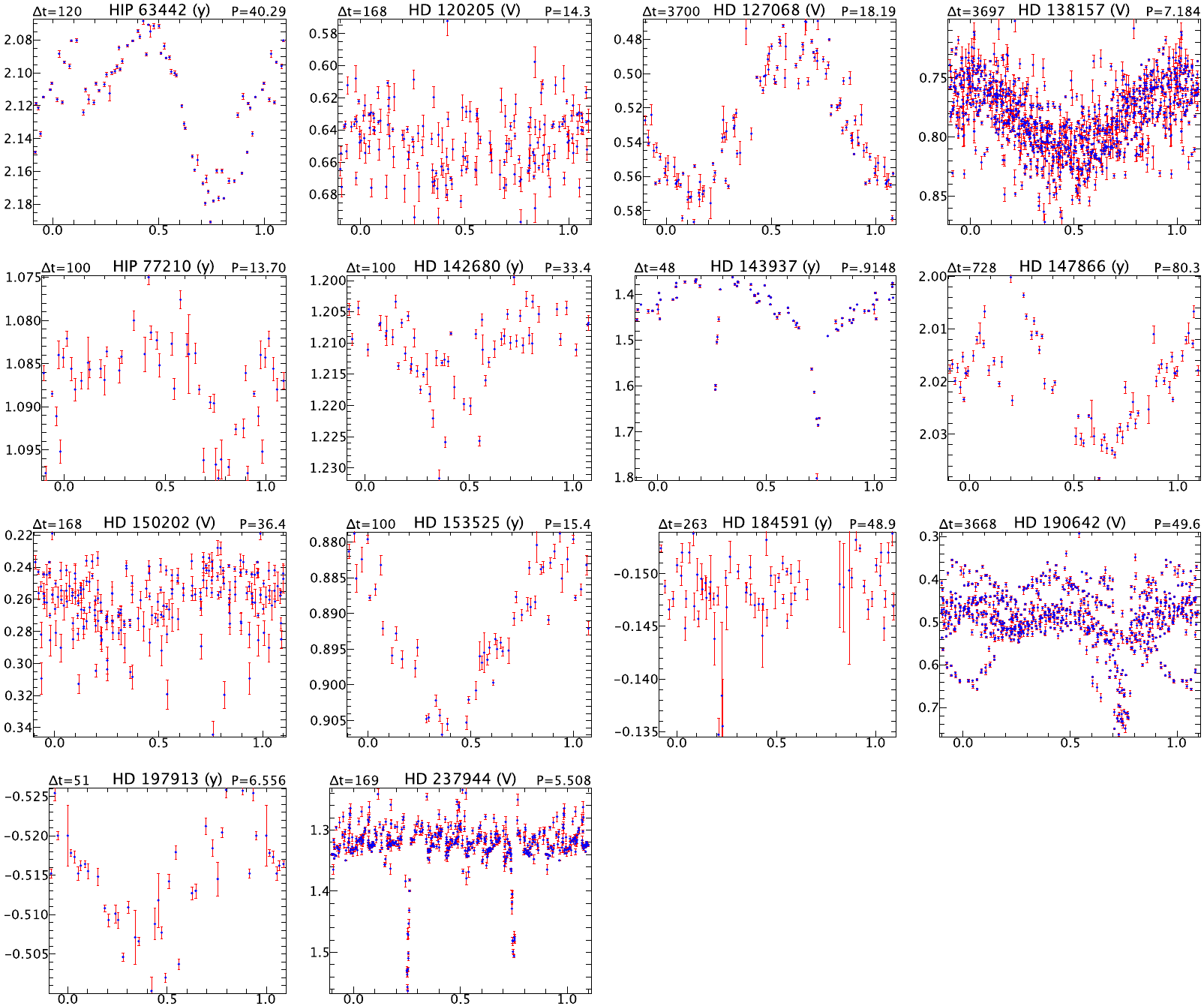}
\caption{(continued)}
\end{figure*}

\subsection{Stellar atmospheric parameters}

Five selected spectral orders (\#28, 29, 33, 37, 39) from the
STELLA/SES spectra are used to determine the stellar effective
temperature, the gravity and the metallicity. These orders cover
the wavelength ranges 549--556, 560--567, 583--591, 608--616, and
615--623~nm. Our numerical tool PARSES (``PARameters from SES''; Allende-Prieto \cite{all04})
is implemented as a suite of Fortran~programs in the STELLA data
analysis pipeline. It is based on the synthetic spectrum fitting
procedure described in Allende-Prieto et al.~(\cite{all}). Model atmospheres and synthetic spectra were taken
from the ATLAS9 CD (Kurucz~\cite{kur}). Synthetic spectra are
pre-tabulated with metallicities between $-2.5$~dex and +0.5~dex in
steps of 0.5~dex, logarithmic gravities between 1.5 and 5 in steps
of 0.5, and temperatures between 3500\,K and 7,250\,K in steps of
250\,K for a wavelength range of 380--920\,nm. All calculations
were done with a microturbulence of 2~\kms. This grid is then used
to compare with the five selected echelle orders of each spectrum.
Free parameters are $T_{\rm eff}$, $\log g$, $[$Fe/H$]$, and line
broadening $\sqrt{(v\sin i)^2+\zeta_{\rm RT}^2}$, where
$\zeta_{\rm RT}$ is the radial-tangential macroturbulence, again
adopted from Gray~(\cite{gray}) and was fixed during the fit.
Internal errors are estimated from the rms of the solutions to the
five echelle orders and were for $T_{\rm eff}$ typically 20--30~K,
for $\log g$ typically 0.06~dex, and for $[$Fe/H$]$ typically
0.03~dex.

Double-lined spectra can not be run automatically through the
PARSES routines. We first apply a similar disentangling technique
as for the automatic radial-velocity measurement of triple-lined
spectra, i.e. we remove once the primary and once the secondary
component from the combined spectrum at well-separated phases in
order to create a single-star spectrum for each component. The
resulting spectra were then subjected to the PARSES
synthetic-spectrum fitting. This was done only for the five
echelle orders that are employed for the PARSES fit (see above)
and not for the entire 80 orders. Errors are generally larger for
SB2s than for SB1 and single systems due to continuum cross talk.
External errors are again estimated from the rms of the solutions
to the five echelle orders and were for $T_{\rm eff}$ typically
100~K, for $\log g$ typically 0.3~dex, and for $[$Fe/H$]$
typically 0.3~dex. However, the range of errors is systematically
larger for the SB2 systems than for the SB1 and single stars
because of the large range of brightness differences between the
components in our sample (ranging from $\approx$1:1, i.e. nearly
equal components, to 12:1 for HIP\,999).

Lithium abundances are determined for all individual spectra that
had sufficient S/N in the respective echelle order. The time
averaged results are listed in Table~\ref{T7} with respect to the
$\log n$(H) = 12.00 scale for hydrogen. A double Gaussian fit to
Li\,{\sc i} 6707.8 \AA \ and the nearby Fe\,{\sc i} 6707.4-\AA \
line yields an individual equivalent width for the lithium line.
Despite that these values are still the combined equivalent width
for the close blend from the $^6$Li and the $^7$Li isotopes, it
effectively removes the Fe\,{\sc i} blend. Weaker CN features are
neglected in this study because of limited S/N and spectral
resolution. The listed equivalent widths are the averages of the
individual measures with unit weight. The errors in Table~\ref{T7}
are still the O--C error from an individual Gaussian fit but again
averaged over all measures in time. Because the rms values of the
equivalent widths of an entire time series per star were never
larger than 3~$\sigma$ of above O--C error, we do not explicitly
list them in Table~\ref{T7}. However, these rms are useful to
judge eventual long-term variability. SB2's are measured only at
phases of quadrature, where blending is minimized. The curves of
growth from Pavlenko \& Magazz\'u (\cite{pav:mag}) are then used
to convert the equivalent width to a logarithmic lithium abundance
using the non-LTE transformation and the effective temperatures
and gravities from Table~\ref{T7}.

%

\begin{table*}
\caption{Determination of astrophysical properties. Column SB
denotes the binary component in case the system is SB2 or even SB3
(S means single star, $a$ means primary component, $b$ secondary
component, $c$ tertiary component). The parallaxes are from the
revised \emph{Hipparcos} catalog (van Leeuwen \cite{nhip}).
Effective temperatures, gravities and metallicities are from the
STELLA PARSES analysis. Note that errors are mean internal errors
from the fits to five spectral orders. The Li abundance is from
equivalent-width measures of the Li\,{\sc i}~670.8nm line
converted with the NLTE calibration of Pavlenko \& Magazz\'u
(\cite{pav:mag}) and the temperatures listed here. $\zeta_{\rm
RT}$ is the radial-tangential macroturbulence adopted. }\label{T7}
\begin{tabular}{llllllllll}
\hline \noalign{\smallskip}
Name  & SB & $\pi$ & $T_{\rm eff}$ & $\log g$ & $[$Fe/H$]$ & $W_{\rm Li}$ &$\log n$(Li) & $\zeta_{\rm RT}$ & $v\sin i$  \\
      &    & (mas)& (K)           & (m\,s$^{-2}$)&        & (m\AA)       &  (H = 12.00)& \multicolumn{2}{c}{(\kms )} \\
\noalign{\smallskip}\hline \noalign{\smallskip}
HD 18645  & S& 8.71  & 5445$\pm$30 & 3.53$\pm$0.04 & --0.19$\pm$0.03 &14$\pm$4 &1.47$\pm$0.15 &3& 11.0$\pm$0.2  \\
HD 23551  & S& 5.00  & 5045$\pm$25 & 3.04$\pm$0.05 & --0.11$\pm$0.02 & 12$\pm$4 & 1.03$\pm$0.16 &3& 6.4$\pm$0.2  \\
HD 24053  & S& 30.74 & 5600$\pm$25 & 4.09$\pm$0.03 & --0.01$\pm$0.02& 49$\pm$8 & 2.26$\pm$0.12 &3& $<$3 \\
SAO150676 & S& \dots & 5750$\pm$70 & 4.54$\pm$0.11 & --0.16$\pm$0.07& 97$\pm$10 & 2.71$\pm$0.10 &3& 23.6$\pm$0.7  \\
HD 43516  & S& 4.10  & 5120$\pm$50 & 2.86$\pm$0.07 & --0.15$\pm$0.04& 39$\pm$6 & 1.72$\pm$0.15 &3& $<$3 \\
HD 76799  & S& 6.13  & 4915$\pm$25 & 3.26$\pm$0.07 & --0.12$\pm$0.03& 12$\pm$4 & 0.86$\pm$0.18 &3& 7.8$\pm$0.3  \\
HIP46634  & S& 27.46 & 5140$\pm$35 & 4.44$\pm$0.04 & --0.01$\pm$0.03& 33$\pm$6 & 1.59$\pm$0.14 &2& $<$3 \\
HD 95188  & S& 27.63 & 5330$\pm$40 & 4.39$\pm$0.06 & --0.18$\pm$0.03& 13$\pm$6 & 1.29$\pm$0.24 &3& 6.9$\pm$0.7  \\
HD 95724  & S& 28.37 & 4960$\pm$20 & 4.45$\pm$0.06 & +0.06$\pm$0.02& 15$\pm$5 & 0.97$\pm$0.16 &2& $<$3 \\
HD 104067 & S& 48.04 & 4820$\pm$20 & 4.37$\pm$0.04 & --0.03$\pm$0.02& 16$\pm$4 & 0.83$\pm$0.15 &2& 8.1$\pm$0.3  \\
HD 108564 & S& 35.30 & 4560$\pm$25 & 4.40$\pm$0.06 & --0.90$\pm$0.03& 21$\pm$7 & 0.66$\pm$0.20 &2& 14.8$\pm$1.4  \\
HD 120205 & S& 31.78 & 5260$\pm$30 & 4.36$\pm$0.05 & +0.03$\pm$0.03& 10$\pm$5 & 1.13$\pm$0.18 &2& $<$3 \\
HD 136655 & S& 24.87 & 5010$\pm$25 & 4.27$\pm$0.03 & +0.09$\pm$0.04& 11$\pm$5 & 0.92$\pm$0.20 &3& $<$3 \\
HD 153525 & S& 57.13 & 4775$\pm$25 & 4.47$\pm$0.05 & --0.12$\pm$0.03& 12$\pm$5 & 0.66$\pm$0.20 &2& $<$3 \\
HD 155802 & S& 34.27 & 5010$\pm$15 & 4.46$\pm$0.04 & --0.15$\pm$0.02& 13$\pm$4 & 0.96$\pm$0.13 &2& 5$\pm$1  \\
HD 171067 & S& 39.73 & 5520$\pm$35 & 4.03$\pm$0.04 & --0.15$\pm$0.03& 12$\pm$4 & 1.43$\pm$0.16 &3& $<$3 \\
HD 184591 & S& 4.58  & 4800$\pm$40 & 2.82$\pm$0.07 & --0.18$\pm$0.03& 14$\pm$6 & 0.80$\pm$0.24 &3& 4.7$\pm$0.8  \\
HD 218739 & S& 34.06 & 5670$\pm$20 & 4.19$\pm$0.05 & --0.07$\pm$0.02& 135$\pm$6 & 2.80$\pm$0.05 &3& 7.6$\pm$0.2  \\
\noalign{\smallskip}
HD 40891  & $a$ & 32.95 & 5140$\pm$25 & 4.37$\pm$0.05 & --0.10$\pm$0.02 & 11$\pm$4 & 1.03$\pm$0.16 &3& \dots  \\
HD 62668  & $a$ & 4.97  & 4660$\pm$40 & 2.92$\pm$0.10 & --0.30$\pm$0.04 & 82$\pm$8 & 1.60$\pm$0.10 &3& 21$\pm$1  \\
HD 66553  & $a$ & 28.03 & 5275$\pm$40 & 4.33$\pm$0.04 & +0.09$\pm$0.02 & 12$\pm$7 & 1.20$\pm$0.27 &2& \dots  \\
HD 82159  & $a$ & 21.11 & 5150$\pm$65 & 4.62$\pm$0.08 & --0.13$\pm$0.03 & 109$\pm$7 & 2.22$\pm$0.10 &2& 12.5$\pm$0.5  \\
HD 82841  & $a$ & 4.09  & 4620$\pm$30 & 2.94$\pm$0.07 & --0.30$\pm$0.03 & 19$\pm$6 & 0.74$\pm$0.19 &3& 8.1$\pm$0.3  \\
HIP 50072 & $a$ & 3.15  & 4615$\pm$39 & 3.08$\pm$0.08 & --0.33$\pm$0.04 & 22$\pm$6 & 0.80$\pm$0.21 &3& 20.8$\pm$0.8  \\
HD 112099 & $a$ & 38.12 & 5095$\pm$20 & 4.37$\pm$0.04 & --0.14$\pm$0.02 & 11$\pm$4 & 1.00$\pm$0.16 &2& 4.1$\pm$0.2  \\
HIP 63322 & $a$ & 26.24 & 4995$\pm$27 & 4.48$\pm$0.07 & --0.23$\pm$0.04 & 142$\pm$8 & 2.19$\pm$0.07 &2& 6$\pm$1  \\
HIP 63442 & $a$ & 3.31  & 4530$\pm$30 & 2.85$\pm$0.08 & --0.35$\pm$0.04 & 41$\pm$6 & 1.07$\pm$0.13 &3& 22.3$\pm$0.6  \\
HD 138157 & $a$ & 5.07  & 4845$\pm$82 & 3.24$\pm$0.12 & --0.36$\pm$0.06 & 20$\pm$5 & 1.03$\pm$0.22 &3& 40$\pm$1  \\
HD 147866 & $a$ & 5.18  & 4580$\pm$15 & 2.71$\pm$0.05 & --0.24$\pm$0.02 & 13$\pm$4 & 0.49$\pm$0.14 &3& 5$\pm$1  \\
HD 150202 & $a$ & 3.75  & 5010$\pm$25 & 3.06$\pm$0.07 & --0.12$\pm$0.03 & 25$\pm$5 & 1.32$\pm$0.16 &3& 7.8$\pm$0.2  \\
$\epsilon$ UMi & $a$ & 9.41 & 5215$\pm$50 & 3.21$\pm$0.08 & --0.25$\pm$0.04 & 38$\pm$6 & 1.80$\pm$0.16 &3& 25.6$\pm$0.2  \\
HD 190642 & $a$ & 4.48  & 4760$\pm$65 & 2.96$\pm$0.11 & --0.47$\pm$0.05 & 19$\pm$6 & 0.90$\pm$0.23 &3& 19.6$\pm$1.0  \\
HD 202109 & $a$ & 21.62 & 4910$\pm$15 & 2.14$\pm$0.04 & --0.08$\pm$0.02 & 21$\pm$5 & 1.13$\pm$0.15 &3& \dots  \\
\noalign{\smallskip}
HD 553   & $a$ & 4.74  & 4645$\pm$90 & 3.38$\pm$0.36 & --0.23$\pm$0.08 & 22$\pm$4 & 0.85$\pm$0.22 &3& 41$\pm$2  \\
         & $b$ &       & 5940$\pm$150& 4.40$\pm$0.77 & --0.41$\pm$0.26 & 21$\pm$6 & 2.07$\pm$0.26 &2& 27$\pm$  \\
HIP 999  & $a$ & 24.69 & 5280$\pm$70 & 4.40$\pm$0.30 & --0.31$\pm$0.04 & 20$\pm$4 & 1.44$\pm$0.19 &2& 27$\pm$1  \\
         & $b$ &       & 4840$\pm$150 & 3.6$\pm$1.2 & +0.3$\pm$0.3 & 15$\pm$8 & 0.83$\pm$0.42 & 3& 50$\pm$2  \\
HD 8997  & $a$ & 43.16 & 5060$\pm$65 & 4.53$\pm$0.25 & --0.12$\pm$0.04 & 10$\pm$6 & 0.91$\pm$0.27 & 2& 3.0$\pm$0.5  \\
         & $b$ &       & 4525$\pm$50 & 4.74$\pm$0.26 & +0.02$\pm$0.18 & 11$\pm$4 & 0.34$\pm$0.21 & 3& 8$\pm$1  \\
HD 9902  & $a$ & \dots & 5130$\pm$80 & 4.0$\pm$0.4 & --0.50$\pm$0.16 & 52$\pm$7 & 1.85$\pm$0.16 & 2& 12$\pm$1  \\
         & $b$ &       & 5715$\pm$150 & 3.37$\pm$0.20 & --0.38$\pm$0.15 & 26$\pm$8 & 1.97$\pm$0.31 & 5& 2.0$\pm$0.5  \\
HD 16884 & $a$ & \dots & 4500$\pm$50 & 2.4$\pm$0.4 & +0.29$\pm$0.11 & 31$\pm$7 & 0.85$\pm$0.20 &3& 6$\pm$1  \\
         & $c$ &       & 4535$\pm$70 & 3.4$\pm$0.5 & --0.26$\pm$0.15 & 10$\pm$9 & 0.36$\pm$0.38 & 3& 44$\pm$2  \\
HD 18955 & $a$ & 20.56 & 5410$\pm$80 & 4.84$\pm$0.10 & +0.03$\pm$0.06 & 9$\pm$2 & 1.25$\pm$0.13 & 2& 9$\pm$1  \\
         & $b$ &       & 5095$\pm$150 & 4.8$\pm$0.3 & +0.22$\pm$0.17 & 8$\pm$2 & 0.90$\pm$0.23 &2 & 5$\pm$1  \\
SAO151224& $a$ & \dots & 4595$\pm$110 & 3.7$\pm$0.4 & --0.18$\pm$0.14 & 45$\pm$7 & 1.20$\pm$0.23 &3 & 45$\pm$2  \\
         & $b$ &       & 5420$\pm$105 & 4.8$\pm$0.3 & --0.09$\pm$0.12 & 22$\pm$6 & 1.67$\pm$0.25 &2 & 6$\pm$2  \\
HD 45762 & $a$ & 3.45  & 4630$\pm$100 & 3.0$\pm$0.7 & --0.48$\pm$0.24 & 31$\pm$5 & 1.02$\pm$0.22 &3 & 60$\pm$2  \\
         & $c$ &       & 6015$\pm$150 & 3.5$\pm$1.0 & --0.19$\pm$0.18 & \dots & \dots &5& 16$\pm$2 \\
HD 50255 & $a$ & 30.26 & 5580$\pm$100 & 4.07$\pm$0.12 & --0.18$\pm$0.11 & 45$\pm$6 & 2.19$\pm$0.18 &2& 3.0$\pm$1.0  \\
         & $b$ &       & 4980$\pm$150 & 3.7$\pm$1.0 & --1.0$\pm$0.8 & 9$\pm$4 & 0.81$\pm$0.30 &3 & 5$\pm$1  \\
\noalign{\smallskip}\hline
\end{tabular}
\end{table*}
\setcounter{table}{6}
\begin{table*}
\caption{(continued)}
\begin{tabular}{llllllllll}
\hline \noalign{\smallskip}
Name  & SB & $\pi$ & $T_{\rm eff}$ & $\log g$ & $[$Fe/H$]$ & $W_{\rm Li}$ &$\log n(Li)$ & $\zeta_{\rm RT}$ & $v\sin i$  \\
      &    & (mas)& (K)           & (m\,s$^{-2}$)&        & (m\AA)       &             & \multicolumn{2}{c}{(\kms )} \\
\noalign{\smallskip}\hline \noalign{\smallskip}
HD 61994 & $a$ & 35.13 & 5630$\pm$150 & 4.13$\pm$0.11 & +0.03$\pm$0.12 & 17$\pm$4 & 1.69$\pm$0.23 &2& 4$\pm$1  \\
         & $b$ &       & 4775$\pm$150 & 4.6$\pm$0.6 & --0.27$\pm$0.33 & 9$\pm$4 & 0.56$\pm$0.33 &3 & 9$\pm$1  \\
HD 73512 & $a$ & 39.28 & 5110$\pm$115 & 4.58$\pm$0.12 & --0.25$\pm$0.04 & 11$\pm$5 & 1.00$\pm$0.17 &2 & 5$\pm$1  \\
         & $b$ &       & 4600$\pm$130 & 4.4$\pm$0.5 & --0.37$\pm$0.17 & 11$\pm$3 & 0.41$\pm$0.27 &3 & 5$\pm$1  \\
HD 82286 & $a$ & 9.57  & 4800$\pm$40 & 3.73$\pm$0.31 & --0.05$\pm$0.09 & 21$\pm$5 & 1.00$\pm$0.17 & 3& 37.5$\pm$1.5  \\
         & $b$ &       & 4785$\pm$150 & 3.92$\pm$0.35 & --0.42$\pm$0.20 & 24$\pm$6 & 1.04$\pm$0.32 &3 & 40.5$\pm$2  \\
HD 93915 & $a$ & 22.78 & 5520$\pm$105 & 4.28$\pm$0.24 & --0.29$\pm$0.09 & 12$\pm$4 & 1.43$\pm$0.22 &2 & 5$\pm$1  \\
         & $b$ &       & 5315$\pm$120 & 4.25$\pm$0.24 & --0.22$\pm$0.09 & 9$\pm$3 & 1.17$\pm$0.22 &2 & 3$\pm$1  \\
HD 95559 & $a$ & 18.43 & 5090$\pm$110 & 4.47$\pm$0.27 & --0.10$\pm$0.12 & 16$\pm$6 & 1.14$\pm$0.28 &2 & 33$\pm$2  \\
         & $b$ &       & 5065$\pm$150 & 4.56$\pm$0.33 & +0.07$\pm$0.12 & 20$\pm$6 & 1.23$\pm$0.32 &2 & 33$\pm$2  \\
HD 105575& $a$ & 19.77 & 5650$\pm$150 & 4.73$\pm$0.11 & --0.46$\pm$0.08 & 14$\pm$2 & 1.61$\pm$0.17  &2& 42$\pm$15  \\
HD 106855& $a$ & 22.88 & 4750$\pm$110 & 4.48$\pm$0.33 & --0.87$\pm$0.19 & 22$\pm$4 & 0.91$\pm$0.24 &2& 30$\pm$2  \\
         & $b$ &       & 4950$\pm$150 & 4.0$\pm$0.8 & --0.5$\pm$0.4 & 11$\pm$4 & 0.85$\pm$0.30 &3 & 5$\pm$2  \\
HD 109011& $a$ & 42.13 & 5030$\pm$75 & 4.59$\pm$0.27 & --0.17$\pm$0.10 & 30$\pm$5 & 1.42$\pm$0.19 & 2& 5$\pm$1  \\
         & $b$ &       & 4900$\pm$150 & 3.7$\pm$1.0 & +0.2$\pm$0.3 & 13$\pm$5 & 0.84$\pm$0.33 & 3& 6$\pm$1  \\
HD 111487& $a$ & \dots & 5500$\pm$150 & 4.47$\pm$0.28 & --0.36$\pm$0.11 & 13$\pm$4 & 1.45$\pm$0.26 & 2& 42$\pm$2  \\
         & $b$ &       & 4700$\pm$150 & 4.6$\pm$0.5 & --0.1$\pm$0.5 & 14$\pm$6 & 0.64$\pm$0.38 &3 & 28$\pm$2  \\
HD 112859& $a$ & 5.24  & 4780$\pm$120 & 3.41$\pm$0.38 & --0.24$\pm$0.15 & 47$\pm$7 & 1.45$\pm$0.23 &3 & 20$\pm$1  \\
         & $b$ &       & 6150$\pm$150 & 2.5$\pm$0.5 & +0.05$\pm$0.3 & \dots & \dots &5 & 5$\pm$1  \\
HD 127068& $a$ & 9.75  & 4850$\pm$135 & 3.61$\pm$0.27 & --0.48$\pm$0.07 & 76$\pm$5 & 1.79$\pm$0.19 &3 & 6$\pm$1 \\
         & $b$ &       & 5560$\pm$150 & 3.9$\pm$0.8 & --0.39$\pm$0.14 & 10$\pm$5 & 1.46$\pm$0.31 &3 & 8$\pm$2  \\
HIP 77210& $a$ & 20.73 & 5190$\pm$120 & 4.41$\pm$0.26 & --0.43$\pm$0.10 & 14$\pm$4 & 1.18$\pm$0.25 &2 & 3$\pm$1  \\
         & $b$ &       & 4720$\pm$150 & 4.74$\pm$0.31 & +0.08$\pm$0.11 & 16$\pm$5 & 0.73$\pm$0.33 &3 & 5$\pm$1  \\
HD 142680& $a$ & 28.23 & 4885$\pm$110 & 4.35$\pm$0.26 & --0.22$\pm$0.05 & 5$\pm$3 & 0.54$\pm$0.24 &3 & 3$\pm$1  \\
         & $b$ &       & 4500$\pm$150 & 4.6$\pm$0.6 & --0.51$\pm$0.36 & 8$\pm$3 & 0.21$\pm$0.29 &3 & 5$\pm$1  \\
HD 143937& $a$ & 23.73 & 5160$\pm$140 & 4.48$\pm$0.16 & --0.30$\pm$0.13 & 12$\pm$6 & 1.11$\pm$0.33 &2 & 55$\pm$2  \\
         & $b$ &       & 4730$\pm$150 & 4.72$\pm$0.17 & --0.01$\pm$0.13 & 13$\pm$6 & 0.65$\pm$0.35 &3 & 45$\pm$2  \\
HD 197913& $a$ & 17$\pm$6$^a$& 5520$\pm$100 & 4.18$\pm$0.32 & --0.06$\pm$0.11 & 35$\pm$4 & 2.00$\pm$0.17 &2 & 5$\pm$1  \\
         & $b$ &       & 5340$\pm$120 & 4.20$\pm$0.45 & --0.08$\pm$0.10 & 19$\pm$6 & 1.47$\pm$0.28 &2 & 6$\pm$1  \\
HD 199967& $a$ & 13.57 & 5850$\pm$100 & 4.26$\pm$0.22 & --0.24$\pm$0.17 & 28$\pm$7 & 2.13$\pm$0.22 &2 & 11$\pm$1  \\
         & $b$ &       & 5835$\pm$150 & 4.4$\pm$0.4 & --0.25$\pm$0.13 & 19$\pm$8 & 1.91$\pm$0.33 &2 & 9$\pm$1  \\
HD 226099& $a$ & 35.46 & 5200$\pm$80 & 4.45$\pm$0.34 & --0.92$\pm$0.23 & 12$\pm$4 & 1.12$\pm$0.21 &2 & 2$\pm$1  \\
         & $b$ &       & 4560$\pm$80 & 4.20$\pm$0.36 & --1.1$\pm$0.1 & 9$\pm$4 & 0.30$\pm$0.25 &3 & 4$\pm$1  \\
HD 237944& $a$ & 10.62 & (5630)$^b$ & \dots & \dots & \dots & \dots & 3& 11$\pm$2  \\
         & $b$ &       & (5560)$^b$ & \dots & \dots & \dots & \dots & 3& 9$\pm$2  \\
         & $c$ &       & (4900)$^b$ & \dots & \dots & \dots & \dots & 3& 11$\pm$3  \\
\noalign{\smallskip}\hline
\end{tabular}

\vspace{1mm}
 $^a$from Jenkins (\cite{jenkins}),\\
 $^b$assumed from spectral classification.

\end{table*}

\begin{figure*}[!tb]
\center
\includegraphics[angle=0,width=160mm]{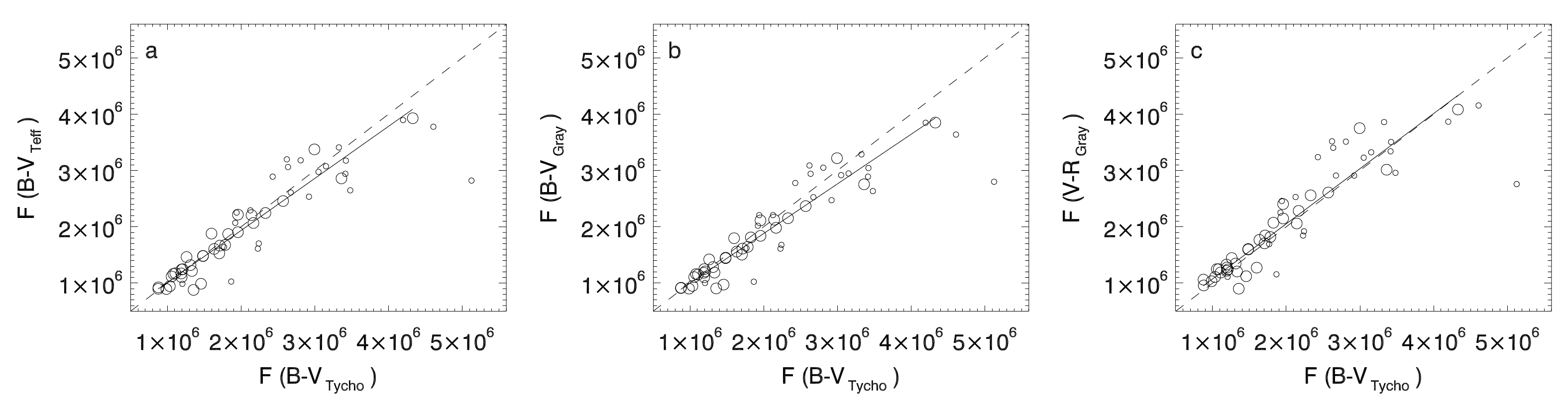}
\caption{A comparison of absolute continuum flux at \Halpha\ from
different calibrations. Only our single and SB1 stars are plotted.
{\bf a} Fluxes based on the effective temperatures from
Table~\ref{T7} converted to $B-V$ with the $T_{\rm eff}$ vs. $B-V$
calibration of Flower (\cite{f96}) versus the flux based on the
$B-V$ listed in the Tycho-2 catalog. {\bf b} Again our effective
temperatures converted to $B-V$ but with the $T_{\rm eff}$
calibration of Gray (\cite{gray}) versus the $B-V$ listed in the
Tycho-2 catalog. {\bf c} Our effective temperatures converted to
$V-R$ with the $T_{\rm eff}$ calibration of Gray (\cite{gray})
versus the $B-V$-based fluxes listed in the Tycho-2 catalog.
Minimized rms errors of the flux residuals are achieved in panel
{\bf a}. The dashed line is just a 45-degree line. The solid lines
are non-weighted regression fits. }\label{F-comp}
\end{figure*}

\subsection{\Halpha - core flux}\label{S4-Ha}

We measure the central 1-\AA \ portion of the \Halpha \ line from
our continuum-normalized spectra and relate it to the absolute
continuum flux, ${\cal F}_{\rm c}$, expected at \Halpha . The latter
is obtained from the relations provided by Hall (\cite{hall}) for
various Morgan-Keenan (MK) classes and color ranges, i.e.
\begin{eqnarray}\label{eq1}
\log {\cal F}_{\rm c} = 7.538 - 1.081 \ (B-V) \\
{\rm for \ MK \ I-V} \ {\rm and} \  0<(B-V)<1.4 \ , \nonumber \\
\log {\cal F}_{\rm c} = 7.518 - 1.236 \ (V-R) \\
{\rm for \ MK \ V} \ {\rm and} \     0<(V-R)<1.4 \ , \nonumber \\
\log {\cal F}_{\rm c} = 7.576 - 1.447 \ (V-R) \\
{\rm for \ MK \ I-IV}\ {\rm and} \   0<(V-R)<1.8 \ . \nonumber
\end{eqnarray}
The stellar surface \Halpha -core flux in erg\,cm$^{-2}$s$^{-1}$
is then computed from the measured 1-\AA \ equivalent width,
$W_{\rm core}$, under the spectrum and zero intensity,
\begin{equation}\label{eq2}
{\cal F}_{{\rm H}\alpha}  = W_{\rm core} \ {\cal F}_{\rm c} \ .
\end{equation}
In this way the equivalent width increases with increased line-core
filling due to chromospheric emission while the continuum flux
reflects the dependency on stellar radius and $T_{\rm eff}$.
Rotational line broadening is accounted for by applying a similar
correction to $W_{\rm core}$ as done by Pasquini \& Pallavicini
(\cite{pas:pal}). We obtain these corrections by artificially
broadening the spectrum of HD\,153525, a K0 dwarf with an otherwise
unresolved $v\sin i$, in steps of 5~\kms\ up to 60~\kms\ and then
measure the \Halpha\ line as done previously. The correction factor,
$\Delta$, for each stellar measurement is obtained from the fit to
these simulations by inserting the stellar values for $v\sin i$ from
Table~\ref{T7};
\begin{equation}
\Delta = A + B \ v\sin i + C \ v\sin i^2 + D \ v\sin i^3 \ \ .
\end{equation}
The coefficients are $A$=--0.00068990582, $B$=0.00029478396,
$C$=0.00011360807, $D$=--9.4974126e-07. Note that the corrections
are negligible for $v\sin i<$10~\kms\ but reach a maximum for
HD\,45762 ($v\sin i$=60~\kms ) of 30\%. Table~\ref{T8} lists the
corrected fluxes.

The line-core fluxes from Eq.~\ref{eq2} are not corrected for
photospheric light. Such a correction for the \Halpha\ line core is
not without problems because several line-formation processes are at
work at the same time and significant star-to-star differences are
expected (see Cram \& Mullan \cite{cra:mul}). There is no readily
available photospheric \Halpha -core flux tabulated for standard
model atmospheres. Therefore, previous attempts relied on empirical
methods, e.g., the subtraction of an observed profile of a bona-fide
inactive star of otherwise similar quantities (e.g. Soderblom et al.
\cite{sod:sta}) or the subtraction of an empirically predetermined
minimum flux from a large sample of stars (e.g. Cincunegui et al.
\cite{cin:dia}). However, as was shown nicely in the work by
Cincunegui et al. (\cite{cin:dia}), this photospheric correction
significantly worsened the relationship with Ca\,{\sc ii} H\&K flux.
The conclusion is that empirical photospheric corrections may not be
fully photospheric in origin. In the present paper, we refrained
from any photospheric correction and give combined
photospheric+chromospheric fluxes as observed.

The largest contribution to the flux error is due to the
continuum-flux calibration and comes from systematic differences
in the $T_{\rm eff}$-color relations and the error in $T_{\rm
eff}$. We compared fluxes based on four different relations. Our
two prime choices for ${\cal F}_{\rm c}$ are based, firstly, on
the $B-V$ from the Tycho-2 catalog (H\o g et al.~\cite{tycho})
and, secondly, on the $B-V$ from the conversion of our measured
$T_{\rm eff}$ in Table~\ref{T7} with the color transformation from
Flower (\cite{f96}). Two secondary choices were fluxes from our
$T_{\rm eff}$ values but converted to $B-V$ and $V-R$ using the
tables in Gray (\cite{gray}). The inter comparison in
Fig.~\ref{F-comp} shows minimized rms errors of the flux residuals
for the two prime choices, i.e. the Tycho-2 $B-V$-based fluxes vs.
the fluxes from measured $T_{\rm eff}$'s with the color conversion
from Flower. It gave an rms of 0.45 in units of
$10^5$~erg\,cm$^{-2}$s$^{-1}$. Comparable rms is obtained for the
Tycho-2 $B-V$-based fluxes vs. Gray's $B-V$ based fluxes but
significantly larger (0.75) if Gray's $V-R$ based fluxes are used
instead. The calibration rms errors per star are listed in
Table~\ref{T8} along with the line-core fluxes based on the
Tycho-2 $B-V$ color for the single and SB1 stars, and on the
Flower (\cite{f96}) $T_{\rm eff}$-($B-V$) transformation for the
SB2s using the conversion in Eq.~\ref{eq1}. These rms values are
useful when judging the reality of time variations, if present. An
exceptionally large rms value indicates most likely that the
calibration assumptions in Eq.~\ref{eq1}--3 were not fully
appropriate, e.g. for SAO150676, which is likely a pre-main
sequence star where the applicability of the $T_{\rm eff}$-($B-V$)
conversion is questionable.

The number of spectra used for the \Halpha\ analysis is smaller than for the radial-velocity curves because we use only the ones with the best S/N and, in case of the SB2 binaries, when the line separation is large enough. We measure single-lined spectra just like single stars, which have
continua presumably unaffected by the companion. The few systems
where we see a \Halpha \ line from the secondary component, but
not its photospheric lines, were also treated like single stars.
The regular SB2 spectra, on the other hand, were first multiplied
by the respective photospheric light ratio, $R$(Fe\,{\sc i}). It
is obtained from line pairs near \Halpha\ (with highest weight
given to the unblended Fe\,{\sc i} 654.6~nm line) before measuring
the \Halpha\ flux for each component. These fluxes are generally
less precise because the available S/N ratio in the continuum is
distributed among the two components. Table~\ref{T8} lists the
measurements for all single stars and single-lined binary stars as
well as for all individual stars in the double-lined binaries. The
triple-lined system HD\,105575 was too
uncertain to be measured reliably and only the primary is included in
Table~\ref{T8}.

\section{Notes to individual systems}\label{S5}

\emph{HD 553 = V741 Cas}. The system was discovered to be an
eclipsing binary by the {\sl Hipparcos} satellite. In paper~I, we
had verified its eclipsing nature and shown that the system is
double lined. A first orbit was published shortly thereafter by
Duemmler et al. (\cite{duemm}) followed by an independent orbit by
Griffin (\cite{griff03}). Both their orbits were circular with a
period of 9.0599~d and a solution rms residual of 0.56 and
1.36~\kms , respectively. The component's rms in the Duemmler et
al. orbit were 0.59 and 0.97~\kms . Our new orbit is also assumed
to be circular. The forced $e$=0 solution gives residuals of 0.41
and 0.39~\kmsﬂ for the two components, respectively. There is no conclusive period in the \Halpha -core
flux but a peak at 4.5~d could be interpreted to be $\approx
P_{\rm orb}/2$. The out-of-eclipse light variations have a $V$
amplitude of 0.10~mag and suggest a photometric period of 9.061~d
(eclipse depth is 0.28~mag in $V$).

\emph{HIP\,999 = LN Peg}.  An initial SB1 orbit was determined by
Latham et al. (\cite{lath88}) while Fekel et al. (\cite{velo})
presented a SB2 orbit and found the star to be at least a triple
system. Our radial-velocity residuals indicate a long-term period of
around 4.2 years with a full amplitude of 12~\kms\and zero eccentricity (see Fig.~\ref{F2}). This verifies
the preference put forward by Fekel et al. (\cite{velo}) that the
long-period orbit is rather 4 years with zero eccentricity than 8
years with 0.5 eccentricity. The long orbit of 1555~d is listed in Table~\ref{T4} and was obtained by including the Fekel
et al. data with a shift of +0.50~\kms . The total time span of data was then 9430~d with 213 velocities. The STELLA data alone
would give elements of $K_3$=5.654$\pm$0.085~\kms ,
$T_0$=2,454,117.4$\pm$8.5, $P$=1310$\pm$16~d. The \Halpha\ line
from the secondary star in the close pair appears in emission. The
third component $c$ is not directly seen in the spectra but is
recognized only due to the systematic radial velocity zero-point
variations of the $ab$ pair. The $ab$-pair orbit is given in Table~\ref{T5} and shown in Fig.~\ref{F3}.

\begin{figure*}[!tb]
\center
\includegraphics[angle=-90,width=150mm]{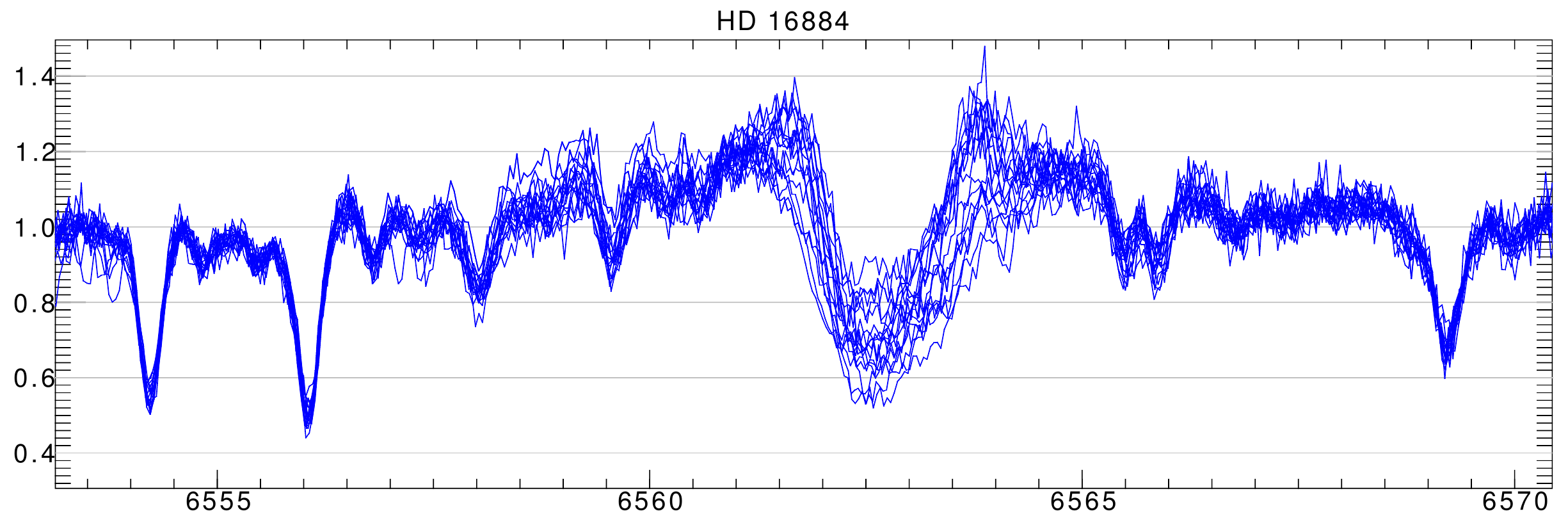}
\caption{\Halpha\ profile variations of HD~16884 in 2007/08. Spectra
are corrected for orbital radial velocity of the primary. The
photospheric absorption lines belong to the primary of the ab
system, a K giant, and agree with the radial velocities from \Halpha
. It is save to conclude that the majority of the \Halpha\ profile
originates from this K giant. The signature from the primary
component of the cd system, i.e. the secondary in this composite
spectrum, is lost in this representation. See text.}\label{F-16884}
\end{figure*}

\emph{HD 8997 = EO Psc}. Griffin's (\cite{griff87}) SB2-orbit is
basically confirmed but now considerably more precise. Our low rms
of $\approx$0.4~\kms\ for the $a$ and $b$ component radial velocities allowed to search for
residual variations and a period of 88.3$\pm$0.5~d is found for
the primary with a full amplitude of 0.20~\kms . No period is seen
in the secondary-star residuals. If above period would be
interpreted to be the rotational period of the primary, and
combined with our $v\sin i$ measures of 3~\kms , the minimum
radius would be 5.2~R$_\odot$. This value is clearly in
disagreement with the distance of 23~pc and a system brightness of
7.74~mag and $B-V$=0.96. This period must be of other origin or is
even spurious. The photometry yields a period of 10.54$\pm$0.11~d,
with a small amplitude of 0.02~mag in $y$. This period suggests
$R\sin i$ of $\approx$0.6~R$_\odot$, in good agreement with the
distance and the spectrum synthesis that indicates $T_{\rm
eff}$=5060$\pm$65~K, a $\log g$ of 4.5, and a metallicity of
--0.12$\pm$0.04, suggesting a K2 dwarf star.

\emph{HD 9902 = BG Psc}. This system contains two almost equally
massive stars with a mass ratio of 0.9750$\pm$0.0026 in a 25.3722-d orbit with
$e$=0.5068. We refer to the slightly more massive and more rapidly
rotating star as the primary. Note that the rms for the orbit for
the secondary (154~\ms ) is lower than for the primary due to its sharper lines. Our photometry of
the combined system reveals a period of 7.44$\pm$0.02~d that we
interpret to be the rotation period of the primary ($v\sin
i$=12~\kms ) rather than that of the secondary ($v\sin i$=2~\kms ). Also note that the $V$ amplitude of 0\fm05 in Table~\ref{T6} is for the combined light and not for, say, the primary.
The expected pseudo-synchronized rotation periods are $\approx$8.9~d
according to Hut~(\cite{hut}). The primary is thus likely rotating
faster than synchronous. 

\emph{HD 16884}. The star is double lined from the
cross-correlation functions but its spectral lines do not appear to originate from a single physical pair. Most likely HD\,16884 is a quadruple system and we see lines
from the two primaries $a$ and $c$, with each system being single
lined, i.e., SB1+SB1. The orbital periods for the two pairs (each 106.63~d)
appear to be identical within its errors, and so does the systemic
velocity (+2.2~\kms ). The K-amplitude of the $a$ and $c$ components
are 30.0~\kms\ and 16.3~\kms , respectively, while their $v\sin i$'s
are 6~\kms\ and $\approx$44~\kms , respectively, which means that
the radial velocity attributed to the $c$ component can not be caused by 
rotationally modulated line-profile variations of the $a$
component, its $v\sin i$ is too small. Not much can be said about
the orbital period $ab$-$cd$ but the two nodal lines appear
co-aligned and did not change recognizably during our observations,
its orbital period must be very long. Griffin (\cite{griff09})
obtained a sparsely sampled single-lined orbit from 14 of his own
velocities, an orbit which is superseded in the present paper. We
also note that the stronger-lined $a$ star's \Halpha\ line profile
appears as a broad and asymmetric emission line (full width at
continuum of up to 400~\kms ) with a deep central absorption that
follows the radial velocities of the $a$ component
(Fig.~\ref{F-16884}). Our photometry reveals a period of
65.44$\pm$0.05~d and an amplitude of 0.16~mag in $V$ for the
combined stars. It is likely that this is the rotation period of the
brighter $a$ component, a K4 giant according to Griffin
(\cite{griff09}). His classification is in rough agreement
with our $T_{\rm eff}$ of 4500~K and $\log g$ of 2.4$\pm$0.4 from
the spectrum synthesis. With a $v\sin i$ of 6~\kms\ its minimum
radius, $R\sin i$, would be 7.8$\pm$1.3~R$_\odot$. However, a K4
giant is expected to have a nominal radius of approximately
25~R$_\odot$. Thus, if indeed a K4 giant, $i$ would be only $\approx 18^\circ$ and together
with an assumed mass of 2.3~M$_\odot$ for the primary, this
contradicts the large mass function of 0.2972$\pm$0.0013, which would
otherwise imply an unreasonable secondary mass of order 7.5~M$_\odot$. Doubling the photometric period to
get the true rotation period would lead to an inclination much
closer to 90$^\circ$ and, lowering the assumed primary mass to
2.0~M$_\odot$ the secondary mass gets closer to a minimum mass of
$\approx$1.6~M$_\odot$, still rather high for an ``unseen'' star.
Certainly, this system warrants further investigation.

\emph{HD 18645 = FU Cet}. Single star. Our photometry suggests a
period of 21.5~d that we interpret to be its rotation period. A
residual period of 3.069~d is seen from the radial velocities but
with a very weak amplitude of 0.1~\kms . Additionally, the \Halpha
-core emission shows a seasonal trend and, after its removal, a
period of $\approx$17~d. Although this period is badly determined
and inconclusive it may suggest that the photometric period of
21.5~d is to be preferred against the 3-d period. Our $v\sin i$
measure of 12.0$\pm$0.2~\kms\ converts then to a minimum stellar
radius of 5.1~R$_\odot$, appropriate for its mid G giant
classification.

\emph{HD 18955 = IR~Eri}. This system consists of two dwarf stars of mass ratio 0.88 in a 43-day orbit with a high eccentricity of
0.7587$\pm$0.0005. The expected pseudo-synchronization period would be
a mere 11\% of the orbital period, i.e. $\approx$5~d. Unfortunately,
our photometry did not allow to obtain a period despite that the
data indicate variability at the $2.3\sigma$ level. Both components
show a weak residual radial-velocity variation with a period of
25--26~d and an amplitude of 0.5~\kms . This could be the stellar
rotation period but the data are too noisy to be conclusive. It
appears that the secondary is cooler than the primary by
$\approx$300~K, and also slower rotating ($v\sin i$=5~\kms\ vs. 9,
respectively).

\begin{figure}[!tb]
\center
\includegraphics[angle=0,width=80mm]{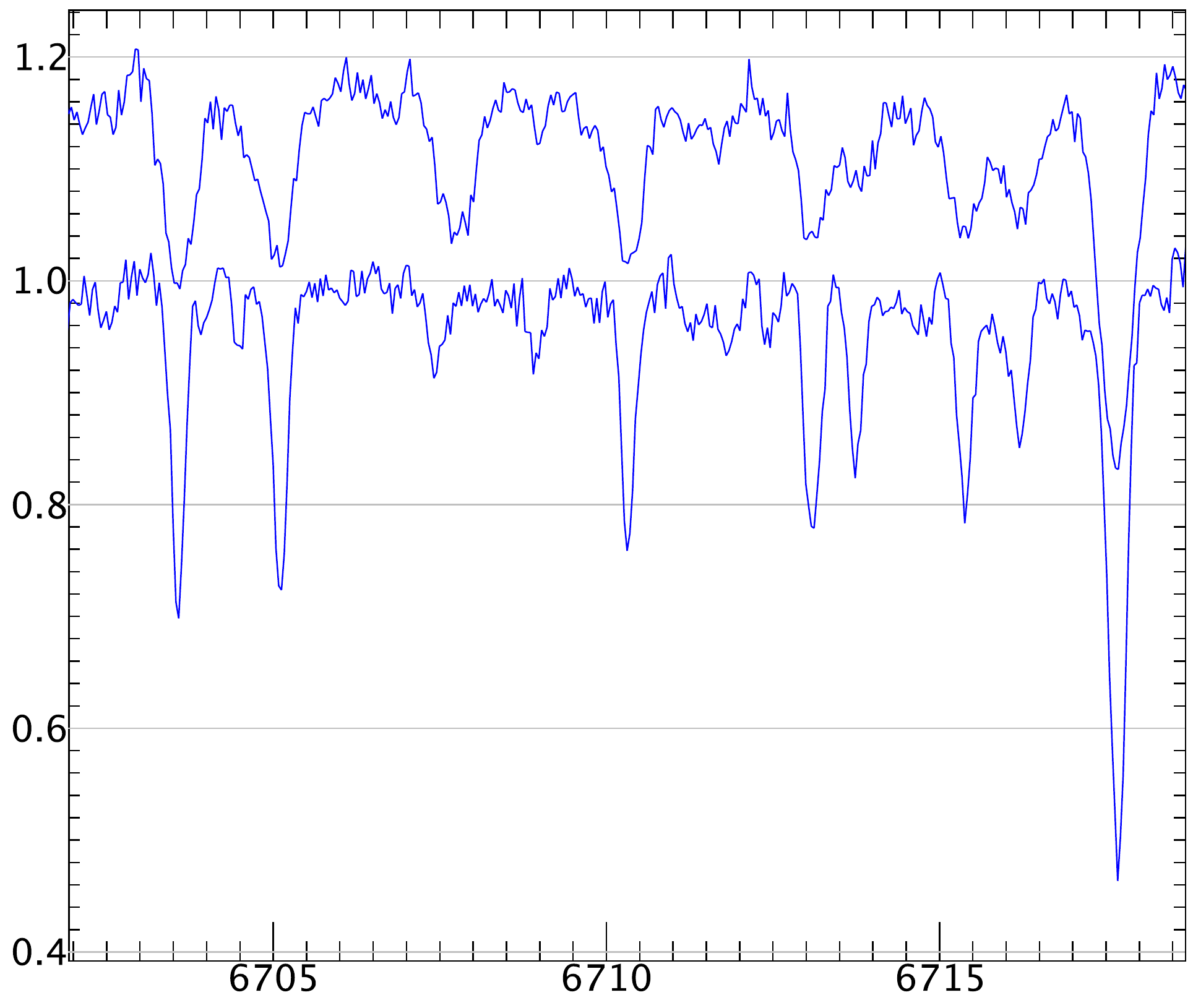}
\caption{Li\,{\sc i} 6707.8-\AA\ line region for the K0III component
of the SB-1 binary HD~62668 (top spectrum; shifted by +0.2) and for
the single K0III star HD~23551 (bottom spectrum). Notice the
broad Li\,{\sc i} absorption line in HD~62668 and its absence in
HD~23551.}\label{Fli}
\end{figure}

\emph{HD 23551 = MM~Cam}. The star appears to be single. A radial
velocity of --7.5$\pm$0.3~\kms\ was reported in the Pulkova survey
by Gontcharov (\cite{pulkova}) in good agreement with the average
$-7.38\pm0.22$ by Famaey et al. (\cite{famaey}). Our two velocities
from KPNO in paper~I were $-9.5$ from the blue spectrum and
$-7.4\pm0.2$ from the red spectrum, also in reasonable agreement. A
peak with a period of $\approx$80~d is existent in the \Halpha\ flux
and the photometry suggests a clear period of 81.69$\pm$0.03~d that
we interpret to be the rotation period of the star. Note that we
observed the star with both APTs (T6 and T7) and that the
independent photometric period from T7 data is 81.2$\pm$0.5~d, in
good agreement with the more precise T6 period. Its $v\sin i$ of
6.4$\pm$0.2~\kms\ converts then to a minimum radius of
10.3~R$_\odot$ in agreement with $\log g$=3.0 from our spectrum
synthesis. We also note its significant metal deficiency of
--0.11$\pm$0.02. No Li\,{\sc i} 6707.8-\AA\ absorption feature is
seen above our detection threshold of $\approx$10~m\AA\
(Fig.~\ref{Fli}).

\emph{HD 24053}. This star first appeared in the literature in the
$uvby$ calibration for FGK stars by Olsen, e.g., Olsen
(\cite{olsen}). Its effective temperature was determined
spectroscopically from spectral line ratios by Kovtyukh et al.
(\cite{kov:sou}) to 5723~K. The revised Geneva-Copenhagen catalog
(Holmberg et al. \cite{holm07}) lists it with a metallicity of
$-0.04$ and 5640~K and suggests an age of 0.2~Gyr. From 2-MASS
colors Masana et al. (\cite{mas:jor}) derived 5711~K and the
N2K-consortium (Robinson et al. \cite{n2k}) derived 5699~K, a
metallicity of $+0.09$, and $\log g$=4.45~cm\,s$^{-2}$. Earlier
velocities were given in our paper~I (5.3$\pm$1.4~\kms\ from a
blue spectrum and 2.9$\pm$0.6~\kms\ from a red spectrum two days
apart) while the catalog by Gontcharov (\cite{pulkova}) lists
4.1$\pm$0.2~\kms. Two spectra are given in the Elodie archive. We
have 77 STELLA spectra and all of them appear to be constant. The
average values from our spectrum synthesis yields 5600$\pm$25~K,
$\log g$=4.09$\pm$0.03, and a metallicity of --0.01$\pm$0.02. No
significant variations are seen in \Halpha .

\emph{SAO\,150676 = AI Lep}. Cutispoto et al. (\cite{cuti1999})
listed the star in their ROSAT follow-up observations. They noted
a spectral classification of G2V but added ``(PMS?)'' indicating
possibly pre main sequence. In the same study a photometric
variability of 0\fm02 in $V$ with a period of 1.78~d was reported.
The authors also measured a $v\sin i$ of 28$\pm$2~\kms\ in
reasonable agreement with 25.6$\pm$2~\kms\ from our paper~I .
Their two radial velocity measures agreed to within their errors
and were averaged to +25.8$\pm$1.6~\kms . In paper~I we had given
velocities of +26.1$\pm$1.6~\kms\  from a single red-wavelength
spectrum and +21.5$\pm$5.3~\kms\ and $-18.5\pm$2.1~\kms\ for two
components from a single Ca\,{\sc ii} H\& K emission-line
spectrum. The latter must have been an erroneous measurement
because it is not verified in the present paper. In paper~I, we
also measured the large equivalent width of 195m\AA\ from the
Li\,{\sc i} 670.8-nm line, in agreement with the suggestion of
being a pre-main sequence star. Our 93 STELLA radial velocities
have an unreasonable high scatter of 0.53~\kms\ which is likely
due to rotational modulation of a spotted surface. A Lomb
periodogram shows two equal peaks at periods of 0.64~d and 1.79~d.
Unfortunately, the \Halpha\ fluxes do not show the same peaks but
favor a (poorly determined) 1.2-day period. However, our
photometry shows a clear 1.79-d period, in good agreement with
Cutispoto et al. (\cite{cuti1999}) and one of the two
radial-velocity periods. We interpret it to be the rotation period
of the star.

\emph{SAO\,151224 = HY CMa}. The star is an eclipsing SB2 binary in
a circular 4.977691-d orbit. Photometry from late 1993 (Cutispoto et al.
\cite{cuti1995}, \cite{cuti1999}) revealed the eclipsing nature while follow-up photometry in 1994 by Cutispoto et al.(\cite{cuti2003}) failed to confirm the eclipses and the author's stated to the contrary that ``No primary eclipses
are apparent.'' Our photometry verifies the eclipsing nature of the
light variations with a period of 4.98~d and an eclipse depth of
0.68~mag in $V$. The residual radial velocities of the primary star
indicate a period of 5.25~d, likely due to rotation of its spotted
surface which partly explains the unexpected high rms of its orbital
fit of 1.13~\kms . The $v\sin i$ of the primary of 45~\kms\ converts
directly into a radius of 4.4~R$_\odot$.

\begin{table*}
\caption{\Halpha\ absolute emission-line fluxes. Column SB denotes
the binary component in case the system is SB2 or even SB3 (S means
single star, $a$ means primary component, $b$ secondary component,
$c$ tertiary component). (B-V) and (V-R)$_C$ are the expected
Johnson-Cousins indices from our $T_{\rm eff}$ and the
transformation from Flower (\cite{f96}) and Gray (\cite{gray}),
respectively, that we adopted for the flux calibration. $N$ is the
number of spectra used. $W_{\rm core}$ is the equivalent width in
m\AA\ in the central 1-\AA\ bandpass and its internal measuring
error. $\langle {\cal F}$(\Halpha )$\rangle$ is the average 1-\AA\
\Halpha -core flux in $10^5$~erg\,cm$^{-2}$s$^{-1}$ and its
propagated error from the equivalent width measure plus the flux
calibration. ``rms'' is its standard deviation from the four
different flux calibrations for comparison. $\delta{\cal F}$ is the
rms residual of ${\cal F}$(\Halpha ) over the time span of our
entire observations. $R$(Fe\,{\sc i}) is the photospheric line ratio
secondary/primary from Fe\,{\sc i} 654.6~nm that is used to weight
the components' \Halpha\ equivalent widths for SB2
systems.}\label{T8}
\begin{tabular}{llllllllll}
\hline \noalign{\smallskip}
Name  & SB   & $N$ & B-V & V-R & $W_{\rm core}$ & $\langle {\cal F}$(\Halpha )$\rangle$ & rms & $\delta{\cal F}$(\Halpha ) & $R$(Fe\,{\sc i})  \\
      &      &   & \multicolumn{2}{c}{(mag)} & (m\AA ) &
      \multicolumn{3}{c}{($\times 10^{5}$ erg\,cm$^{-2}$s$^{-1}$)}& (--) \\
\noalign{\smallskip} \hline \noalign{\smallskip}
 HD 18645   &   S &  64 & 0.75 & 0.60 &   421$\pm$19 &   21.4$\pm$2.7 &    0.6 &    1.0 & \dots \\
 HD 23551   &   S &  98 & 0.91 & 0.70 &   345$\pm$12 &   12.1$\pm$1.4 &    0.2 &    0.5 & \dots \\
 HD 24053   &   S &  71 & 0.70 & 0.55 &   349$\pm$10 &   19.6$\pm$2.1 &    1.7 &    0.6 & \dots \\
 SAO 150676 &   S &  78 & 0.65 & 0.53 &   617$\pm$25 &   43.3$\pm$5.4 &    1.9 &    1.9 & \dots \\
 HD 43516   &   S &  52 & 0.88 & 0.67 &   334$\pm$13 &   13.1$\pm$1.6 &    0.3 &    0.6 & \dots \\
 HD 76799   &   S &  69 & 0.97 & 0.73 &   385$\pm$17 &   10.8$\pm$1.4 &    0.6 &    0.5 & \dots \\
 HIP 46634  &   S &  88 & 0.87 & 0.70 &   461$\pm$10 &   18.3$\pm$1.9 &    1.1 &    0.4 & \dots \\
 HD 95188   &   S &  82 & 0.80 & 0.62 &   460$\pm$11 &   23.3$\pm$2.5 &    1.6 &    0.6 & \dots \\
 HD 95724   &   S &  69 & 0.95 & 0.78 &   446$\pm$11 &   14.8$\pm$1.6 &    0.6 &    0.4 & \dots \\
 HD 104067  &   S &  74 & 1.02 & 0.87 &   444$\pm$9  &   13.3$\pm$1.4 &    0.6 &    0.3 & \dots \\
 HD 108564  &   S &  69 & 1.15 & 0.99 &   482$\pm$11 &   13.6$\pm$1.4 &    2.0 &    0.4 & \dots \\
 HD 120205  &   S &  88 & 0.82 & 0.65 &   414$\pm$11 &   19.6$\pm$2.1 &    1.2 &    0.5 & \dots \\
 HD 136655  &   S & 103 & 0.93 & 0.76 &   424$\pm$12 &   14.8$\pm$1.6 &    0.6 &    0.5 & \dots \\
 HD 153525  &   S & 185 & 1.04 & 0.90 &   480$\pm$12 &   11.9$\pm$1.3 &    0.3 &    0.3 & \dots \\
 HD 155802  &   S & 111 & 0.93 & 0.76 &   483$\pm$13 &   17.8$\pm$1.9 &    0.8 &    0.5 & \dots \\
 HD 171067  &   S & 135 & 0.73 & 0.57 &   351$\pm$8  &   21.7$\pm$2.2 &    1.2 &    0.5 & \dots \\
 HD 184591  &   S & 145 & 1.03 & 0.75 &   366$\pm$8  &   14.6$\pm$1.5 &    2.0 &    0.4 & \dots \\
 HD 218739  &   S &  96 & 0.68 & 0.54 &   376$\pm$12 &   25.7$\pm$2.9 &    1.0 &    0.9 & \dots \\
 \noalign{\smallskip}
 HD 40891   & $a$ &  82 & 0.87 & 0.70 &   410$\pm$10 &   17.1$\pm$1.8 &    0.9 &    0.4 & \dots \\
 HD 62668   & $a$ &  99 & 1.10 & 0.81 &   571$\pm$34 &   11.9$\pm$1.8 &    0.7 &    0.8 & \dots \\
 HD 66553   & $a$ &  69 & 0.82 & 0.82 &   399$\pm$15 &   16.0$\pm$1.9 &    2.4 &    0.6 & \dots \\
 HD 82159   & $a$ &  97 & 0.87 & 0.69 &   830$\pm$46 &   30.0$\pm$4.1 &    2.8 &    1.7 & \dots \\
 HD 82841   & $a$ &  59 & 1.12 & 0.82 &   430$\pm$25 &    9.9$\pm$1.4 &    0.6 &    0.6 & \dots \\
 HIP 50072  & $a$ &  56 & 1.12 & 0.82 &   566$\pm$53 &   11.9$\pm$2.2 &    0.8 &    1.3 & \dots \\
 HD 112099  & $a$ &  73 & 0.89 & 0.72 &   419$\pm$9  &   16.4$\pm$1.7 &    0.8 &    0.4 & \dots \\
 HIP 63322  & $a$ &  98 & 0.94 & 0.77 &   823$\pm$36 &   33.6$\pm$4.2 &    2.3 &    1.5 & \dots \\
 HIP 63442  & $a$ &  79 & 1.16 & 0.85 &   531$\pm$57 &    8.8$\pm$1.8 &    0.8 &    1.1 & \dots \\
 HD 138157  & $a$ & 149 & 1.00 & 0.74 &   522$\pm$29 &   10.6$\pm$1.7 &    0.8 &    0.8 & \dots \\
 HD 147866  & $a$ & 114 & 1.14 & 0.83 &   471$\pm$20 &   10.3$\pm$1.3 &    0.7 &    0.5 & \dots \\
 HD 150202  & $a$ & 176 & 0.93 & 0.71 &   344$\pm$10 &   11.1$\pm$1.2 &    0.3 &    0.3 & \dots \\
 $\epsilon$ UMi& $a$ & 110 & 0.84 & 0.65& 400$\pm$14 &   12.6$\pm$1.6 &    0.9 &    0.6 & \dots \\
 HD 190642  & $a$ & 180 & 1.05 & 0.77 &   631$\pm$55 &   17.1$\pm$3.0 &    1.0 &    1.6 & \dots \\
 HD 202109  & $a$ & 200 & 0.97 & 0.74 &   299$\pm$5  &    8.8$\pm$0.9 &    0.3 &    0.2 & \dots \\
 \noalign{\smallskip}
 HD 553     & $a$ &  68 & 1.10 & 0.82 &   590$\pm$25 &    9.9$\pm$1.4 &    0.6 &    0.6 & 0.62 \\
            & $b$ &     & 0.69 & 0.50 &   908$\pm$21 &   41.1$\pm$4.9 &    4.2 &    1.7 & \\
 HIP 999    & $a$ &  38 & 0.81 & 0.64 &   793$\pm$33 &   34.0$\pm$4.3 &    2.5 &    1.6 & 0.21 \\
            & $b$ &     & 1.00 & 0.74 &   929$\pm$16 &    4.4$\pm$0.8 &    0.3 &    0.5 & \\
 HD 8997    & $a$ & 100 & 0.91 & 0.74 &   614$\pm$12 &   22.5$\pm$2.3 &    1.1 &    0.5 & 0.42 \\
            & $b$ &     & 1.16 & 1.00 &   844$\pm$11 &    6.4$\pm$0.8 &    0.2 &    0.2 & \\
 HD 9902    & $a$ &  16 & 0.88 & 0.70 &   883$\pm$11 &   35.1$\pm$3.2 &    2.3 &    0.5 & 1.23 \\
            & $b$ &     & 0.66 & 0.54 &   632$\pm$21 &   53.3$\pm$5.6 &    2.1 &    1.4 & \\
 HD 16884   & $a$ &  29 & 1.18 & 0.87 &   636$\pm$95 &   11.3$\pm$2.6 &    0.9 &    1.7 & 0.19 \\
            & $c$ &     & 1.16 & 0.84 &  1378$\pm$120&    4.5$\pm$2.6 &    0.5 &    2.3 & \\
 HD 18955   & $a$ &  31 & 0.77 & 0.60 &   557$\pm$34 &   29.2$\pm$4.1 &    2.0 &    1.8 & 0.67 \\
            & $b$ &     & 0.89 & 0.72 &   851$\pm$30 &   21.8$\pm$2.9 &    1.3 &    1.2 & \\
 SAO 151224 & $a$ &  21 & 1.13 & 0.83 &   934$\pm$103&   16.1$\pm$3.5 &    1.1 &    2.2 & 0.98 \\
            & $b$ &     & 0.76 & 0.60 &   913$\pm$43 &   47.6$\pm$6.1 &    3.2 &    2.3 & \\
 HD 45762   & $a$ &  26 & 1.11 & 0.82 &   501$\pm$29 &   10.3$\pm$1.5 &    0.7 &    0.7 & 0.35 \\
            & $c$ &     & 0.56 & 0.47 &   775$\pm$50 &   16.4$\pm$5.5 &    0.6 &    4.2 & \\
 HD 50255   & $a$ &   9 & 0.71 & 0.56 &   498$\pm$17 &   30.8$\pm$3.5 &    1.6 &    1.0 & 0.26 \\
            & $b$ &     & 0.94 & 0.71 &   846$\pm$26 &    7.4$\pm$1.5 &    0.2 &    0.9 & \\
\noalign{\smallskip}\hline
\end{tabular}
\end{table*}
\setcounter{table}{7}
\begin{table*}[t]
\caption{(continued)}
\begin{tabular}{llllllllll}
\hline \noalign{\smallskip}
Name  & SB   & $N$ & B-V & V-R & $W_{\rm core}$ & $\langle {\cal F}$(\Halpha )$\rangle$ & rms & $\delta{\cal F}$(\Halpha ) & $R$(Fe\,{\sc i})  \\
      &      &   & \multicolumn{2}{c}{(mag)} & (m\AA ) &
      \multicolumn{3}{c}{($\times 10^{5}$ erg\,cm$^{-2}$s$^{-1}$)}& (--) \\
\noalign{\smallskip} \hline \noalign{\smallskip}
 HD 61994   & $a$ &  12 & 0.69 & 0.55 &   474$\pm$9  &   30.0$\pm$3.0 &    1.4 &    0.6 & 0.27 \\
            & $b$ &     & 1.04 & 0.90 &   823$\pm$23 &    5.7$\pm$1.1 &    0.1 &    0.6 & \\
 HD 73512   & $a$ &  12 & 0.88 & 0.71 &   582$\pm$20 &   22.9$\pm$2.6 &    1.4 &    0.8 & 0.57 \\
            & $b$ &     & 1.13 & 0.97 &   805$\pm$22 &    9.5$\pm$1.2 &    0.1 &    0.5 & \\
 HD 82286   & $a$ &  33 & 1.03 & 0.75 &  1197$\pm$73 &   29.3$\pm$4.3 &    2.1 &    2.0 & 0.58 \\
            & $b$ &     & 1.03 & 0.76 &  1114$\pm$66 &   15.1$\pm$3.0 &    1.0 &    1.8 & \\
 HD 93915   & $a$ &  31 & 0.73 & 0.57 &   565$\pm$45 &   33.1$\pm$5.3 &    2.1 &    2.7 & 0.89 \\
            & $b$ &     & 0.80 & 0.63 &   801$\pm$36 &   34.5$\pm$4.5 &    2.5 &    1.8 & \\
 HD 95559   & $a$ &  66 & 0.89 & 0.72 &   928$\pm$30 &   32.0$\pm$3.7 &    1.9 &    1.2 & 1.09 \\
            & $b$ &     & 0.90 & 0.73 &   886$\pm$21 &   32.2$\pm$3.4 &    1.8 &    0.8 & \\
 HD 105575  & $a$ &  21 & 0.68 & 0.55 &   775$\pm$39 &   41.2$\pm$5.8 &    1.5 &    2.6 & \dots \\
 HD 106855  & $a$ &  35 & 1.05 & 0.91 &  1198$\pm$64 &   28.2$\pm$3.9 &    0.3 &    1.6 & 0.34 \\
            & $b$ &     & 0.95 & 0.79 &  1061$\pm$23 &   12.1$\pm$1.7 &    0.4 &    0.8 & \\
 HD 109011  & $a$ &  23 & 0.92 & 0.75 &   581$\pm$17 &   20.7$\pm$2.3 &    1.0 &    0.6 & 0.34 \\
            & $b$ &     & 0.98 & 0.74 &   926$\pm$24 &    9.7$\pm$1.5 &    0.3 &    0.8 & \\
 HD 111487  & $a$ &  34 & 0.74 & 0.58 &   799$\pm$23 &   37.8$\pm$4.3 &    2.2 &    1.3 & 0.60 \\
            & $b$ &     & 1.08 & 0.93 &  1005$\pm$15 &   13.2$\pm$1.4 &    0.1 &    0.4 & \\
 HD 112859  & $a$ &  21 & 1.04 & 0.76 &   684$\pm$34 &   17.0$\pm$2.3 &    1.1 &    0.9 & 0.43 \\
            & $b$ &     & 0.52 & 0.44 &   825$\pm$36 &   32.4$\pm$5.8 &    1.2 &    3.3 & \\
 HD 127068  & $a$ &  18 & 1.00 & 0.74 &   930$\pm$25 &   26.5$\pm$2.8 &    1.5 &    0.8 & 0.12 \\
            & $b$ &     & 0.71 & 0.57 &   996$\pm$66 &    7.1$\pm$4.5 &    0.2 &    4.0 & \\
 HIP 77210  & $a$ &  58 & 0.85 & 0.68 &   613$\pm$19 &   26.1$\pm$2.9 &    1.7 &    0.8 & 0.33 \\
            & $b$ &     & 1.07 & 0.92 &   876$\pm$13 &    6.9$\pm$0.9 &    0.1 &    0.3 & \\
 HD 142680  & $a$ &  24 & 0.99 & 0.83 &   546$\pm$11 &   16.4$\pm$1.7 &    0.4 &    0.4 & 0.31 \\
            & $b$ &     & 1.18 & 1.02 &   883$\pm$14 &    4.9$\pm$0.7 &    0.1 &    0.3 & \\
 HD 143937  & $a$ &  47 & 0.87 & 0.69 &   937$\pm$38 &   30.6$\pm$4.0 &    2.0 &    1.6 & 0.79 \\
            & $b$ &     & 1.06 & 0.92 &  1016$\pm$29 &   16.7$\pm$2.1 &    0.2 &    0.8 & \\
 HD 197913  & $a$ &  52 & 0.73 & 0.57 &   543$\pm$14 &   31.8$\pm$3.4 &    2.0 &    0.9 & 0.86 \\
            & $b$ &     & 0.79 & 0.62 &   830$\pm$20 &   35.2$\pm$3.8 &    2.7 &    1.0 & \\
 HD 199967  & $a$ &  70 & 0.62 & 0.52 &   531$\pm$23 &   39.0$\pm$4.8 &    0.2 &    1.8 & 0.57 \\
            & $b$ &     & 0.62 & 0.53 &   804$\pm$19 &   33.3$\pm$4.1 &    0.2 &    1.4 & \\
 HD 226099  & $a$ &  94 & 0.85 & 0.67 &   595$\pm$19 &   25.4$\pm$2.8 &    2.0 &    0.8 & 0.65 \\
            & $b$ &     & 1.15 & 0.99 &   759$\pm$23 &    9.4$\pm$1.2 &    0.2 &    0.5 & \\
 HD 237944  & $a$ &  14 & 0.69 & 0.55 &   721$\pm$29 &   44.8$\pm$5.4 &    2.1 &    1.9 & 1.07 \\
            & $b$ &     & 0.71 & 0.56 &   747$\pm$26 &   47.7$\pm$5.3 &    2.8 &    1.6 & 0.61 \\
            & $c$ &     & 0.98 & 0.82 &   870$\pm$26 &   16.2$\pm$2.1 &    0.4 &    0.9 & \\
\noalign{\smallskip}\hline
\end{tabular}
\end{table*}

\emph{HD 40891}. We see no traces of Ca\,{\sc ii} H\&K emission in
our echelle spectra. Therefore, its emission can be classified at
best ``very weak''. Note that in paper~I, we listed the star in
the table for stars with emission, but did not show a plot of the
spectrum. This might indicate a logistics error in our paper~I.
The star is an SB1 in an eccentric orbit with
$e$=0.5105$\pm$0.0007 and a period of 49.0823~d. The expected
pseudo-synchronization period would be $\approx$17~d.

\emph{HD 45762}. Griffin (\cite{griff10}) noted in his Table~1
``Binary, P$\sim$60 days'' but otherwise gave no information. We
find the star to be a double-lined triple system with the more
massive component being an SB1 with a period of $\approx$1/3 that
of the wider $c$ component. The $(ab)-c$ orbit has a period of
59.9363$\pm$0.0016~d and a small but non-zero eccentricity of
0.0150$\pm$0.0009. Its rms of the residuals is 0.533~\kms\ which
increases to 0.88~\kms\ if an $e$=0 solution is enforced. The mass
ratio $c$\,/$(a+b)$ is 0.23 from an averaged $ac$ solution (shown
as a dotted line in the respective panel in Fig.~\ref{F3}). The
close SB1 orbit has only a small amplitude and we find the
following preliminary elements ($e$=0 assumed);
$K$=7.78$\pm$0.77~\kms , $T_0$=2,454,116.84$\pm$0.49, and
$P(ab)$=19.933$\pm$0.014~d. Note that this star's $v\sin i$ is
60~\kms , as compared to the $c$-component's 16~\kms , and is
likely an asynchronous rotator because a $P_{\rm rot}$=19.933~d
would suggest a large minimum radius of 23.6~R$_\odot$. Our photometry shows a double-humped light curve with a clear 62-day periodicity but must be partly due to an ellipticity effect because of its phase coherence with the orbit. The $c$
component has an expected minimum radius of 19.6~R$_\odot$ if
$P_{\rm rot}$=62~d. Obviously, both stars are cool giants, as
already indicated by the small Hipparcos parallax and the
$V$=8.3-mag brightness.

\emph{HD 50255}. The star is a SB2 in an eccentric 455-d orbit. The scatter of the residual radial velocities of
the secondary component (0.167~\kms ) is larger than expected for a star
with $v\sin i$=5~\kms . A Lomb periodogram shows significant excess
power at 485~d with an amplitude of 0.65~\kms, close to the orbital
period of 455~d, but given the comparable short coverage of 1471~d
of the entire data set it could be also a sampling effect and
remains to be confirmed. Tentatively, we interpret it to be the rotation period of the secondary. Note that our photometry did not reveal any clear light variations above, say, 0.05~mag.

\emph{HD 61994}. There are 102 secondary-star velocities detected
from the STELLA cross-correlation functions and an eccentric SB2 orbit is
given ($P$=552.8~d, $e$=0.423). Duquennoy \& Mayor
(\cite{duq:may}) had discovered its binarity and also presented a
(marginal) SB2 orbit based on 9 CORAVEL measurements. The visual
brightness difference between the two components of 2.6~mag is
close to its detection limit. The average line ratio in our
spectra is 8.5. The expected pseudo-synchronized rotation period
is $\approx$250~d, but neither the velocity residuals nor the
\Halpha -core flux show traces of any significant period.

\begin{figure*}[!tb]
\center
\includegraphics[angle=0,width=8cm]{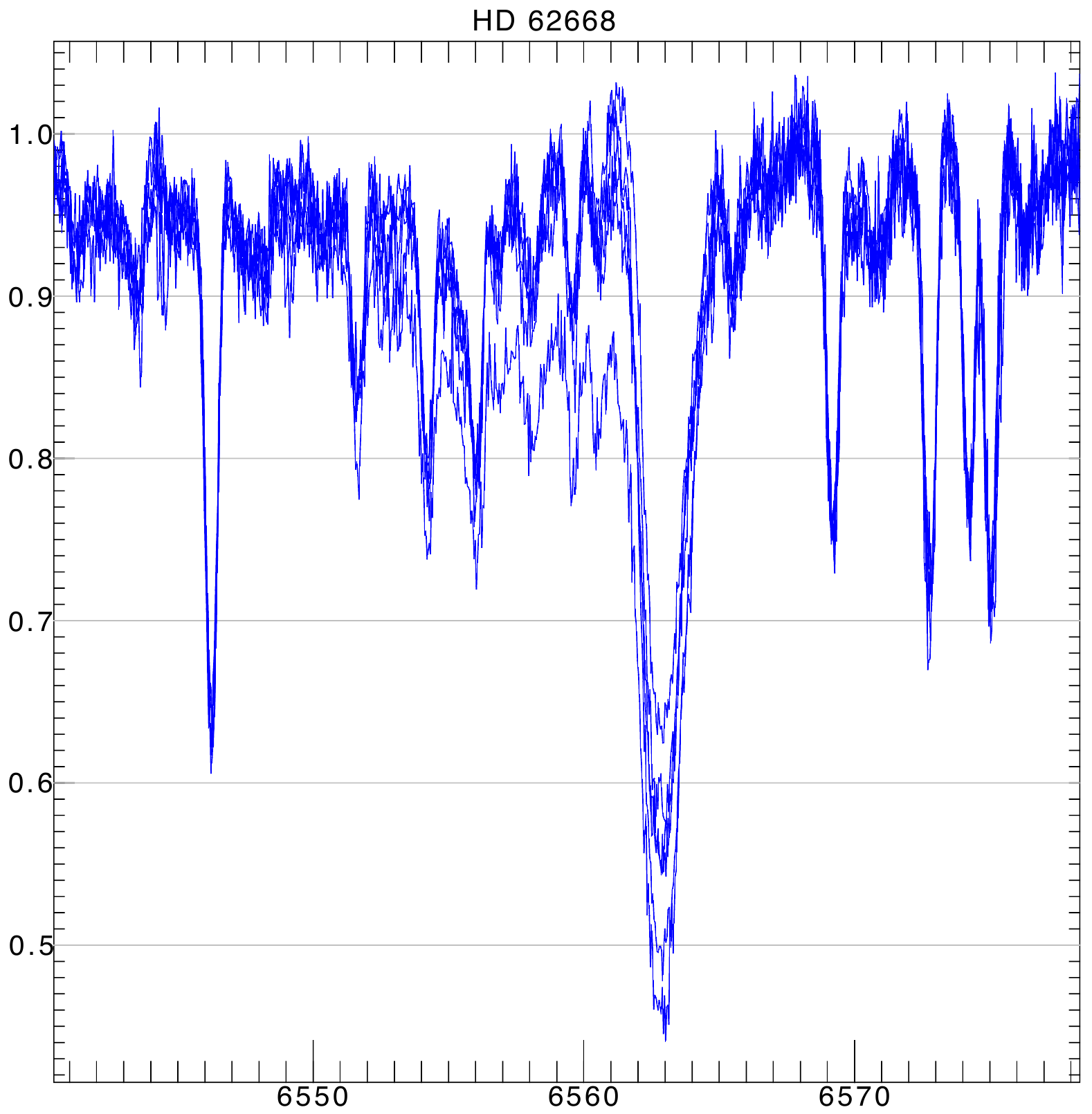}\hspace{4mm}
\includegraphics[angle=0,width=8cm]{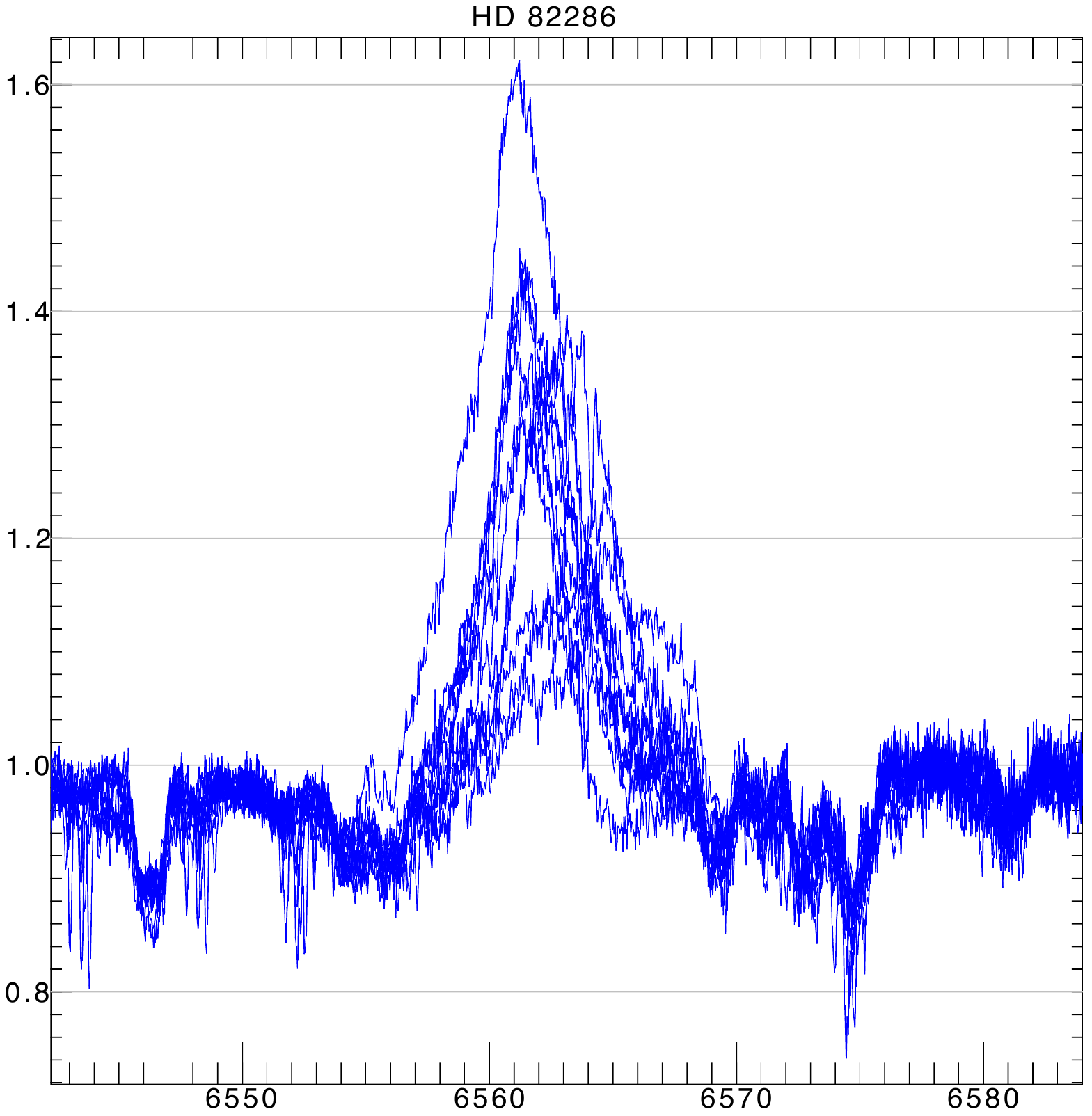}
\caption{\Halpha\ profile variations of, left, HD~62668 (SB1) in
2007 and, right, HD~82286 (SB2) in 2010. Spectra are corrected for
orbital radial velocity. Note that the spectral lines from the
secondary star in HD~82286 appear as apparently increased scatter
around the primary-star lines while the sharp lines are actually
water-vapor lines. }\label{F-Ha}
\end{figure*}

\emph{HD 62668 = BM Lyn}. Our orbit is assumed circular with a period of
69.3023~d. A formal eccentricity of 0.0043$\pm$0.0011 is judged too uncertain. The photometry is best fit with a period of
67.470$\pm$0.007~d, close but not identical to the orbital period.
The \Halpha\ absorption line also shows a clear time-dependent
variability (Fig.~\ref{F-Ha}) but no period. The residual radial
velocities indicate a period of 22.35~d, which we believe is an
alias of the 69-d period. A weak lithium line at 6707.8~\AA \ is
seen (a representative plot is shown in Fig.~\ref{Fli}).

\emph{HD 66553}. There are some secondary-star velocities detected
from our cross-correlation functions but close-up inspection of
the spectra did not unambiguously show spectral lines from it.
Therefore, we refrained from giving an SB2 orbit. This system
shows the most eccentric orbit in our sample
($e$=0.8719$\pm$0.0028) at a period of 75.8996~d. Latham et al.
(\cite{lath}) discovered the radial-velocity variations and
determined an SB1 orbit. ``The ninth catalogue of spectroscopic
binary orbits'' (Pourbaix et al.~\cite{SB9}) lists another
independent SB-1 orbit from Imbert (\cite{imb}), in agreement with
the Latham et al. orbit. The expected pseudo-synchronization
period would be $\approx$3.4~d, i.e. just $\approx$4\% of the
orbital period. No signs of periodic velocity residuals or \Halpha
-flux variations are evident. Thus, the rotation period of the
star still remains to be determined. Our spectrum synthesis gives
$T_{\rm eff}$=5275$\pm$41~K, $\log g$=4.33$\pm$0.04, and a higher
than usual metallicity of +0.09$\pm$0.02.

\emph{HD 73512}. A SB2 orbit with a period of 128.241~d and
$e$=0.26240$\pm$0.00045 is given. The mass ratio is 0.88 while the component's line-depth ratio is 2.2
with both stars having the same $v\sin i$ of $\approx$5~\kms . The expected
pseudo-synchronized rotation period is $\approx$90~d. The star appears constant in our photometry.

\emph{HD 76799}. The star shows radial-velocity variations with an
amplitude of 0.23~\kms , peak to valley, with an uncertainty of
$\pm$0.04~\kms\ if phased with the most-likely period of
50.13$\pm$0.15~d from a Lomb-Scargle analysis. A periodogram from
four years of \Halpha\ line fluxes shows two peaks with periods of
$\approx$64~d and 55~d. Although not quite in agreement with the
photospheric radial-velocity period, they are in the same range.
We interpret both variations to be due to stellar rotation. The
star is otherwise a single star. Our spectrum synthesis gave
$T_{\rm eff}$=4915$\pm$24~K, $\log g$=3.26$\pm$0.07, and a
significantly sub-solar metallicity of --0.12$\pm$0.03. The $v\sin
i$ of 7.8$\pm$0.3~\kms\ suggests a minimum radius of
7.7~R$_\odot$, in agreement with the low gravity and the small
parallax.

\emph{HIP\,46634 = BD+11\,2052\,B}. This star is the B component of
the visual binary ADS\,7406, about 14\arcsec\ away from the brighter
A component. G\'alvez et al. (\cite{galvez05}) found it to exhibit a
constant radial velocity of +27.07$\pm$0.04~\kms\ within their time
of observation in 2004. The A component is HD\,82159 and was
identified in our KPNO H\&K survey as an active single-lined
spectroscopic binary. Our current STELLA/SES spectra show the B
component to have a constant radial velocity of
+27.927$\pm$0.047~\kms . It is apparently single. There are two
equally strong peaks of 372~d and 185~d in the radial velocities
with an amplitude of 0.2~\kms . The \Halpha -line fluxes show peaks
at 120~d and 298~d indicating some long-term variations. None of the
periods is conclusive though and the true rotation period remains to
be determined.

\emph{HD 82159 = GS~Leo = BD+11\,2052\,A}. This is the bright
component of the double star ADS\,7406, denoted component A. Its
orbital elements with a period of 4.8122 days and $e$=0.176 were
given by G\'alvez et al.~(\cite{galvez05}, \cite{galvez06}) from
data taken within nine nights in March/April 2004. Griffin \&
Filiz Ak (\cite{gri:fil}) had revised this orbit from 30 new
Cambridge CORAVEL measurements from 2008/09. Our SB1 orbit is
computed from 107 STELLA measures spread over four years with a
period of 3.855871$\pm$0.0000066~d and an eccentricity of
0.2640$\pm$0.0012 and supersedes yet the Griffin orbit. The
velocity residuals are systematic and indicate a period of 3.50~d,
significantly longer than our photometric period of 3.054~d. Note
that our photometry is polluted by the B component HIP46634 which, however, should not affect
the period determination. The \Halpha -core emission fluxes
indicate, besides an annual trend that must be removed, a residual
period of 2.97~d, in good agreement with the photometric period.
Therefore, we feel confident to interpret the photometric period
to be the rotation period of the primary star. Its expected
pseudo-synchronization period is $\approx$2.7~d. Obviously, the
system did not reach equilibrium yet and the primary is still
rotating asynchronously.

\emph{HD 82286 = FF~UMa}. A double-lined binary with a short 3.275-d
orbit and a very precise mass ratio of 0.4632$\pm$0.0002 is given. Both component's spectral lines are rotationally broadened to
nearly the same value (38~\kms )\footnote{In paper~I we measured
$v\sin i$ for the two components of 35 and 29~\kms , respectively,
but G\'alvez et al. (\cite{galvez07}) quoted us with values of 17
and 16~\kms\ which we do not know where they came from.} and appear
variably asymmetric, most likely due to cool surface spots. Our
photometry shows an amplitude of 0.16~mag with a clear and precise
period of 3.27629$\pm$0.00003~d. We attribute it to the brighter
primary component (line ratio of 2.9). The spots are the largest
contributor to the radial velocity error budget in our orbital
solution, in particular for the secondary component. Residual
periods of 1.945~d and 8.714~d are evident from its radial velocity.
Given the 9.6-mas parallax and the 7.8-mag system brightness, we
expect two subgiant stars, already noted by Henry et al.
(\cite{hen:fek}) and in our paper~I, while G\'alvez et al.
(\cite{galvez07}) favored the odd K0V+K1IV classification with the
less-massive component to be the evolved star. Two subgiants are
more likely and would favor the 1.945-d period to be the true
rotation period of the secondary (suggesting a minimum radius of
1.45~R$_\odot$). The \Halpha\ line appears in emission and with a
strongly variable profile (Fig.~\ref{F-Ha}).

\begin{figure}[!tb]
\center
\includegraphics[angle=0,width=8cm,clip]{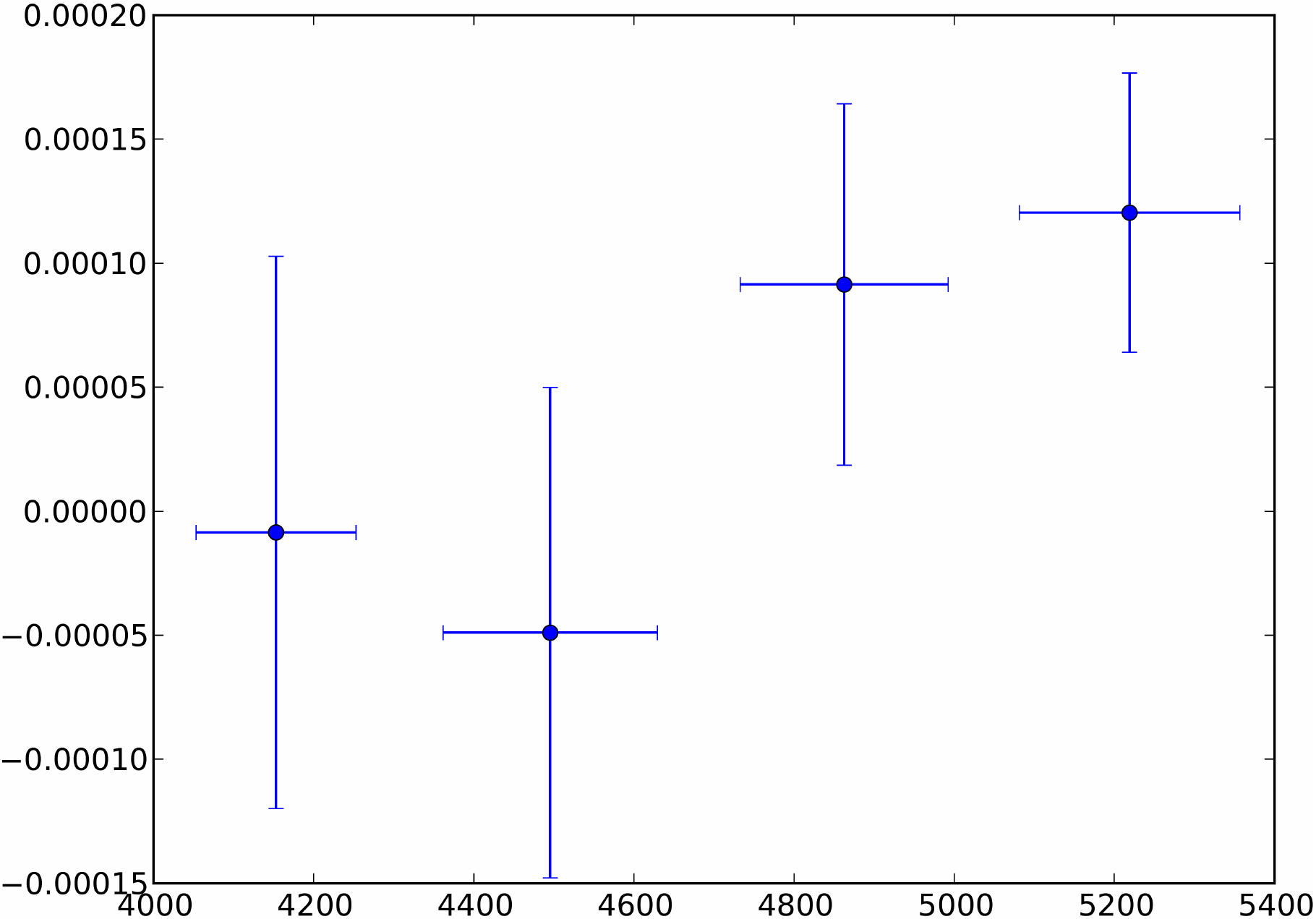}
\caption{Seasonal orbital periods for HD\,82286 (FF~UMa). Shown is
the period difference in days versus truncated Julian date for
four independent orbital solutions with respect to the period that
is obtained from the full data set (i.e. 3.2751924~d; see
Table~\ref{T5}). }\label{F-pvar}
\end{figure}

G\'alvez et al. (\cite{galvez07}) claimed to have detected an
orbital-period variation on the order of $dP/P=$5\,$10^{-4}$, but
Griffin (\cite{gri:ffuma}) could not confirm this variation from a
reanalysis of the same data. Due to this situation our STELLA data received a special treatment of the systematic errors. We followed the procedure outlined by Weber \& Strassmeier (\cite{capella}) for Capella. Briefly, we compute synthetic spectra for both components and combine them according to their real brightness ratio and then extract a radial velocity with the same 2d-cross correlation as for the real data. This enables the removal of the systematic blending effects and converges on rms residuals smaller than the precision of a single data point. With this, our four consecutive
years of observations do vaguely hint towards such an
orbital-period variation but unfortunately remain still
inconclusive (Fig.~\ref{F-pvar}). We computed orbits for all four observing seasons
separately as well as for the entire data set and derived upper
and lower limits for $dP/dt$ between 2.76\,$10^{-7}$ and
0.2\,$10^{-7}$. In case we take an equal-weight approach for the
four solutions the most likely value for $dP/P$ would be
5.5\,$10^{-5}$, with an upper limit of 1.2\,$10^{-4}$ and a lower
limit of 8.5\,$10^{-6}$, i.e. close to zero; altogether not quite
consistent with above claim by G\'alvez et al. (\cite{galvez07}).
Fig.~\ref{F-pvar} shows the orbital periods of the four STELLA
observing seasons over time.

\emph{HD\,82841 = OS Hya}. Our photometry shows a period of
51.54$\pm$0.02~d while the SB1 orbital solution gives a period of
56.7316~d with a slightly eccentric orbit with $e$=0.087. The
pseudo-synchronous rotation period would be expected to be near
53.9~d. As the photometric period is even shorter than this, the
star is likely an asynchronous rotator. Our $v\sin i$ measure of
8~\kms\ converts to a minimum radius of 8.1~R$_\odot$ and, together
with the temperature and gravity from our spectrum synthesis,
suggests a K2 giant. We note that the metallicity of --0.30$\pm$0.03
is significantly subsolar.

\emph{HIP\,50072 = EQ Leo}. We give an SB1 orbit with a formally
but marginally significant eccentricity of 0.014$\pm$0.0043 and a
period of 34.3157~d. An $e$=0 solution has an rms of the residuals
worse by $\approx$10~\% . The \Halpha -line profile appears as an
asymmetric and variable absorption line, indicating chromospheric
activity. The \Halpha -core flux shows a period of 34.8~d, very
close to the orbital period while the photometry gives a 33.36~d
as the best-fit period. The latter is better defined and we
interpret it to be closest to the true rotation period of the
star. The system is then in or close to dynamical equilibrium.

\emph{HD 93915}. Our SB2 orbit from STELLA data is in good
agreement with the Cambridge orbit from Griffin (\cite{griff09}).
As our baseline is more than five times longer and better sampled,
our orbit clearly supersedes the Cambridge orbit. The orbital period is
near 223~d and $e$=0.3795$\pm$0.0009. The expected
pseudo-synchronization period is $\approx$111~d but
our photometry does not show any type of periodic behavior despite
of having a dispersion of up to 0\fm2 mag in $V$ over 138 days.

\emph{HD 95188 = XZ LMi}. We judge this star to be single. The
radial velocities possibly show small-amplitude variations of
$\pm$0.1~\kms\ but no clear period while the \Halpha -core flux
indicates a period of 190~d, most-likely spurious. Our photometry
revealed an uncertain period of 7.00$\pm$0.06~d but with a small
amplitude of just 0.025~mag in $V$, that we nevertheless interpret
to be the rotation period.

\emph{HD 95559 = GZ~Leo}. The star was found to be a double-lined
spectroscopic binary by Jeffries et al. (\cite{jef}) who also
predicted its photometric variability. The orbit was later refined
and the photometric variability discovered by Fekel \& Henry
(\cite{fek:hen}). Our new circular orbit is very well determined despite the rotationally broadened line
profiles of $v\sin i$ of 33~\kms\ for both components. In our
original survey paper (paper~I), we noted that the star may be
triple-lined based on a single coud\'e spectrum. We find
no evidence for a third component in our new echelle spectra. However, the
secondary's residual radial velocities of 0.165~\kms\ indicate
a period of 0.592~d that appears to be an alias of the 1.517-d
period obtained from our photometry. The true rotation period is
likely 1.517~d as it also agrees with the orbital period of 1.526~d. Obviously, the system is well synchronized.

\emph{HD 95724 = YY LMi}. Its light variability was discovered in
our paper~I and accordingly named YY~LMi by Kazarovets et al.
(\cite{kaz:sam}). The star has a constant radial velocity from 76
STELLA measurements and is likely single. Our APT photometry shows
a period of 11.49~d and confirms that YY~LMi is a spotted and
active star.

\emph{HD 104067}. Single star with no obvious periodicity down to
the detection limit of $\approx$100~\ms\ for the given $v\sin i$ of
8~\kms . However, there are two equally significant peaks in the
periodogram from the \Halpha\ flux of 34~d and 102~d, one the alias
of the other. We had no photometry for this target.

\emph{HD 105575 = QY Hya}. Szalai et al. (\cite{szalai}) presented
a photometric and spectroscopic study of this 0.29-d eclipsing
binary and derived a complete system's solution from $BV$ light
curves and radial velocities. These authors also announced the
discovery of a tertiary component in the cross correlation
function and listed absolute parameters appropriate for a
(K5/M1)/G4 system. We dare to note that they (and the referee) had missed
our paper~I in A\&A from 2000 where we gave radial velocities of the
binary plus the tertiary and determined \Halpha\ emission-line
fluxes for all three components. We then noted that the primary
had moderately strong and sharp Ca\,{\sc ii} H\& K emission and
showed that the resonance lines appeared to be asymmetric due to
the resonance lines of a third component. The Ca\,{\sc ii}
emission lines even appeared with sharp and red shifted
interstellar absorption. Because our new STELLA data sample only
one fourth of the tertiary orbit, we added our earlier paper-I KPNO
coud\'e-feed spectra, two HARPS and one FEROS velocity from the
ESO archive. With these, we extent the time coverage to 4489~d and obtain an orbital period for the
tertiary around the eclipsing pair of 4550$\pm$52~d and give a
preliminary orbit with $e$=0.391$\pm$0.006 for it. Our close-pair orbit is
SB2 and circular with a period of 0.2923378~d. Its mass ratio is 0.457. The secondary's K amplitude is 158~\kms\ and both stars have a $v\sin i$ of around 40~\kms ; and accordingly large are the orbital residuals. The disagreement with the elements given by Szalai et al. (\cite{szalai}) possibly comes from the different ways the cross correlation is determined. Our photometric
orbital period is 0.292386~d and agrees with the orbital period
within its uncertainties. The eclipse depth is 0.36~mag in $V$. No
rotational modulation is detected from the radial velocities. 

\emph{HD 106855 = UV~Crv}. The catalogue of composites of double and
multiple stars (Dommanget \& Nys \cite{ccdm}) identifies HD\,106855
as the double star Don\,520 = CCDM\,J12174-2104AB. The B component
is just 1.3\arcsec\ away from the A component at a position angle of
38\degr\ (both values measured in 1959 by Donner) and appears
approximately 1.7~mag fainter in $V$ than the A component. Worley
(\cite{worley}) gave a more recent measurement of 1.64\arcsec\ and
31.6\degr\ in 1972.1. The Hipparcos/Tycho photometry lists $V_{\rm
T}$=9.76$\pm$0.03 and ($B-V$)$_{\rm T}$=0.95 for the A component
(Fabricius \& Makarov \cite{tycho2}). The fiber-core diameter of our
STELLA/SES projected on the sky is 3.0\arcsec\ centered on the
brighter A component. Therefore, the B-component is just at the
rim of the field of view and could have been within the diaphragm from time to time. Our cross-correlation function shows a broad, sometimes doubled, main peak and a third narrow peak which is sometimes stronger than the main peak.

The orbital period from the main peak, i.e. presumably from the Aa component, is 0.67~d with strong aliasing
at $\approx$2~d. There are 11 phases where the main peak is split into two peaks and possibly represent detections of the Ab component (plotted in Fig.~\ref{F3}). However, these velocities have unreasonable high scatter (9.9~\kms ) and a comparably small $K$-amplitude (37.7~\kms ). Possibly, the Ab component is part of another spectroscopic binary. Its velocities alone would suggest a 0.2-d period. A tentative 0.85-d period is found from the third, narrow peak of the cross correlation function but we can not claim it conclusively. However, it clearly appears sometimes on the one or the other side of the main peak, but not always. One possibility is that the visual B component is only sometimes within the STELLA fiber aperture \emph{and} is itself a spectroscopic binary. It would make HD\,106855 a quadruple system.

\emph{HD 108564}. Single star with a high average radial velocity
of +111.363$\pm$0.009~\kms . A periodogram of the radial
velocities hint at a period of 9.58~d with a small amplitude of
0.1~\kms\ while the \Halpha\ flux favors a (formally even more
uncertain) period of 3.63~d. The $v\sin i$ of 14.8~\kms\ would
convert to a minimum radius of $\approx$1~R$_\odot$ if 3.63~d is
interpreted as the rotation period. The 9.58-d period would result
in a minimum radius in disagreement with the Hipparcos distance of
28~pc. Our spectrum synthesis clearly indicates a late K dwarf
with $T_{\rm eff}$=4560$\pm$24~K, $\log g$=4.40$\pm$0.06 and a
significant sub-solar metallicity of --0.90$\pm$0.03 which implies
a radius of $\approx$0.7~R$_\odot$.

\emph{HD 109011 = NO~UMa}. Griffin (\cite{griff10}) noted for this
star ``Slow changes in velocities, SB2''. We verify the system to
be a SB2 and find an orbital period of 1274.7$\pm$0.9~d. Our
photometry indicates variations with a period of 8.4$\pm$0.2~d and
an amplitude of 0.02~mag in $V$, which we interpret to be due to
rotational modulation of the primary component. This period is
similar but not completely in agreement with the values of 8.56
and 9.19~d found by Gaidos et al. (\cite{gai:hen}). But in any
case, the system contains at least one very asynchronously
rotating component because the expected pseudo-synchronization
period would be $\approx$440~d ($e$=0.507). The orbital solution
shows radial-velocity residuals of 0.17 and 0.46~\kms ,
respectively, which are larger than expected for the $v\sin i$
given. A Lomb-Scargle periodogram of these residual velocities
indicate periods of 639~d for the primary and 246~d for the
secondary, the former roughly half the orbital period. The large
Hipparcos parallax and a $V$=8.1~mag system brightness suggests
two dwarfs rather than giants. If interpreted due to rotation, the
periods of $\approx$639 and 246~d would indicate minimal radii
appropriate for supergiants, totally impossible for a distance of
23.7~pc. If these periods are real at all, they are possibly
connected with surface activity, e.g. due to a spot cycle or
active longitudes. The adopted rotation period of 8.4~d and the
$v\sin i$ of 5~\kms\ for the primary suggests a minimum radius of
0.83~R$_\odot$.

\begin{figure}[!tb]
\center
\includegraphics[angle=0,width=80mm]{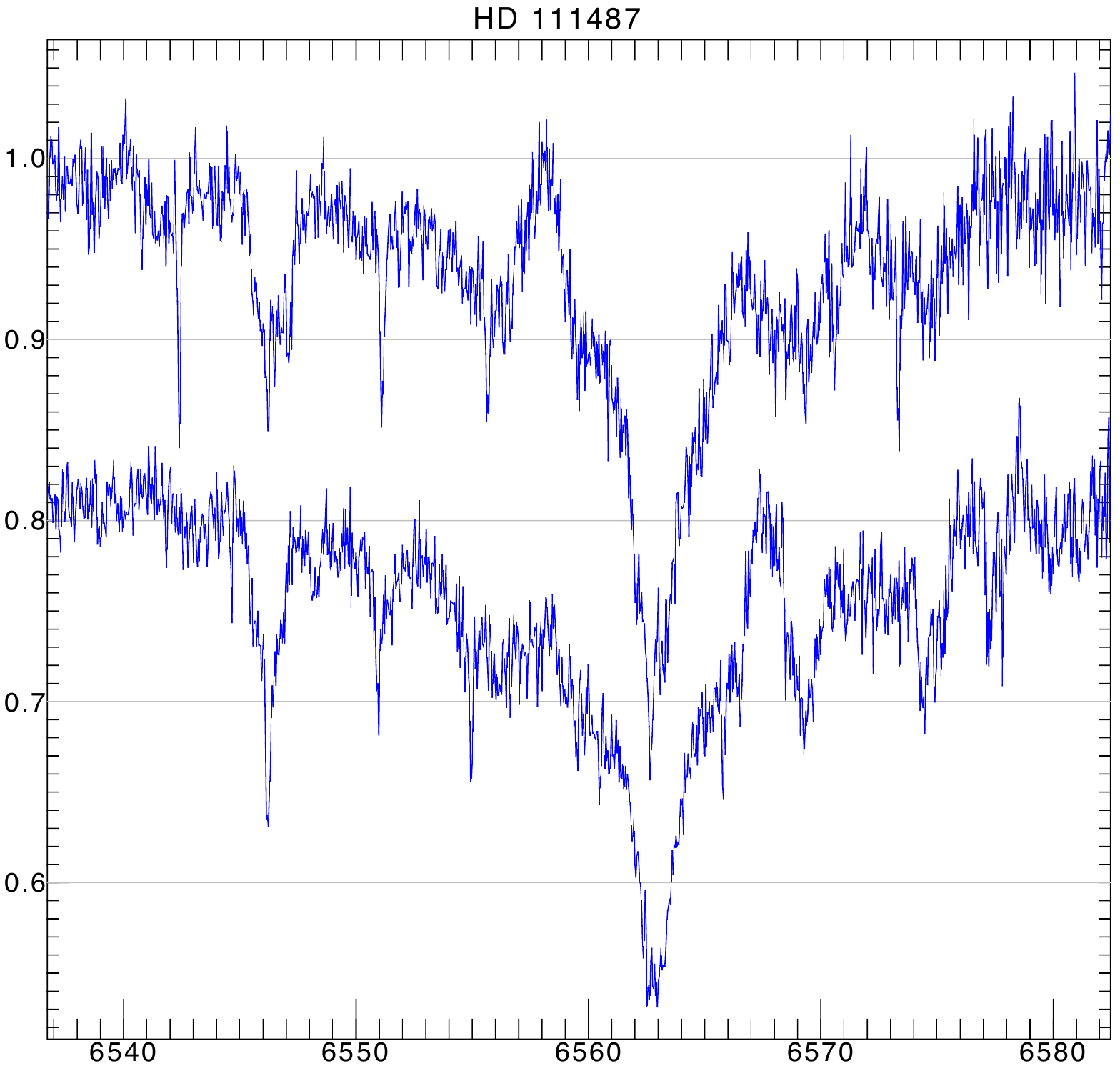}
\caption{\Halpha\ line profile of HD~111487 at two orbital phases
180$^\circ$ apart. Its secondary component is detectable only in
\Halpha\ where it is seen as an emission line (appearing bluewards
of the primary's absorption line in the top spectrum, and redwards
in the lower spectrum).}\label{F-111487}
\end{figure}

\emph{HD 111487 = IM~Vir}. An SB2 orbit with a period of 1.3086147~d
is given. Our photometry confirms the eclipsing nature of the system
with a period of 1.3085900$\pm$0.000007~d and an eclipse depth of
0.76~mag in $V$. A rms of the residual velocities of the primary of
0.268~\kms\ is appropriate for rotationally broadened spectra of
$v\sin i$=42~\kms\ but the residuals for the secondary-star
velocities are anomalously high (1.8~\kms ; $v\sin i$=28~\kms ),
indicating an asymmetrically spotted star. However, residual periods
of 3.45~d or 1.41~d (one the alias of the other) are seen for the
primary but none for the secondary. Their amplitude is
$\approx$0.6~\kms . The secondary shows strong \Halpha\ emission
while the primary appears with \Halpha\ in absorption (see
Fig.~\ref{F-111487}). The photometry reveals a period of 1.309~d from
data outside of eclipse and have a full amplitude of 0.15~mag in $V$
that we interpret to be from rotational modulation synchronized to
the orbital motion.

\emph{HD\,112099}. A single-lined eccentric orbit with a period of
23.50447$\pm$0.00013~d and an rms of 67~\ms\ is obtained. The
expected pseudo-synchronization period is $\approx$9.6~d
($e$=0.447). A residual period of 4.6~d is seen in a Lomb-Scargle
periodogram but is not conclusive. Together with our $v\sin i$
measure of 4.2~\kms , it would indicate a minimum radius of
0.38~R$_\odot$, at least in principle agreement with the Hipparcos
parallax. Our spectrum synthesis confines this to a K1V
classification, in agreement with its $B-V$ color.

\emph{HIP\,63322 = BD+39\,2587\,B}. B component of a visual pair
with HD\,112733 (HIP63317) as the primary. The spectrum appears at
almost all wavelengths single lined but shows doubled Ca\,{\sc ii}
H\&K lines (Strassmeier et al.~\cite{ca2}). G\'alvez et
al.~(\cite{galvez06}) presented an SB2 orbit from data taken during
nine nights in March/April 2004. They found an eccentricity of 0.3
and noted that the spectrum is best fit with a G8V primary and an
M0V secondary star. An SB1 orbit with zero eccentricity and
$P$=4.593843~d is derived from STELLA data. Our spectra show the
secondary with \Halpha\ in emission and the primary with \Halpha\ in
absorption, similar to HD~111487. The \Halpha -line flux of the
primary shows a clear periodicity of 2.30~d, exactly half the
orbital period within the uncertainties, which is simply due to the
fact that the secondary-star emission is sweeping through the
primary-star line. The photometry reveals a period of
2.227$\pm$0.003~d, assumed to be from a double-humped light curve,
and we adopt twice its value to be the true rotation period of the
star.

\emph{HD\,112859 = BQ CVn}. Just recently, Griffin
(\cite{griff09}) obtained an orbit very similar to ours and we
refer the reader to his discussion on the various stellar
properties. Our orbit was determined from 93 velocities for the
primary and 87 for the secondary over a time range of 1249~d and
achieved rms residuals of 0.30~\kms\ and 0.26~\kms\ for the
primary and secondary, respectively. Our photometry indicates a
rotational period that is synchronized to the orbital revolution.

\emph{HIP\,63442 = CD CVn}. A single-lined orbit with $P$=38.7156~d
and $e$=0 is given. We see many detections of a second line system
in the cross-correlation functions but with such a large scatter
(9.9~\kms ) and small amplitude (0.5~\kms ) that we can not be sure
of its reality. Its marginally significant period is in agreement
with the orbital period of the primary but in phase with all its
measurements. It resembles HD\,16884, possibly being a pair of SB1
binaries in a bound orbit. The residuals from the primary orbit have
an rms of 0.27~\kms\ and show a periodicity of 19.14~d, nearly
exactly half the orbital period. The \Halpha -core flux varies with
a period of 39.2~d and our photometry gives a period of 40.3~d,
which we interpret to be the true rotational period of the star.
Thus, CD~CVn is rotating not quite but close to synchronous.


\emph{HD\,127068 = HK Boo}. We present a new SB2 orbit with a period of 15.15011$\pm$0.00011 from 98
velocities for the primary and 99 for the secondary spanning a time
range of 1257 days. The residual rms is 0.19~\kms\ for the primary
and 0.36~\kms\ for the secondary. There are periods in the residuals 
of both stars of 18.6~d and 24.3~d for the primary and secondary,
respectively. The primary's value is in agreement with our
photometric period of 18.187~d in this paper and suggests that it is
likely the rotational period of the star. The secondary's period is
maybe spurious as its amplitude is also very low, 0.3~\kms .  From
photometry in paper~I we found 17.62~d. In any case, this would make
the HK~Boo primary an asynchronously rotating star. Griffin
(\cite{griff09}) gives a discussion of the star's properties and
cited a photometric period from Hipparcos data of 6.953~d that we
can not verify.

\emph{HD 136655}. In paper~I, we had given two velocities of
$-29.9\pm 4.1$ and $+17.9\pm 1.7$~\kms\ from a single
blue-wavelength spectrum for two components. The latter must have
been an erroneous measurement because it is not verified in the
present paper. Table~\ref{T3} lists an average STELLA velocity of
$-31.921\pm 0.035$~\kms\ from 122 spectra. Its Ca\,{\sc ii} H\& K
emission appears very weak, if detectable at all. Our photometry
shows the star to be constant.

\emph{HD\,138157 = OX Ser}. Our SB1 orbit suggests a very small
eccentricity of 0.0051$\pm$0.0009 from a total of 148 velocities.
The rms of the residuals is 0.22~\kms ,  an excellent value for
the large line broadening of $v\sin i$=40~\kms , and somewhat
better than if a circular orbit is enforced (rms of 0.25~\kms ). A
periodogram from the \Halpha -line fluxes gives the strongest peak
at a period of 14.38~d, very close to the orbital period. Our
photometry shows clear variations but suggests a period of 7.18~d,
pretty much half the period of the orbital revolution. Its phase
stability indicates that the amplitude is dominated by the
ellipticity effect but annual variations indicate that part of it
is due to spots. We conclude that the rotation period of the star
is very near the orbital period and thus synchronized with the
orbital motion.

\emph{HIP 77210 = V381 Ser}. The star's alternative name is G16-9. An initial SB2 orbit with a period of
9.94319~d was determined by Goldberg et al. (\cite{gold}). Our SB2 orbit is among our most
precise with an rms of the residuals of the primary of 52~\ms\
from 96 velocities (but four times larger rms for the
secondary). Our photometry shows variations with a period of
13.7$\pm$0.1~d that we interpret to be the rotation period of the
primary, thus, significantly asynchronous with respect to the
orbital period. The secondary's velocity residuals show a
systematic rms with a period of 4.79~d, a little less than half
the orbital period. The orbital eccentricity of
0.06032$\pm$0.00024 suggests a formal pseudo-synchronization
period of $\approx$9.7~d, which is close to twice the secondary's
residual period. We conclude that 9.58~d is likely the true
rotation period of the secondary and thus synchronized to the
orbital revolution.

\emph{HD 142680 = V383~Ser}. The photometric period from data in
1999 in paper~I was 33.52~d. Our new photometry principally
verifies this period and gives 33.4$\pm$1.5~d. The orbital
eccentricity is 0.3158 from our new STELLA SB2 orbit and is in agreement
with the eccentricity of 0.3139 from Griffin's (\cite{griff09})
SB1 orbit. The pseudo-synchronous rotation period would be around
15.3~d, which is suspiciously close to 1/2 of the photometric
period. However, the photometric period can not be halved if it is
due to spots rotating in and out of view and we conclude that the
system is rotating super synchronously. The systemic velocity is
relatively high with $-83.83$~\kms , making this system a
high-velocity binary.

\emph{HD 143937 = V1044~Sco}. This is the A component of the visual
binary GJ~9539. The B component is 10.3\arcsec\ away and
approximately 2.7~mag fainter than component A in the Hipparcos bandpass. It
is therefore included in our photometry with a 30\arcsec\ diaphragm
but well separated for the STELLA spectroscopy. Cutispoto et al.
(\cite{cuti1999}) found the A component to be an SB2 binary (it is
listed as an eclipsing binary in Malkov et al. \cite{malkov} based
on a note in Cutispoto et al. \cite{cuti1999}). We verify its
eclipsing nature and find an orbital period from photometry of
0.91479$\pm$0.00003~d, in excellent agreement with the
radial-velocity orbit of 0.91483076~d. The eclipse depth is 0.41~mag in $V$ while
the spot wave shows an amplitude of 0.10~mag. The rms of our orbital
residuals for both stars are comparably high, 0.28~\kms\ and
1.26~\kms, but appropriate for the $v\sin i$ of 55 and 45~\kms\ for primary and
secondary, respectively. <sucvh large residuals are related to the short orbital
period of 0.9~d and the corresponding degree of line blending.

\emph{HD 147866 = V894 Her}. We present an SB1 orbit with a period
of 114.050~d and an eccentricity of 0.5141. The $K$ amplitude is
just 6.71~\kms\ and the mass function accordingly small. Our
photometry reveals a likely rotation period of 80.3$\pm$2.2~d,
while pseudo synchronization would be expected at a period of
$\approx$40~d. It appears that the system had not had enough time
yet to obtain orbital and rotational equilibrium. With our $v\sin
i$ measure of 5$\pm$1~\kms\ the minimum radius is 7.93~R$_\odot$.
The Hipparcos parallax places the star at a distance of 193~pc
which implies an absolute magnitude of +1.67~mag (for V=8.1~mag).
Our spectrum synthesis gave $T_{\rm eff}$=4580~K and $\log g$=2.7
with a significant sub-solar metallicity of --0.24$\pm$0.02. The
star is most likely a K1III giant.

\emph{HD 150202 = GI Dra}. We confirm Griffin's (\cite{griff09})
orbital period with a value of 68.47160$\pm$0.00097~d. The STELLA
orbit's rms of the residuals is just 83~\ms \ from 179 velocities over 1112~d.
Our spectrum synthesis yielded $T_{\rm eff}$=5010$\pm$25~K and a
$\log g$=3.06$\pm$0.07, suggesting more a G8III star than a K0III as
put forward by Griffin (\cite{griff09}). The periodogram from our
photometry shows the strongest peak at a period of 36.4$\pm$0.5~d.
This is likely just one-half of the true rotation period (72.8~d)
and is then in rough agreement with the period of 76.70~d from
Hipparcos photometry. Thus, the star is an asynchronous rotator with
$P_{\rm rot}>P_{\rm orb}$.

\emph{HD 153525 = V1089 Her}. From 196 STELLA spectra we obtain a
mean radial velocity of --6.844$\pm$0.039~\kms . The star is
single but at least moderately active judging from its Ca\,{\sc
ii} H\&K emission-line strength (c/o paper~I). The new photometry
shows variations with a period of 15.4$\pm$0.1~d that we
tentatively interpret to be the rotational period. A distance of
just 17.5~pc and our values for $T_{\rm eff}$=4775$\pm$23~K and
$\log g$=4.47$\pm$0.05 suggest a K dwarf with a $v\sin i<$3~\kms ,
in agreement with our upper limit of 3~\kms.

\emph{HD 153751 = $\epsilon$ UMi}. The system is a single lined
spectroscopic binary and has an orbit determination dating back now
60~yrs (Climenhaga et al. \cite{cli:mck}). This is certainly because
of the high declination of +82\degr\ despite that Climenhaga et al.
already noted the Ca\,{\sc ii} H\&K emission and thus put the system
out for further observation. Its eclipsing nature was discovered by
Paul Guthnick in 1946 and later verified by Hinderer (\cite{hind})
who concluded on a pair of gG1 and dA8-dF0 stars for the primary and
secondary, respectively. Today's accepted classification is G5III
for the primary (e.g. Keenan \& York \cite{kee:yor}). Our spectrum
synthesis gives $T_{\rm eff}$=5215$\pm$47~K, $\log g$=3.21$\pm$0.08
and a metallicity of --0.25$\pm$0.04, i.e. significantly subsolar
but altogether in agreement with a G5III classification. Our new SB1
orbit has a period of 39.48042$\pm$0.00012~d and zero eccentricity.
Its rms of 121~\ms\ from 220 velocities spanning 2144 days is mostly due to the
moderately large rotational velocity of 25.6~\kms\ and only less
from the jitter from starspots because no convincing rotation period is found
from the residuals.

\emph{HD 155802}. In paper~I, we had given velocities of
--33.5$\pm$1.2 and --33.3$\pm$0.5~\kms\ from a blue-wavelength and
a red-wavelength spectrum, respectively, and +24.7$\pm$2.0~\kms\
for another component from the one blue-wavelength spectrum. The
latter must have been an erroneous measurement because it is not
seen in the spectra in the present paper. The average velocity
from 114 STELLA spectra is --32.711$\pm$0.033~\kms . Its rms is
practically the precision of the STELLA/SES system and thus the
star is constant down to our detection limit of $\approx$30~\ms .
However, its chromospheric activity is verified by the presence of
weak Ca\,{\sc ii} H\&K emission lines.

\emph{HD 184591}. The star is single according to our 155 STELLA
velocities. Its mean is --37.635$\pm$0.044~\kms . The photometry
indicates just a weakly determined period of P=48.9$\pm$2.2~d with
an amplitude of below 0.01~mag in $y$. Our spectrum synthesis
resulted in $T_{\rm eff}$=4800$\pm$40~K, $\log g$=2.82$\pm$0.07
and a metallicity of --0.18$\pm$0.03. If the 48.9-d period is real
and the true rotational period of the star, the $v\sin i$ of
4.7$\pm$0.8~\kms\ confines the radius to be at least
4.5~R$_\odot$, in principal agreement with above gravity and the
Hipparcos parallax.

\emph{HD 190642 = V4429 Sgr}. Strohmeier (\cite{stroh}) discovered
this star on his Bamberg plates to be variable with an amplitude
of 0.25~mag. No other information was given. But based on the
Hipparcos satellite data, Kazarovets et al. (\cite{kaz:sam})
assigned a variable star designation. We find the star to be an
eclipsing system and detect a photometric period of
24.8339$\pm$0.0009~d after excluding the eclipse points which, if
doubled, is very close to the orbital period from radial
velocities (49.5969$\pm$0.0012~d, $e$=0.0126$\pm$0.0011). The eclipse depth is
0.43~mag. The rms (0.445~\kms ) of the velocity residuals is
larger than expected for $v\sin i$=19.6~\kms\ and a periodogram
analysis shows a convincing residual periodicity of
24.72$\pm$0.02~d with an amplitude of 1.2~\kms . The \Halpha -core
flux shows also a significant period at 24.9~d. All of these
periods must be doubled to obtain the true rotation period and
then suggest that the primary's rotation is synchronized to the
orbital motion, or at least nearly so. Our spectrum synthesis
gives values of $T_{\rm eff}$=4760$\pm$63~K and $\log
g$=2.96$\pm$0.11 with a significant sub-solar metallicity of
--0.47$\pm$0.05. The Hipparcos parallax suggests an absolute
system magnitude of +1.34~mag ($V$=8.08~mag, distance of 223~pc,
no extinction). Altogether these values are compatible with a
K1III-IV giant. Knowing $i\approx$90\degr\ gives directly a radius
of 19.6~R$_\odot$ and a rotation period of $\approx$50~d, in good
agreement with our various observations. Assuming a mass of two
solar masses for the primary, the mass function of 0.330 demands a
secondary mass of 1.22~M$_\odot$; it suggests a late F
main-sequence star, maybe F6-8.

\emph{HD 197913 = OR~Del}. This is the A component of the visual
binary ADS~14270. Griffin (\cite{griff05}) gave an overview of its
observational history and presented a full SB2 orbit in very good
agreement with our present determination. Our rms residuals are very low though, 100~\ms\ for the primary and 161~\ms\ for the secondary. In paper~I, we presented
double-lined Li spectra for the A component and a single-lined
spectrum for the B component (erroneously identified with
lower-case a and b in that paper; \emph{pace} Griffin). The
fainter B component appeared with somewhat stronger Ca\,{\sc ii}
H\&K emission as the (blended) A components. The 6.56-d
photometric period that we derived from our APT time series from
1999 was assigned to HD~197913, without an A nor a B
specification. This was because our APT diaphragm was 30\arcsec\
and the AB separation roughly 11\arcsec . Consequently, it
prompted Kazarovetz et al. (\cite{kaz:sam}) to assign the variable
star designation OR~Del to the AB system but Simbad resolves it
for the A component of which, however, it does not give the HD
number. Variability from the B component can not be excluded but
it is indeed more likely that the photometric variability stems
from the 0.8-mag brighter (combined) A components. Our current
photometry has a best-fit period of 6.56$\pm$0.07~d. The AB
period is suggested by Griffin (\cite{griff05}) to be around
``thousand years''. Our new STELLA orbit gives a period of
1778.5$\pm$1.3~d for the A component from basically a single
revolution (time span of data is 1848~d) but still better confined than the published value of
1790$\pm$13~d.

\emph{HD\,199967}. We find it to be an SB2 in a 6.100213-day orbit
with a very low but significant eccentricity of 0.0024$\pm$0.0002. An $e$=0 solution has an rms of the residuals
worse by more than 10~\% for each component. No photometric period is
detectable but such a short orbital period makes it likely a
synchronized system. Our $v\sin i$ measures of 11 and 9~\kms\ for
the primary and the secondary, respectively, then lead to minimum
radii of 1.3 and 1.1~R$_\odot$. The effective line ratio is 3.2 to
1. The Hipparcos parallax gives a distance of 74~pc and an
absolute system magnitude of +3.67~mag, basically excluding two
main-sequence stars. The system's $B-V$ is 0.58~mag and, with a
mass ratio of 0.776, does suggest two early G-type dwarf-subgiants.

\emph{HD 202109 = $\zeta$~Cyg}. The primary is a mild barium star
with a white-dwarf companion. Griffin \& Keenan (\cite{gri:kee})
discussed its spectral classification in detail and presented an
orbit with a period of 6489$\pm$31~d, $e$=0.22$\pm$0.03 and an
amplitude of $K$=3.31$\pm$0.12~\kms . More recently, Eaton \&
Williams (\cite{tsu}) listed a mean velocity of
19.55$\pm$0.25~\kms\ from 74 observations but did not give the
individual values. Our own radial velocities show a clear
non-linear trend with a full range of 2.8~\kms\ during the 3.5~yrs
of observation. We present a new orbit from the combined values
from CORAVEL, the Cambridge spectrometer, STELLA and the literature cited in Griffin \& Keenan (\cite{gri:kee}) in Fig.~\ref{F2} and list it in Table~\ref{T4}. It covers a time range of 114.8 years! We shifted the
Griffin \& Keenan (\cite{gri:kee}) points by +0.50~\kms\ to bring
them from the IAU to the STELLA system. The STELLA data in the
panel are the ones between phase 0.1--0.3 and an average rms of
42~\ms . The refined period is now 6446$\pm$14~d and $e$=0.244$\pm$0.016,
in fair agreement with the Griffin \& Keenan orbit.

\emph{HD 218739}. Visual component to HD\,218738 = KZ And, itself
a cataloged RS~CVn binary 15.6\arcsec\  away. HD\,218739 appears
with a constant radial velocity of $-5.201\pm$0.040~\kms .

\emph{HD 226099}. Duflot et al. (\cite{duf:feh}) listed six
velocities between +22 and +4~\kms\ which were the first
measurements that showed the variable velocity. In our paper~I, we
gave velocities for both stellar components from a single
blue-wavelength spectrum but did not find significant Ca\,{\sc ii}
H\&K emission lines. In this paper, we present its first orbit (SB2) with
a period of 18.78209~d and an eccentricity of 0.3096
(Fig.~\ref{F3}). The two components are of comparable mass with a
ratio of 0.897. The $v\sin i$ for the primary component is below our
resolving limit and estimated to be 2~\kms , but twice as larger for
the cool secondary. The rms residuals for the two components are
85~\ms\ and 182~\ms , possible due to the sharp lines.
The orbital eccentricity suggests a formal pseudo-synchronization
period of $\approx$11.6~d but no we have no photometry for this star. However, the radial-velocity residuals of the secondary indicate a well-defined period of 3.131$\pm$0.008~d with a full amplitude of up to 1.6~\kms , and no power for periods longer than $\approx$4.5~d. We conclude that the 3.1 days is the rotation period of the  secondary and that it rotates significantly asynchronous.

\emph{HD 237944}. The system has a triple-lined spectrum due to Aab and the visual B component and, according to Otero \& Dubovsky (\cite{otero}), is also an eclipsing system ($ab$
components). The two stronger lines from component A make up the $ab$ SB2 in
our Table~\ref{T5}. The third, visual B or spectral $c$ component's average radial
velocity is --12.25$\pm$1.3~\kms , different by 3.0~\kms\ from the
systemic velocity of $ab$, in accordance with an expected orbital
period of $\approx$2500 years noted by Griffin (\cite{griff09}). Our rotational velocities are from a single (well-exposed) spectrum and gave $v\sin i$ of Aa, Ab, and B of 11~\kms , 9~\kms , and 11~\kms, respectively, with errors of at least $\pm$2~\kms. Note that the SB2 orbital period is just 5.5~d
with a very small but significant eccentricity. Our photometry confirms the eclipses of the
$ab$ pair (Fig.~\ref{F4}) and also shows a sinusoidal light
variation with the orbital period indicating that a spotted star
is rotating synchronously with the orbit. Therefore, and assuming $i=90^\circ$, radii are 1.19~R$_\odot$ and 0.98~R$_\odot$ and masses are 1.042~M$_\odot$ and 1.027~M$_\odot$, respectively. For our Li abundances in Table~\ref{T7}, we assume effective temperatures based on Gray's \cite{gray} table B.1 and Griffin's (\cite{griff09}) individual absolute magnitudes.

\section{Analysis}\label{S6}

\subsection{Rotational synchronization}

Before we try to derive a rotation-activity relation, we inspect
our binary sample in the rotational-period versus orbital-period
frame and identify its distribution of synchronicity, or the
deviation thereof. Asynchronously rotating binaries, or
pseudo-synchronously rotating binaries in eccentric orbits, could
experience enhanced activity due to stronger shearing forces in
their convective envelopes (Schrijver \& Zwaan \cite{sch:zwa}).
This may possibly be coupled with enhanced deep convective mixing
as in Hertzsprung-gap giants (e.g. B\"ohm-Vitense \cite{boehm}).
It may thus be rewarding to visualize the $P_{\rm rot}$-$P_{\rm
orb}$ plane.

\begin{figure}
\center
\includegraphics[angle=0,width=80mm,clip]{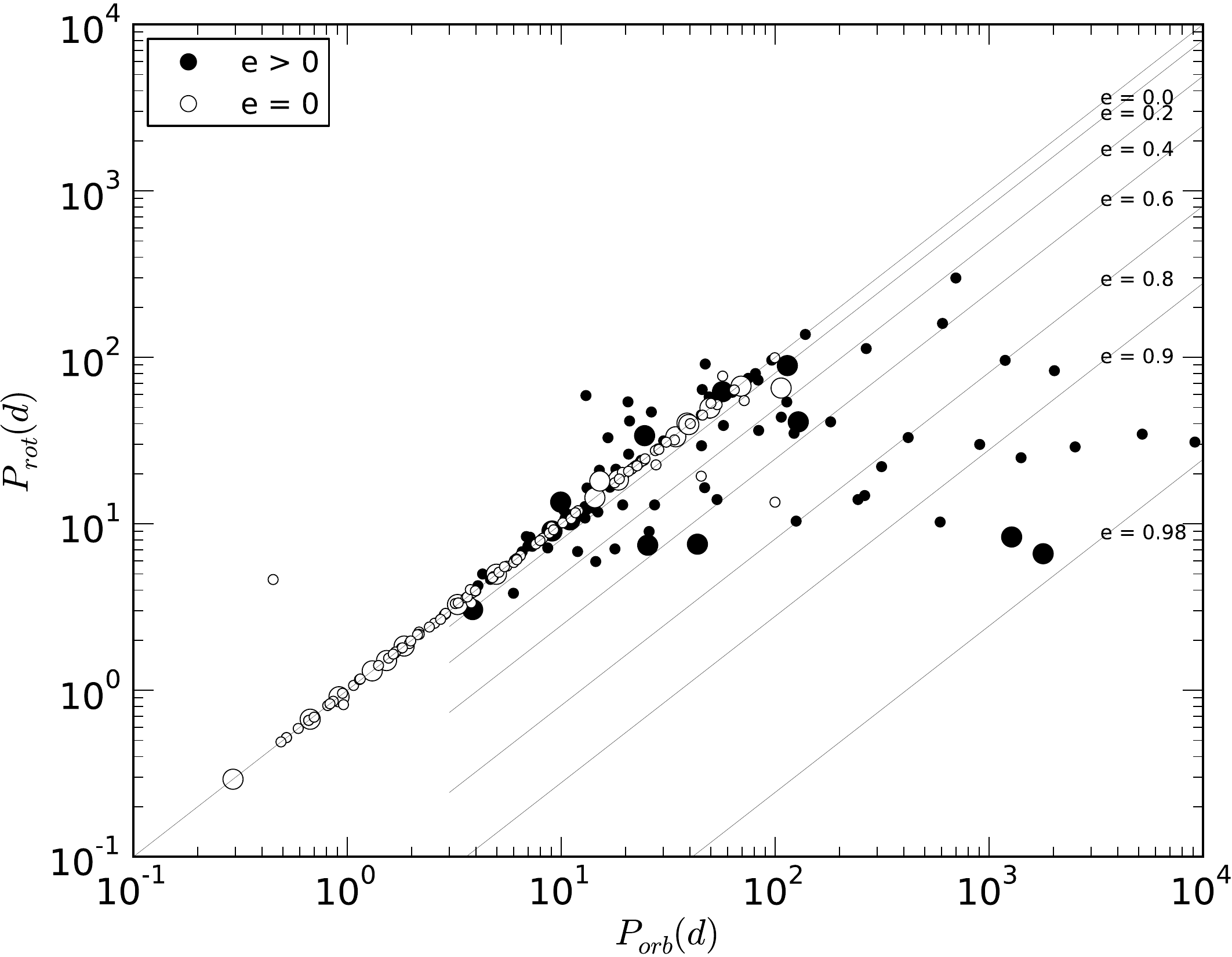}
\caption{Rotational period versus orbital period for the binary
stars from Table~1 (big dots) and other RS-CVn and BY-Dra binaries
from the CABS catalog (small dots). Open symbols indicate circular
orbits, a filled symbol indicates an eccentricity larger than
zero. The lines depict the (pseudo)-synchronization periods for
various values of the orbital eccentricity, $e$, according to Hut
(\cite{hut}). Our sample mainly consists of synchronized binaries
with circular orbits but 26\% are asynchronous rotators. See text.
}\label{Fx}
\end{figure}

Fig.~\ref{Fx} shows the distribution of synchronism in our active
binary sample. The full sample comprises of the entries in the
``Chromospherically Active Binary Star (CABS)'' catalog
(Strassmeier et al. \cite{cabs1}, \cite{cabs2}, Eker et al.
\cite{cabs3}), expanded by the stars from the present paper. Of
the approximately 410 CABS binaries only 230 are known with both a
rotational period and an orbital period (150 have no rotational,
30 have no orbital period). We note that care must be taken when
using CABS-III because, e.g., the photometric periods listed are
mixed values of rotational periods and $P_{\rm orb}/2$-periods due
to the ellipticity effect or are simply orbital periods in case
the system is eclipsing. Besides, some of the photometric periods
are derived from $v\sin i$ measurements and radius estimates (e.g.
for HD\,37171 of $P_{\rm rot}$=300~d) and are accordingly
uncertain. From Fig.~\ref{Fx} we see that the bulk of active
binaries is synchronized and in circular orbits but from the 47
systems that have rotational periods considerably shorter than the
orbital period 41 are with $e>0$ and six with $e$=0 (HD\,7205,
HD\,16884ab, see Figs.~\ref{F2} and \ref{F4}; HD\,72688 HD\,90385,
HD\,108102, and 93~Leo). These six are definitely sub-synchronous
rotators and all have orbital periods less than approximately
100~d. Of the 41 systems with $e>0$, only seven (HD\,54563,
HD\,118981, and HD\,133822=HS\,Lup if its photometric period is
exactly doubled, plus the four systems already listed in
Hall~\cite{hall86}) are rotating at or near the expected
pseudo-synchronized value according to the criteria of Hut
(\cite{hut}), leaving 34 more asynchronous rotators in the sample.
Among these, the two most extreme cases are HD\,109011 ($P_{\rm
rot}$=8.34~d, $P_{\rm orb}$=1275~d, $e$=0.508) and HD\,197913
($P_{\rm rot}$=6.62~d, $P_{\rm orb}$=1790~d, $e$=0.265), rotating
54 times and 190 times faster than pseudo synchronization,
respectively. If real -- see the individual note for HD\,197913 --
these systems either had significant mass transfer in the past or
are so young that magnetic braking was not effective yet or they
experienced a yet unconsidered spin-up mechanism in the past, e.g.
the engulfing of Jupiter-sized planets or brown dwarfs (see e.g.
Livio \& Soker \cite{liv:sok}, Carney et al. \cite{car:lat}). We
will come back to this thought later in the section on lithium in
active stars (Sect.~\ref{S6c}).

Stars where the rotational period is longer than the orbital
period are in any case asynchronous rotators, independent of their
orbital eccentricity. The early discoveries were summarized in the
paper by Hall (\cite{hall86}), e.g. $\lambda$~And and
39~Cet=AY~Cet. Three more systems stick out suspiciously above the
45-degree line in our Fig.~\ref{Fx} though. HD\,31738 is a
double-lined short-period binary with a recently revised orbital
period of just 0.45~d (and $e$=0; Griffin~\cite{griff09}). Its
photometric period of 4.55~d was derived from an earlier APT
campaign (Strassmeier et al.~\cite{str:hal89}) which, in the light
of the new orbital period, was likely an alias of the true orbital
period and needs revision. HD\,193216 (a quintuple visual system)
is a new entry in CABS-III and its A component a single-lined
binary with $P_{\rm orb}$=418.77~d and $e$=0.08 according to
Griffin~(\cite{griff02}). In CABS-III it is erroneously listed
with the orbital period of 3.62~d of its visual B component,
another SB according to Griffin~(\cite{griff02}). The rotation
period given by Wright et al. (\cite{wright}) was determined for
the brighter A-component indirectly from two measures of its
Ca\,{\sc ii} H\&K $S$-index and an empirical rotation-activity
relation, thus, lacks a real observation. The third system is
HD\,181809 ($P_{\rm rot}$=59.85~d, $P_{\rm orb}$=13.045~d,
$e$=0.04; Fekel \& Henry \cite{fek:hen05}) and is a well-studied
asynchronously rotating K0 giant/subgiant. In total there are 21
rotators with $P_{\rm rot}>P_{\rm orb}$ in Fig.~\ref{Fx}. The more
significant ones are HD\,217188 (K0III, $P_{\rm orb}$=47~d,
$e$=0.47), HD\,170829 (G8IV, 26~d, 0.18), HD\,33363 (K0III, 21~d,
0.07), HD\,71071 (K1IV, 16~d, 0.13), and the four stars HD\,82841
(K1III, 56~d, 0.09), HD\,127068 (G5-8IV, 15~d, 0), HIP\,77210
(K2V, 10~d, 0.06) and HD\,142680 (K0-2V/K7V, 24~d, 0.31) from this
paper (see Tables~\ref{T4} and \ref{T5}). Despite that these
systems are a mixture of giants and dwarfs, all of them have
orbital periods shorter than 60~d with HIP\,77210 as short as
9.94~d (and $P_{\rm rot}$=13.52~d). This subclass does not quite
follow Middlekoop \& Zwaan's (\cite{mid:zwa}) argument that giants
in binaries with periods less than 120--200 days will be
synchronized, nor Boffin et al.'s (\cite{boff}) suggestion that
they should be generally circularized if the orbital period is
shorter than $\approx$50~d. Our active binaries with $P_{\rm
rot}>P_{\rm orb}$ would need an extra rotational acceleration in
order to get synchronized and, at the same time, need to
redistribute orbital momentum in order to get circularized (but see Verbunt \& Phinney \cite{ver:phi} for the interaction of tidal circularization and orbital eccentricity). Note
that the very old (metal poor) binaries with dwarf components in
the sample of Latham et al. (\cite{lath}) appeared to reach
circularization at a transition period near 20~d. As rotational
synchronization is predicted to occur before orbital
circularization (Hut~\cite{hut}), all of above systems should have
$e>0$, which is actually the case, except for HD\,127068 (this
paper) and 39~Cet. To summarize, our full sample includes
$\approx$26\%\ asynchronous rotators (61 stars) of which about
half are rotating slower than in orbital equilibrium and can not
be explained by being too young to be synchronized yet or by
uncertainties in the measured quantities. We conclude that active
binaries must have gone through an extra spin-down besides tidal
dissipation and suggest this to be most likely due to a
magnetically channeled wind with its subsequent braking torque on
the star as described already, e.g., by Mestel~(\cite{mestel}).

\begin{figure}
\center
\includegraphics[angle=0,width=80mm,clip]{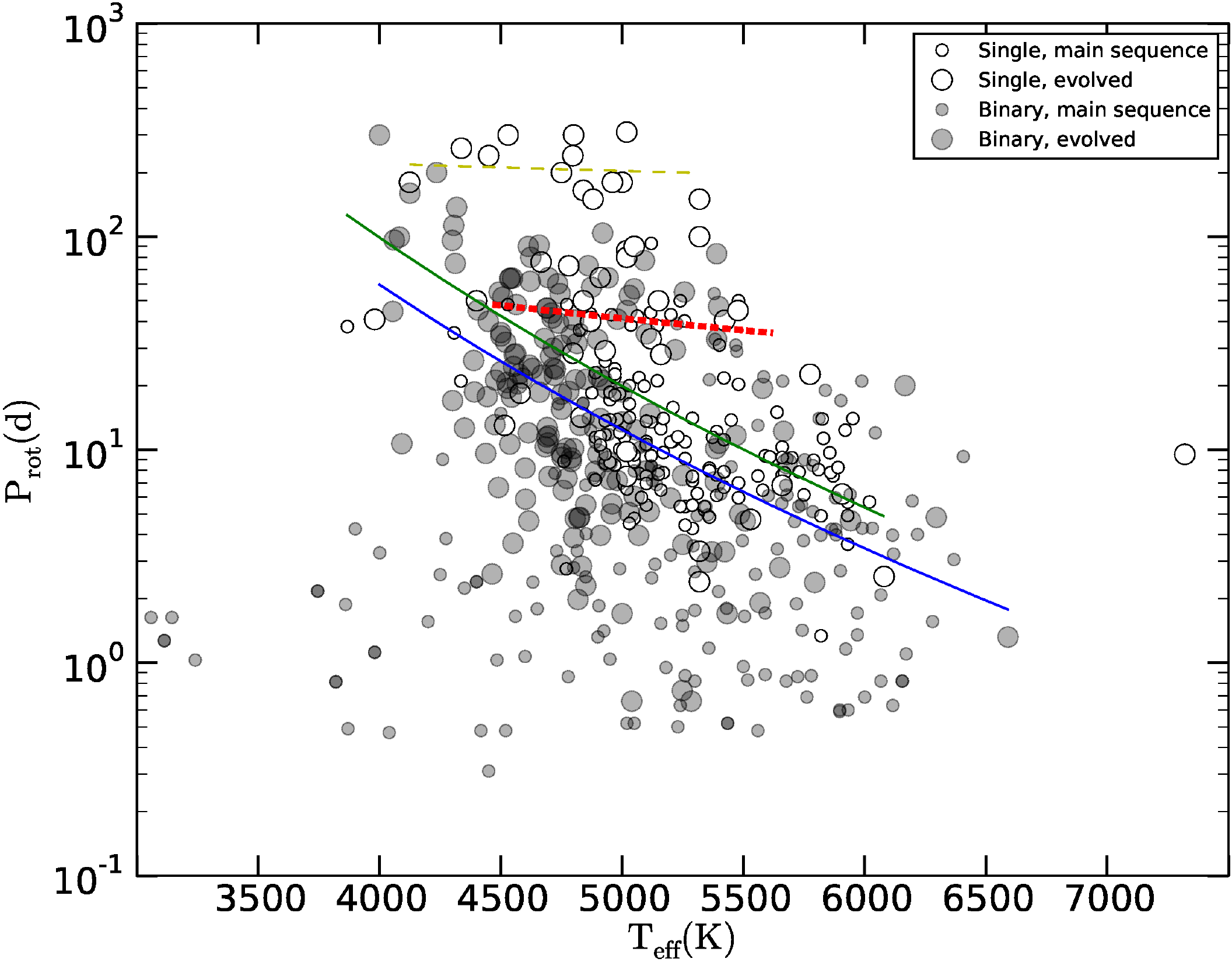}
\caption{The relation between rotation period and effective
temperature for active single stars (open symbols) and for binaries
(filled symbols). Our program stars are now mixed with other stars
and are not highlighted. The two symbol sizes indicate a star's
luminosity class. Larger symbols are for evolved stars of class III
and IV, smaller symbols are for main sequence stars of class V. The
short dashed line is the fit to the I-sequence dwarf stars from
B\"ohm-Vitense (\cite{boehm07}) while the long dashed line is our
fit to the weakly active (single) giant stars. The two full lines
indicate the $T_{\rm eff}^{-X}$ fits to the single active stars
(upper line) and the evolved binaries (lower line). See text.
}\label{Fy}
\end{figure}

\subsection{The dependency of rotation on effective temperature}

Most stars in this paper are within the temperature range
4000--6000~K, i.e. roughly M0 to G0. Such stars undergo the more
magnetic braking the stronger their initial magnetic field (c/o
Gray \cite{gray} and the abundant literature cited therein). To
first order, this should be proportional to the effective
temperature. Our first step is thus to extract the relation
between rotation period and effective temperature, in particular
for the active binary sample where we could expect a
qualitatively different behavior due to the
spin-orbit coupling. A similar attempt for single main-sequence G
and K stars led B\"ohm-Vitense (\cite{boehm07}) to find a strong
and systematic increase of $P_{\rm rot}$ with decreasing $T_{\rm
eff}$ for ``active stars'' but at most a very weak increase for
the ``inactive stars'' of form $T_{\rm eff}^{-1.33}$. Its cause is
not so clear but ascribed to an angular momentum transfer from the
surface to the interior due to some deep mixing.

Fig.~\ref{Fy} shows the distribution of $P_{\rm rot}$ versus $T_{\rm
eff}$ in our full active-star sample. Also shown are the single
dwarf stars used by B\"ohm-Vitense (\cite{boehm07}), drawn from the
Saar \& Brandenburg (\cite{saa:bra}) sample, and the young solar
analogs by Gaidos et al. (\cite{gai:hen}). We also add the single
stars with H\&K emission from our paper~I where we had a photometric
period determined as well as the single giants from Strassmeier et
al. (\cite{str:han}). From the latter sample of 12 single giants,
only two had a photometric (or S-index) period that we interpreted
to be the rotation period ($\delta$~CrB, HD\,160952), the others
were based on $v\sin i$ measures and an estimate of the radius from
the revised Hipparcos parallax. Note that we dropped HD\,181943 from
the sample (single K1V star seen almost pole on) because the
photometry likely shows a spot-variability time scale rather than
the rotational period (e.g. Fekel \& Henry \cite{fek:hen95}).

\begin{figure}
\center
\includegraphics[angle=0,width=80mm,clip]{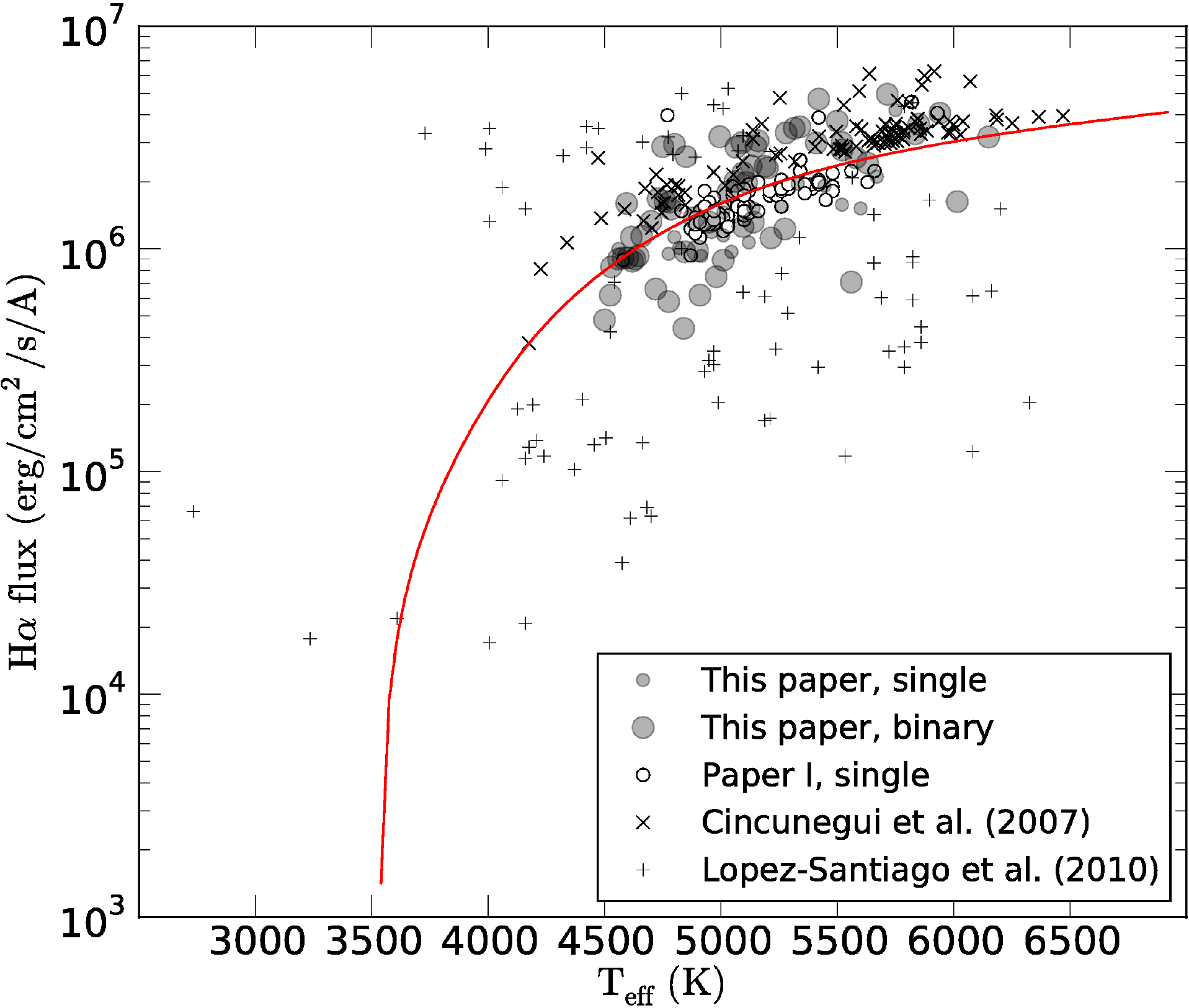}
\caption{Average \Halpha -core flux as a function of effective
temperature. Our binary sample is shown as large grey dots while the
single stars are shown as small grey dots. The single stars
from our paper~I are shown as open circles. The single-star sample
from Cincunegui et al. (\cite{cin:dia}) is shown as crosses while
the stars in the L\'opez-Santiago et al. (\cite{lop:mon}) catalog
are shown as pluses. The line is the minimum-flux envelope derived by
Cincunegui et al. (\cite{cin:dia}.}\label{Fxx}
\end{figure}

In Fig.~\ref{Fy} the four subgroups -- single stars on the main
sequence, evolved single stars, binary components on the main
sequence, and evolved binary components -- appear well separated
from each other, as expected due to the different radii and the
tendency towards synchronization in binaries. The longest rotation
periods are found for giants, the shortest periods for dwarfs, while
the binary components are generally more rapidly rotating than
single stars. For stars with $T_{\rm eff}<$4000~K, one has only
dwarf stars to work with because M giants exhibit mostly pulsation
periods rather than rotational periods. However, the strong increase
of $P_{\rm rot}$ with decreasing $T_{\rm eff}$ is obvious in
our sample. Two simple functional fits to the single stars and the
binaries, respectively, indicate that the steep increase is not only
seen in single stars but also in binary components. Their slopes are
similar or possibly even identical, as expected if there is a common
mechanism. We find
\begin{eqnarray}
 P_{\rm rot} &=& (27.9\pm 3.2) \ \ T_{\rm eff}^{-7.2\pm 0.9} \ \ \ \  {\rm for \ singles,}\\
 P_{\rm rot} &=& (27.1\pm 3.5) \ \ T_{\rm eff}^{-7.0\pm 0.9} \ \ \ \  {\rm for \ binaries.}
\end{eqnarray}
This is to be compared with the active (A) sequence stars in
B\"ohm-Vitense (\cite{boehm07}).

A separate fit to the inactive, or only very weakly active, single
giants and subgiants with $P_{\rm rot}>$100~d yields a slope of
$T_{\rm eff}^{-1.12}$, very similar to the inactive (I) sequence fit
for dwarf stars in the B\"ohm-Vitense (\cite{boehm07}) sample, just
offset towards longer periods. Note that we plot her fit for
inactive dwarfs as a dashed line in Fig.~\ref{Fy} for comparison.
Our sample also shows a void of single stars for a period range of
$\approx$15--30~d for $T_{\rm eff}>$5250~K, although not overly
striking.

\begin{figure}[!tb]
\center
\includegraphics[angle=0,width=70mm]{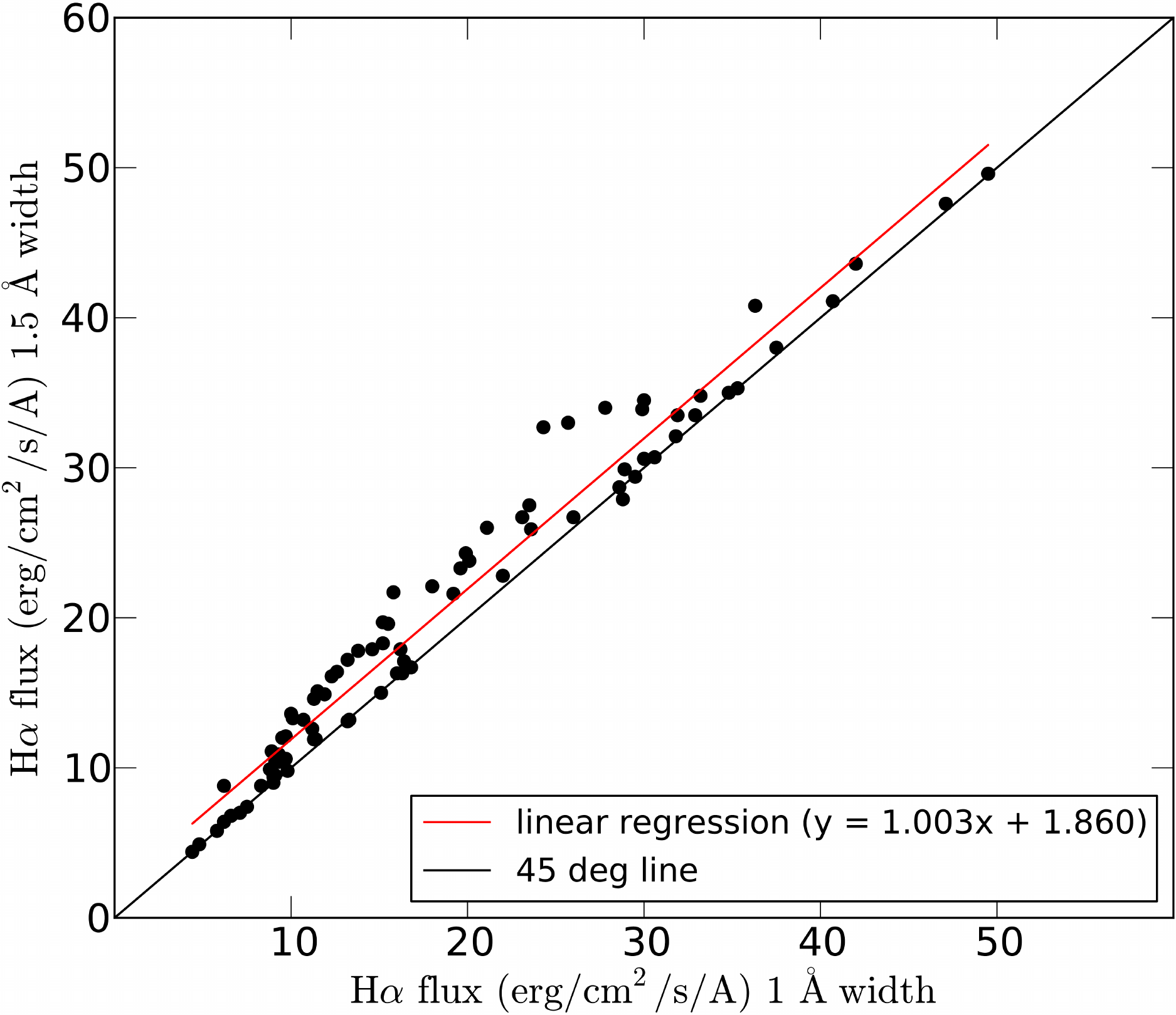}
\caption{\Halpha\ flux measurements of our sample with two
different central bandwidths (1~\AA\ and 1.5~\AA ). The short line
is a linear regression fit, the other line is just a
45-deg line to guide the eye. Both slopes are identical but the
offset is 2\,10$^5$ erg\,cm$^{-2}$s$^{-1}$\AA$^{-1}$.
}\label{F-ff}
\end{figure}

\subsection{The \Halpha -line core as an activity indicator}

The \Halpha -line profile is influenced by a large range of
atmospheric optical thicknesses, Doppler widths, and photospheric
line strengths, thereby stretching its formation from the
photosphere and the chromosphere to the transition region with the
corona. In normal FGK stars, \Halpha\ is dominated by
photoionization. Collisions with free electrons are a significant
source for the cooler stars and the line becomes particularly
sensitive for mass motions and all sorts of (non-thermal) velocity
fields (e.g., Cram \& Mullan \cite{cra:mul}). Moreover,
circumstellar hydrogen could add an extra component to the net
\Halpha\ absorption profile. For the average quiet Sun the core of
the line forms at about 1500~km and is thus chromospheric in origin.
However, extracting just the non-thermal (=activity) contribution
from the line core is not unique and neither is its interpretation.
Particular contributors are chromospheric inhomogeneities like
bright plages, chromospheric network variations, localized in- and
outflows associated with flares (nano, micro and macro versions)
and, of course, global winds. Cincunegui et al. (\cite{cin:dia})
pointed out that despite they had found a strong relationship
between the mean fluxes in \Halpha\ and Ca\,{\sc ii} H\&K the
relation can break down for individual stars. They suggested that
the relationship is the product of the dependence on stellar color
rather than on similar activity phenomena.

\begin{figure*}
{{\bf a} \hspace{80mm} {\bf b}}\\
\includegraphics[angle=0,width=80mm,clip]{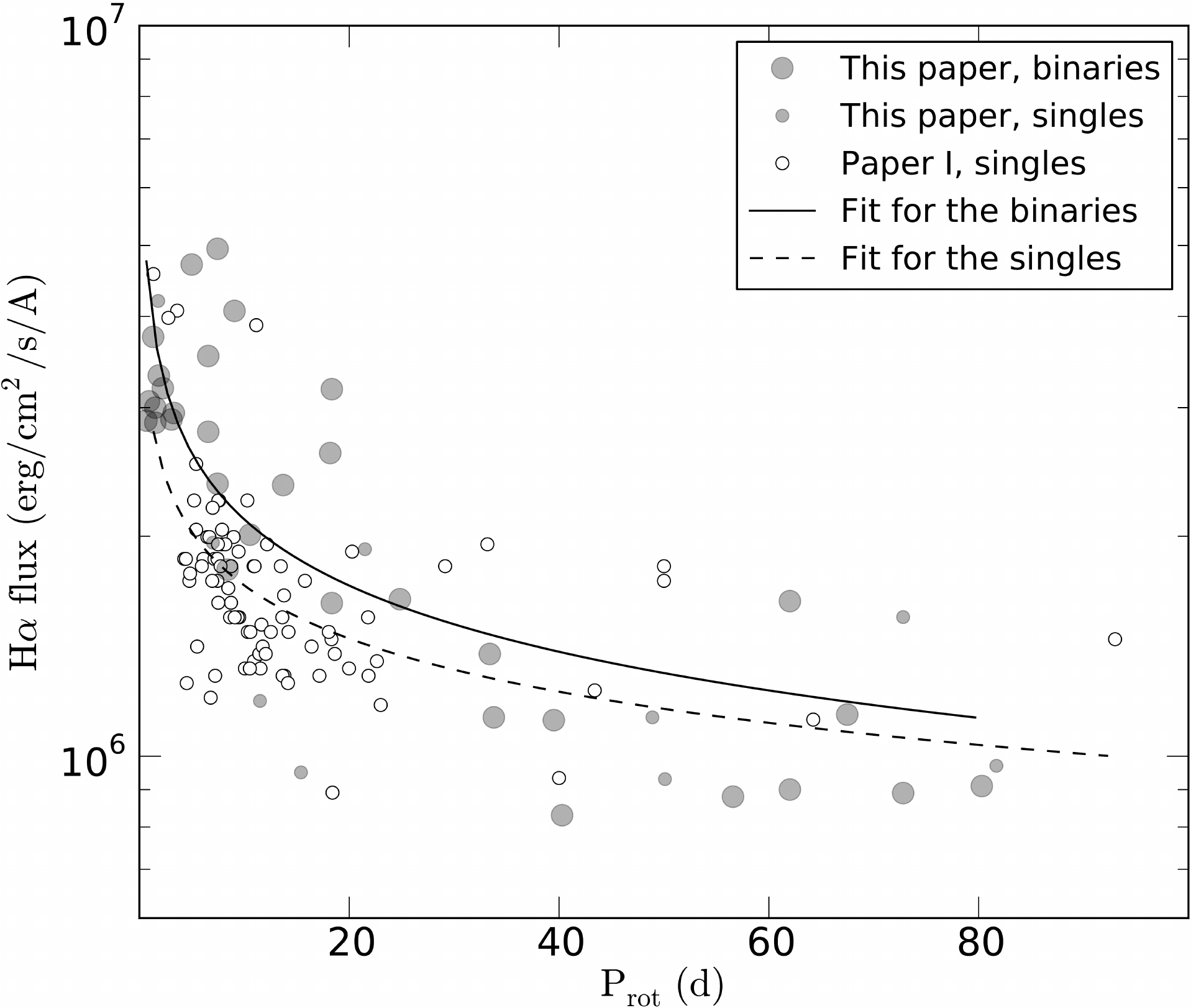}\hspace{5mm}
\includegraphics[angle=0,width=80mm,clip]{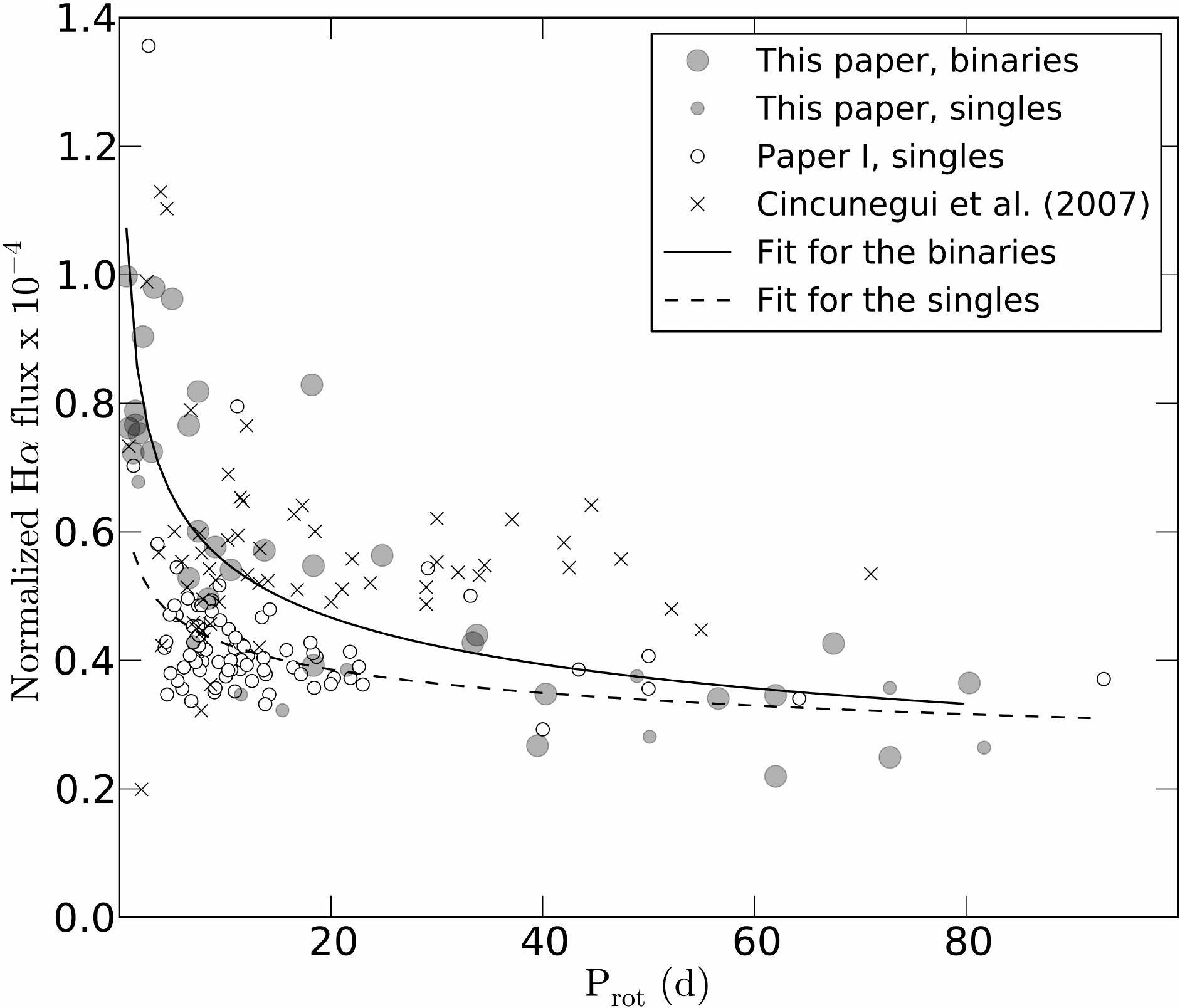}
\caption{{\bf a} Average \Halpha -core flux versus rotation period. {\bf b} The same flux but normalized to the bolometric luminosity $\sigma T_{\rm eff}^4$. The
grey scale dots are the singles and binary components from this paper. The other
symbols are single stars from sources as indicated. The lines are power-law fits as described in the text.}\label{Fz}
\end{figure*}

In Fig.~\ref{Fxx} we plot the \Halpha -core flux, obtained as
described in Sect.~\ref{S4-Ha}, versus effective temperature. Our
sample is made up of the 71 components of active binaries and 18
single stars from Table~\ref{T1} with temperatures obtained from
spectrum synthesis. These entries are complemented by the stars
from paper~I, the 109 southern stars in the study by Cincunegui et
al. (\cite{cin:dia}), and the averages for 86 stars from the
survey of L\'opez-Santiago et al. (\cite{lop:mon}). Unfortunately,
we have no stars in common with Cincunegui et al. and
L\'opez-Santiago et al. which makes an empirical determination of
the zero-point shift between the samples merely impossible. The
same is true for the observations of F8-K5 dwarf stars by Pasquini
\& Pallavicini (\cite{pas:pal}), also done in the southern
hemisphere, while our survey was done in the northern hemisphere.
Cincunegui et al. (\cite{cin:dia}) were able to connect their zero
point with that in Pasquini \& Pallavicini (\cite{pas:pal}) for
the 24 stars in common and found excellent agreement, suggesting
fluxes good to within 20\%. Nevertheless, we did not add the
Pasquini \& Pallavicini (\cite{pas:pal}) sample in our
Fig.~\ref{Fxx} because it adds additional uncertainty due to them
using a mix of $V-R$ and $B-V$ colors. A zero-point difference
between our fluxes and those of Cincunegui et al. (\cite{cin:dia})
is obvious for the single stars in the sense that our fluxes
appear lower by approximately 50\%. We note that Cincunegui et al.
(\cite{cin:dia}) and Pasquini \& Pallavicini (\cite{pas:pal})
measured \Halpha -core fluxes from a bandwidth of 1.5~\AA\ and
1.7~\AA , respectively, while we used a 1.0-\AA \ bandwidth. It
affects how much basal flux is being included in the measurement,
which could be significantly different for the cooler stars as
well as for the subgiants and giants. To quantify this, we have
remeasured our spectra with a 1.5-\AA\ bandwidth and compared them
to the 1.0-\AA\ fluxes (Fig.~\ref{F-ff}). We found a basically
constant offset of 2$\pm$2\,10$^5$
erg\,cm$^{-2}$s$^{-1}$\AA$^{-1}$ (where the error is the rms of
the linear fit) and no dependency on color nor on gravity. A
correction of this amount was then applied to the Cincunegui et
al. fluxes before plotting in Fig.~\ref{Fxx}.

Quite different is the case for the stars observed by
L\'opez-Santiago et al. (\cite{lop:mon}). Their fluxes were
obtained after the subtraction of bona-fide inactive standard-star
spectra and likely represent only the active part of the
chromosphere. This is markedly different to all other fluxes which
contain the photospheric flux but also the basal chromospheric
flux. We converted the $B-V$s in their tables to an effective
temperature with the calibration of Flower (\cite{f96}) and plot
them as small pluses in Fig.~\ref{Fxx} for comparison. The
L\'opez-Santiago et al. fluxes per temperature bin are on average
an order of magnitude lower than the other measures, as expected.

Fig.~\ref{Fxx} also shows that the binary components follow
basically the same trend as single stars but exhibit twice as much
scatter as the single stars. This is expected because the
temperature errors for the binary components are typically also
twice as large, in particular for the SB2s. The line in
Fig.~\ref{Fxx} is the minimum \Halpha\ flux for single stars found
by Cincunegui et al. (\cite{cin:dia}) from a quadratic polynomial.
It essentially represents the photospheric contribution. Our
measurements have a number of stars with fluxes that fall below
this line. These are almost exclusively the weak-lined secondary
components in SB2s with comparably uncertain flux measurements.

\subsection{An activity-rotation relation for magnetically active
binaries}

Would a binary component with an equal rotational period of a single
star be more active just because it is a component in a tidally
interacting binary? Or is rapid rotation in binaries just an unique
evolutionary situation because they are speed up to stellar
rotations that otherwise would never be reached during regular
single-star evolution? Tidal coupling, expressed as the ratio of
stellar to Roche radii, was ruled out as a direct factor for
increased activity in binaries by Basri (\cite{basri87}) despite
that the evolved binary components in his sample were always more
active for a given period than a single star (Basri et al.
\cite{basri85}). Yet other authors had claimed such a dependency
(e.g. Glebocki \& Stawikowski \cite{gle:sta2}). More recently, Dall
et al. (\cite{dall1}) tried to find a relation between binarity,
magnetic activity, and chemical surface abundances by comparing
observations of two selected cool stars in great detail. What they
found was a correlation between the bisector inverse slope and the
activity index $\log R_\mathrm{HK}$, which both vary in phase with
tiny (few \ms ) radial velocity variations. In that paper, Dall et
al. (\cite{dall1}) speculated towards the existence of an unseen
very-low mass star or even planetary companion as the cause of the
low-amplitude velocity variations, but in their follow-up paper of
the target in question, Dall et al. (\cite{dall2}) unambiguously
verified these to be due to stellar oscillations with an
$\approx$50-min period. We conclude that a straightforward
comparison of two well-selected stars, even with (nearly) equal
rotation periods, may not be conclusive.

From statistical attempts, weak activity correlations with
rotation for binaries were reported in the literature. Simon \&
Fekel (\cite{sim:fek}) found a marginally decreasing UV flux with
increasing period for evolved binaries. Strassmeier et al.
(\cite{str90}) claimed a slight trend for Ca\,{\sc ii} H\&K, while
Dempsey et al. (\cite{demp}) saw a similar trend for the Ca\,{\sc
ii} infrared triplet. Strassmeier et al. (\cite{str:han})
determined a relation for Ca\,{\sc ii} H\&K flux in giants versus
rotational velocity and temperature and emphasized the large range
of fluxes for a given rotation rate. Also, the overall consensus
was that the flux for evolved binaries with periods longer than
10~d is fairly constant and only increases significantly for stars
with periods below $\approx$10~d.


In Fig.~\ref{Fz}a, we employ the \Halpha -core flux, ${\cal
F}_{\rm H\alpha}$, as an activity measure and plot our program
stars in the log-normal activity-rotation plane. The sample is
expanded by those stars from paper~I that had a photometrically
determined rotation period. The figure shows that the \Halpha
-core flux also levels out with longer periods; for single stars
near a period of 10~d, for binaries more near periods of 20~d,
comparable to what has been seen from other activity indicators.
It also shows that binary components appear generally more active
than single stars for a given period bin. Notice that in
Fig.~\ref{Fz} only those SB2s are plotted where the photometric
rotation period can be assigned unambiguously to the primary star.
The secondary stars were not included in the fit for the
spectroscopic binaries either because their rotation periods are
assumed from $v\sin i$ and a radius instead of observed, as for
the primaries. The lines in Fig.~\ref{Fz}a are simple power-law
fits of form ${\cal F} = a \ P^x$.

Fig.~\ref{Fz}b shows the same flux as in panel~a but normalized to
the bolometric luminosity $\sigma T_{\rm eff}^4$,
\begin{equation}
R_{\rm H\alpha}  =  {\cal F}_{\rm H\alpha} / \sigma \ T_{\rm eff}^4 \ .
\end{equation}
Effective temperatures for
the majority of stars were determined spectroscopically from our
SES spectra (see Table~\ref{T7}) or, for the single
stars from paper~I, from the color-temperature calibration from
Flower (\cite{f96}). We also include in the single-star fit the
targets from Cincunegui et al. (\cite{cin:dia}) whenever a
rotation period from photometry was known and an effective
temperature given. These data are plotted as crosses. Note that
the CABS stars have no \Halpha -flux determination as an ensemble
but just individual measurements collected from the literature
with a variety of calibration issues and are not included here.
Again, the lines are power-law fits of the form $R = a \ P^x$. The
results from Fig.~\ref{Fz}a and Fig.~\ref{Fz}b are
\begin{eqnarray}\label{EFP}
{\cal F}_{\rm H\alpha} & = (4.22^{+0.43}_{-0.39}\, 10^{+6}) \ P_{\rm rot}^{-0.30\pm0.05} \ \ \ \  {\rm binaries,}\\
{\cal F}_{\rm H\alpha} & = (2.99^{+0.23}_{-0.22}\, 10^{+6}) \ P_{\rm rot}^{-0.24\pm0.04} \ \ \  {\rm singles,}\\
R_{\rm H\alpha}        & = (9.72^{+0.67}_{-0.22}\, 10^{-5}) \ P_{\rm rot}^{-0.24\pm0.03} \ \ \ \  {\rm binaries,}\\
R_{\rm H\alpha}        & = (5.92^{+0.34}_{-0.32}\, 10^{-5}) \ P_{\rm rot}^{-0.14\pm0.03} \ \ \  {\rm singles .}
\end{eqnarray}

The rms values for the fits in Eq.~\ref{EFP}--12 are, from top to
bottom, 0.920e+6, 0.522e+6, 1.353e-5, and 1.105e-5, respectively.
We see that the fluxes are only weakly dependent on the rotation
period, in particular when compared to the $P^{-1.33}$ dependency
for inactive single stars from Ca\,{\sc ii} H\&K (B\"ohm-Vitense
\cite{boehm07}). This is partly because the magnetic flux seen in
the \Halpha\ core is dispersed over a larger range of optical
depths at which a variety of (local) velocities prevail. However,
our fit demonstrates the existence of a period-activity relation
also for \Halpha\ and that binary components are stronger
dependent on period than single stars.

\subsection{An activity-Li abundance relation?}\label{S6c}

\begin{figure*}
{{\bf a} \hspace{80mm} {\bf b}}\\
\includegraphics[angle=0,width=80mm,clip]{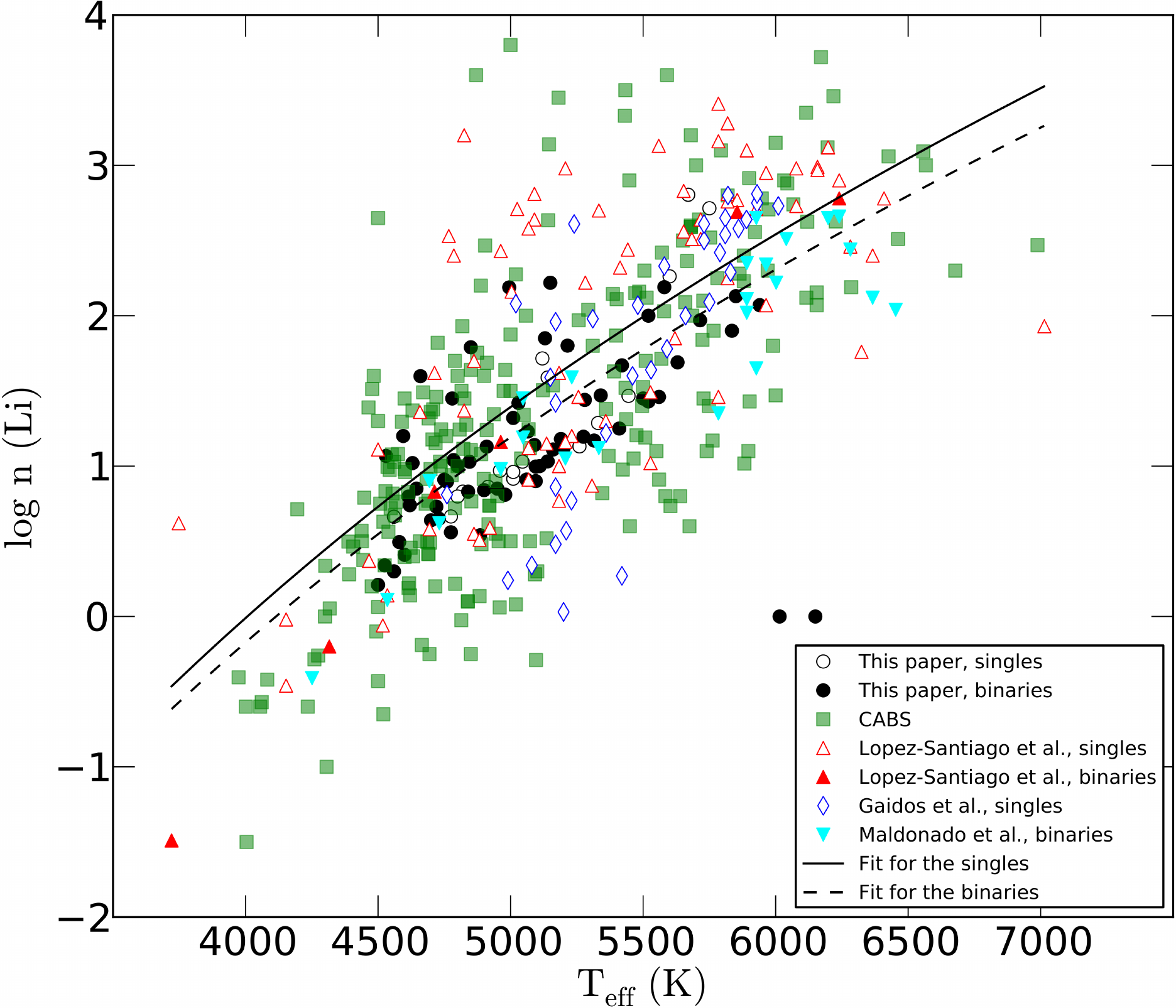}\hspace{5mm}
\includegraphics[angle=0,width=80mm,clip]{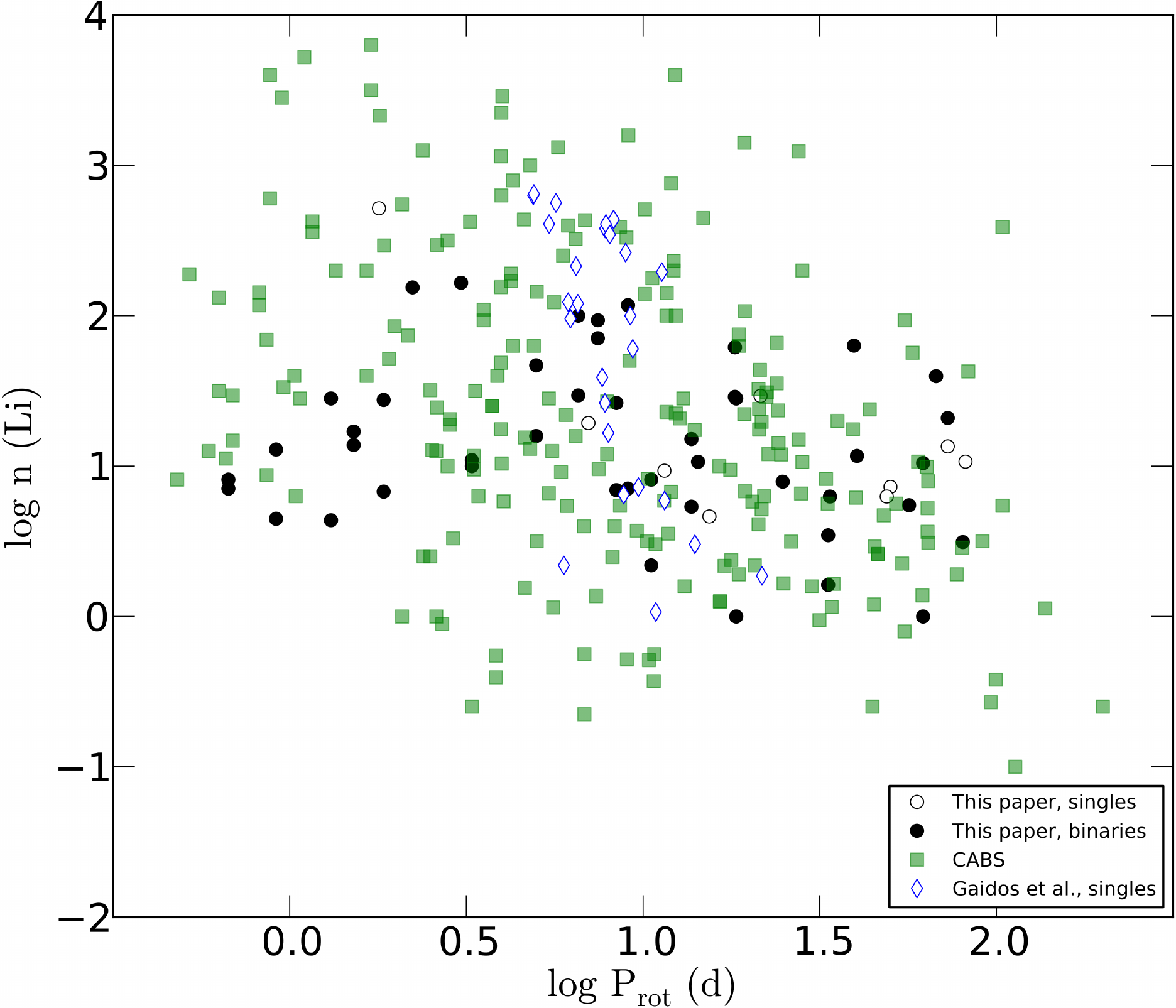}
\caption{Average lithium abundances {\bf a} versus effective
temperature,  {\bf b} versus (logarithmic) rotation period. Filled symbols refer
to binaries, open symbols to single stars. The symbols and
the sample sources are identified in the inserts.
Active binaries show generally lower Li abundances than comparable singles.
Our sample of binaries and single (field) stars exhibits
a dispersion of up to 2--3 orders of magnitude in the full range 4500--6000~K
that peaks at $\approx$5000~K. When plotted versus (logarithmic) rotation period, the dispersion is even 3--4 orders of magnitude.}\label{Frotli}
\end{figure*}

In Fig.~\ref{Frotli}a and Fig.~\ref{Frotli}b the lithium abundance
relative to hydrogen is shown as a function of effective
temperature and as a function of rotational period, respectively.
The sample consist of our 18 singles and 42 binaries from
Table~\ref{T7} (the triple systems were omitted), 236 of the 408 CABS binaries that have a Li
abundance, plus 75 out of the 109 single stars from paper~I.  Also
included in Fig.~\ref{Frotli} are the young solar analogs from
Gaidos et al. (\cite{gai:hen}). Seven
stars in the CABS catalog were in common with the more recent Li
study by Maldonado et al.~(\cite{mal:mar}) but only one of them
had a reasonable lithium equivalent-width measure (HIP\,78843),
which we converted to a logarithmic abundance of 0.72 with $T_{\rm
eff}$=4790K and the NLTE tables in Pavlenko \& Magazz\'u
(\cite{pav:mag}). The same conversion was applied to the averaged
equivalent-width measures for 78 single stars and 6 binaries from
Lopez-Santiago et al. (\cite{lop:mon}) and 153 singles and 24
binaries from Maldonado et al.~(\cite{mal:mar}) and added to the
sample in Fig.~\ref{Frotli}a. For these stars, only $B-V$ colors
are listed and $T_{\rm eff}$'s are consequently obtained from the
$B-V$ calibration of Flower (\cite{f96}). Despite that $B-V$
colors are affected by metallicity (see, e.g.,
Gray~\cite{gray94}), and that at least Maldonado et
al.~(\cite{mal:mar}) obtained metallicities from their spectra for
most of their targets, the typical metallicity range for our stars
is at most $\pm$0.3~dex with the one or other outlier. This would
affect the temperature scale at most by $\pm$50~K and therefore,
we decided not to correct for metallicity simply because
metallicities were not known for all stars in our samples.

\subsubsection{The lithium-temperature dependency}

Figure~\ref{Frotli}a shows a generally increasing Li abundance
for binaries with increasing effective temperature (and
therefore likely with increasing mass). Such a result was found
earlier for single stars in various open clusters, most noteworthy
in the Hyades, Pleiades, Praesepe, M34, and M67 (see, e.g., Jones et
al.~\cite{jon:fis}). Our binary sample shows increasing
dispersion of Li abundance with increasing temperature, opposite
to what was found for Pleiades stars (e.g. Soderblom et
al.~\cite{sod:jon}) but in agreement with the Hyades FGK stars
(e.g. Thorburn et al. \cite{tho:hob}). It reaches up to
three orders of magnitude for binaries with $T_{\rm
eff}\approx$5000~K, i.e. 15--20$\sigma$ of the observational
uncertainty, and still two orders of magnitude between
4500--5000~K. The dispersion is still one order of magnitude at the cool end at
$T_{\rm eff}<4500$~K. For the binaries hotter than 5000~K, the
maximum observed Li abundances are primordial
($\approx 3.5$) and the dispersion decreases due to the generally
higher abundances with temperature. Part of the explanation for
such a drastic dispersion is certainly that our binary sample is a
mix of ages, most likely ranging from ZAMS binaries to maybe 10~Gyr systems.
This is evidenced by the fact that our single-star sample in
Fig.~\ref{Frotli}a shows a qualitatively comparable behavior but the dispersion never reaches much above two
orders of magnitude.

Binaries have larger systematic errors of
the equivalent-width measures than singles, and that certainly also adds to
the dispersion. It is particularly due to the double-lined nature for
some binaries and their relative continuum uncertainties and
associated larger errors for $T_{\rm eff}$. However, we expect equivalent-width
errors never larger than 10\%\ for stars with equivalent widths
$>50$~m\AA\ and at most $\approx$30\%\ for equivalent widths
$<20$~m\AA\ while effective temperatures are good to within, say,
$\pm$50--70~K for singles and SB1s, and $\pm$100--150~K for SB2s. We
also note that we use Li detections and never just upper limits, and
use temperatures determined from spectrum synthesis rather than
photometry. Our Li abundances are therefore generally good to within
$\pm$0.2~dex.  Maybe other -- extrinsic -- factors may affect the
Li abundance to an unknown amount but were deemed unreasonable by
Soderblom et al.~(\cite{sod:jon}); c/o their discussion. We
conclude that a significant part of the Li-abundance dispersion in active
binaries must be intrinsic.

\subsubsection{The lithium-rotation dependency}

If the depletion mechanism(s) were dependent on stellar rotation,
the expectation is that tidally locked binaries would exhibit
lower surface abundances than single stars (of same mass and age).
If independent of rotation, one could expect higher Li abundance
in binaries because of less depletion. There is evidence in Fig.~\ref{Frotli}a that binaries exhibit on average less surface lithium than singles.  Log-linear fits to the data of the form $\log n (\mathrm{Li}) = a T_{\rm eff}^b + c$  provide the following average relations;
\begin{eqnarray}\label{LT}
\log n (\mathrm{Li})  & = a \ T_{\rm eff}^{+0.0010405} \ - \ b \ \ \ \  {\rm for \ binaries,}\\
\log n (\mathrm{Li})  & = c \ T_{\rm eff}^{+0.0076945} \ - \ d \ \ \ \  {\rm for \ singles,}
\end{eqnarray}
where $a,b,c,d$ are the constants 58.29306\,$10^{2}$,  58.8000$\,10^{2}$, 7.66085\,$10^{2}$, 8.16580\,$10^{2}$, respectively. The difference between the two fits is significant for our sample and amounts to $\approx$0.25~dex at 5500~K. However, the fits should not to be taken too literally and act just like a guideline.

The general picture is that
the Li depletion is related to the star's angular momentum loss,
itself related to internal flows and shear instabilities, mixing,
as well as external winds (e.g.  Pinsonneault et al.
\cite{pin:kaw}). A crystal-clear picture still has to emerge but
that magnetic fields and winds play a major role has been accepted
widely and new numerical simulations are going in this direction
(e.g. Matt et al.~\cite{matt}). The rotational angular momentum
must be thought of being coupled to the orbital angular momentum
in binaries where the depletion mechanism(s) can chew on
the orbital energy. There is some observational evidence that
tidally locked binary systems in the Hyades and M67 have larger
lithium abundances than single stars though (see discussion in
Barrado~y~Navascues et al. \cite{bar:del}) but, so far, did not
amount to a conclusive (quantitative) relation.

In Fig.~\ref{Frotli}b, we plot Li abundance versus rotational period. Rotation periods range between 0.5~d and 200~d. Because the vast majority of the binaries in our sample cluster below or near periods of $\approx$10~d, we had to choose a logarithmic presentation for the period axis as well. Nevertheless, the log-log plot shows a clear trend of higher Li abundance with faster rotation, for binaries as well as for singles, main sequence or evolved. Again, as in the Li-$T_{\rm eff}$ plane, the dispersion is very large, peaking at 3--4 orders of magnitude at around a rotation period of 10~d. This effectively camouflages a clear Li-$P_{\rm rot}$ relation, in particular because we do not know ages for our systems as one does for open clusters. As an averaged guideline, we fit linears to the binary data in the log-log plane and obtain
\begin{eqnarray}\label{LP}
\log n (\mathrm{Li})  & = a \ \log P_{\rm rot} \ + \ b \ \ \ \  {\rm for \ binaries,}
\end{eqnarray}
where $a,b$ are the constants $-0.60067$ and $1.84783$, respectively. No fit is given for the single stars because our sample is not complete due to a general lack of rotation periods. Jones et al. (\cite{jon:fis}), together with parallel and subsequent work by
several other groups, presented clear evidence that the (single) rapid
rotators in the Pleiades (age $\approx$100~Myr) have generally
also higher Li abundances than their slowly rotating cousins, at least
for the temperature range 4400--5400~K. A similar trend appeared
in the data for M34 (age $\approx$300~Gyrs) but only for stars of
$\approx$4500~K, while the Hyades (625~Myr) shows no rapid rotators
anymore and thus no Li relation with rotation.

Several mechanisms were suggested to explain the high Li
abundances in rapidly rotating, active stars. Among them the
production of Li during coronal flares through spallation
reactions, or through an underestimated surface temperature due to
large amounts of cool starspots (see, e.g., Pallavicini et al.
\cite{pal:cut}) or simply through the explanation that active
stars are generally being young (Randich et al. \cite{ran:gra}).
Another interesting explanation for tidally coupled binary
components is through a lack of rotationally induced mixing due to
the little magnetic braking compared to single stars (Pinsonneault
et al. \cite{pin:kaw}). B\"ohm-Vitense (\cite{boehm}) noted that
for giants with $T_{\rm eff}$ less than $\approx$6300~K the
reduced lithium surface abundance appears related with the
decreased surface rotation. She attributes this to effects from
deep mixing. Just recently, Takeda et al. (\cite{tak:hon})
announced evidence for a (positive) correlation of Li abundance
with rotational velocity in solar-analog stars. A depletion of two
orders of magnitude in Li abundance is seen in the range of $v\sin
i$ from 4~\kms\ to 1~\kms\ according to their Fig.~5. It calls for
a \emph{very} efficient envelope mixing process, possibly also
related to the exterior angular momentum carried by planets.

Accretion of already a few Earth masses during early main-sequence
phase would bring up the $^7$Li abundance in these stars and a
proper amount of mixing may prevent $^6$Li from destruction (Murray
et al. \cite{murr}). Such planet engulfing would be in agreement
with the observation of increased Li abundance in rapidly-rotating
active stars but requires the detection of $^6$Li in active stars.
Such a detection was reported by Israelian et al. (\cite{isr}) for
the planet-hosting solar-type star HD\,82943 or the active subgiant binary HD\,123531 by Strassmeier et al. (\cite{hd123351}) (but see Cayrel et al.
\cite{cayrel} for a warning of such detections).

\section{Summary and conclusions}\label{S7}

We present and analyzed high-resolution spectra from the STELLA
robotic telescope and determined orbital elements for 45 binaries
in the period range 0.29--6450~d, many for the first time but all
with substantially increased precision and likely also increased accuracy.
Additional photometric monitoring of many of these stars with our APTs allowed the
determination of precise rotation periods.

We summarize our findings as follows.
\begin{itemize}
\item STELLA/SES achieves a radial-velocity precision for an individual
radial-velocity observation of $\approx$30~\ms\ without an iodine cell or
a simultaneously recorded comparison spectrum. It enables rms values
of as small as $\approx$40~\ms\ for fits of orbital elements. For
one of the favorable cases in this paper (the SB2 binary HIP~77210 with
$P_{\rm orb}$=9.9~d) the minimum primary mass is precise to 0.12\%\
and the secondary mass to even 0.07\% .
\item The radial-velocity zero-point difference between STELLA and
CORAVEL measurements is +0.503~\kms .
\item The bulk (74\%) of rapidly-rotating active stars in binaries are
synchronized and in circular orbits.
\item About 26\%\ (61 targets) of the binaries in our full sample are
asynchronous rotators. About half of them have $P_{\rm rot}>P_{\rm
orb}$ and of these, all but two have $e>0$. As rotational
synchronization is predicted to occur before orbital
circularization active binaries must have gone through an extra
spin-down phase besides tidal dissipation and we suggest this to
be due to a magnetically channeled wind with its subsequent
braking torque. Such braking depends on the stellar magnetic-field
geometry and would result in very different braking efficiencies
for stars of otherwise nearly identical physical parameters.
\item We find a steep increase of $P_{\rm rot}$ for
lower $T_{\rm eff}$ in both single stars and binary components.
This verifies earlier claims for active single stars with
known cycles and shows that it is also the case in our
active binaries. A functional dependence of $P_{\rm rot}\propto
T_{\rm eff}^{-X}$ is suggested, where $X=7.2$ for single stars and
$X=7.0$ for binary stars. The difference in $X$ is not significant
though.
\item A relation between \Halpha -core flux and rotational period
is evident for both single stars and binaries. Power-law fits
suggest $R_{\rm H\alpha} \propto P_{\rm rot}^{-0.24}$ for binary
stars and $R_{\rm H\alpha} \propto P_{\rm rot}^{-0.14}$ for single
stars, where $R$ is \Halpha -core flux normalized to $\sigma
T_{\rm eff}^4$. The difference of the exponents is only weakly significant on the 3$\sigma$ level.
\item Our data suggest that Li abundances of active binary components also increase with effective temperatures, as known for single stars. However, binaries appear to have on average 0.25~dex lower Li abundances than single stars of same effective temperature, which is what we expect if the depletion mechanism is rotationally enhanced in binary stars. The dispersion for binaries
appears to peak around $T_{\rm eff}\approx$5000~K,
amounting to three orders of magnitude. This dispersion decreases
towards both ends of the temperature range between 4500~K to
6500~K. Our single-star sample follows this trend as well but with
an order of magnitude smaller dispersion. We can not
separate age effects in our sample but attribute a significant
fraction of the dispersion to it.
\item Rotational dependency of Li abundance in binaries is also evident and suggests a law of form $\log n (\mathrm{Li}) \propto -0.6 \ \log P_{\rm rot}$ as a guideline. However, a dispersion of up to 3--4 orders of magnitude explains why no quantitative relations were found so far. The dispersion appears largest for binaries with periods shorter than $\approx$10~d and still amounts to 2--3 orders of magnitude for periods larger than $\approx$10~d.
\end{itemize}

\acknowledgements We acknowledge the extremely useful services of
ADS/SAO-NASA and CDS/Strasbourg. The STELLA project was funded by
the Science and Culture Ministry of the German State of
Brandenburg (MWFK) and the German Federal Ministry for Education
and Research (BMBF). Its a pleasure to thank Carlos Allende-Prieto
and Lars Koesterke for their help implementing PARSES and Janos
Bartus for his help with some of the figures. Discussions with Sydney Barnes are also appreciated. We also thank an anonymous referee for the many comments that made this paper a better one.

\end{document}